\documentclass[a4paper,12pt]{article}

\usepackage{a4}
\usepackage{graphics}
\usepackage{epsfig}
\usepackage{amssymb}
\usepackage{color}
\usepackage{cite}
\usepackage{epsf}
\usepackage{axodraw}
\usepackage{wasysym}

\definecolor{green3}{rgb}{0.,0.7,0.0}
\definecolor{red1}{rgb}{0.9,0,0}

\def\lsim{\raise0.3ex\hbox{$\;<$\kern-0.75em\raise-1.1ex\hbox{$\sim\;$}}}
\def\gsim{\raise0.3ex\hbox{$\;>$\kern-0.75em\raise-1.1ex\hbox{$\sim\;$}}}
\DeclareMathAlphabet{\scr}{U}{rsfs}{m}{n}
\begin{document}
\hspace*{\fill} MAN/HEP/2009/17\\
%-----------------------------------------------------------------------------
\vspace{1.5cm}
\begin{center}
{\Large\bf 
Probing SUSY CP Violation in\\[2mm]Two-Body Stop Decays at the LHC
}                       
\vspace{1.5cm}

{\sc Frank~F.~Deppisch\footnote{Email: frank.deppisch@manchester.ac.uk}}\\
\vspace*{.2cm} 
{\small \it School of Physics and Astronomy, University of Manchester, \\
Manchester M13 9PL, United Kingdom
}
\vspace*{1.0cm}
            
{\sc Olaf~Kittel\footnote{Email: kittel@th.physik.uni-bonn.de}}\\
\vspace*{.2cm} 
{\small \it 
Departamento de F\'isica Te\'orica y del Cosmos and CAFPE, \\
Universidad de Granada, E-18071 Granada, Spain
}
\vspace*{.5cm}
\end{center}

%------------------------------------------------------------------------------
\begin{abstract}\noindent
We study CP asymmetries in two-body decays of top squarks into neutralinos and sleptons at the LHC. These asymmetries are used to probe the CP phases possibly present in the stop and neutralino sector of the Minimal Supersymmetric Standard Model. Taking into account bounds from experimental electric dipole moment searches, we identify areas in the mSUGRA parameter space where CP asymmetries can be sizeable and discuss the feasibility of their observation at the LHC.  As a result, potentially detectable CP asymmetries in stop decays at the LHC are found, motivating further detailed experimental studies for probing SUSY CP phases.
\end{abstract}
%------------------------------------------------------------------------------
\newpage
%------------------------------------------------------------------------------

\section{Introduction}
\label{sec:Introduction}
%------------------------------------------------------------------------------

Supersymmetry~(SUSY)~\cite{mssm} is a  well motivated theory to extend the Standard Model~(SM) of particle physics. SUSY models are not only favored by gauge coupling unification and naturalness  considerations, but are also attractive from the cosmological point of view. For instance the lightest SUSY particle~(LSP) is a good dark matter candidate if it is stable, massive and weakly interacting~\cite{Goldberg:1983nd,RelicCP}.  Most interestingly SUSY models provide a number of new parameters, among some having physical phases which cause manifest CP violation~\cite{Haber:1997if}. In the SM, the CP phase in the quark mixing matrix, which is currently confirmed in $B$ meson experiments~\cite{Amsler:2008zz,Buras:2005xt}, cannot explain the observed baryon asymmetry of the  universe~\cite{Gavela:1993ts}, and additional sources of CP violation in models beyond the SM are required~\cite{Riotto:1998bt}.

In the Minimal Supersymmetric Standard Model (MSSM)~\cite{mssm},  a set of remaining complex parameters is obtained,  after absorbing non-physical phases by redefining particle fields. The complex parameters are conventionally chosen to be the Higgsino mass parameter $\mu$, the  ${\rm U(1)}$ and ${\rm SU(3)}$  gaugino mass parameters $M_1$ and $M_3$, respectively, and the trilinear scalar coupling parameters $A_f$ of the third generation sfermions ($f=b,t,\tau$),
\begin{eqnarray}
\label{eq:phases}
	\mu = |\mu| e^{i \phi_\mu}, \quad  
	M_1 = |M_1| e^{i \phi_{1}}, \quad
	M_3 = |M_3| e^{i \phi_3},   \quad
	A_f = |A_f| e^{i \phi_{A_f}}.
\end{eqnarray}
These phases contribute to the electric dipole moments (EDMs) of the electron, neutron and deuteron, which can be beyond their current experimental upper bounds~\cite{Amsler:2008zz,Regan:2002ta, Baker:2006ts,Griffith:2009zz,Semertzidis:2003iq}. However, the extent to which the EDMs can constrain the SUSY  phases strongly depends on the considered model and its parameters~\cite{CP-Problem,EDM,Barger:2001nu,Bartl:2003ju,Ellis:2008zy,Choi:2004rf}.
Thus measurements of SUSY CP observables outside the low energy EDM sector are necessary to independently determine or constrain the phases.   

For example the phases of the trilinear scalar coupling parameters $A_f$ have a significant impact on the MSSM Higgs sector~\cite{Pilaftsis}. Loop effects, dominantly mediated by third generation squarks, can generate large CP-violating scalar-pseudoscalar transitions among the neutral Higgs bosons~\cite{HA}. As a result the lightest Higgs boson with a mass of order $10$~GeV or $45$~GeV \cite{Carena:2000ks} cannot be excluded by measurements at LEP~\cite{LEPbounds}. The fundamental properties and the  phenomenology of CP-violating neutral Higgs boson mixings have been investigated in detail in the literature~\cite{Hreview}.

In addition, the phases of the  coupling parameters ${A_{b,t,\tau}}$ in the third generation sfermion sector are rather unconstrained by the EDMs~\cite{Ellis:2008zy,Choi:2004rf}. Therefore it looks appealing to find the adequate CP observables in sfermion production and decays at high energy colliders like the LHC~\cite{LHC} or ILC~\cite{ILC}, which are sensitive to the eventually large CP phases of $ A_{b,t, \tau}$. Third generation sfermions also have a rich phenomenology due to a sizable mixing of left and right states. 

The phases can drastically change SUSY particle masses, cross sections, branching ratios~\cite{Bartl:2003he,Bartl:2002uy,Alan:2007rp}, and longitudinal polarizations of final fermions~\cite{Gajdosik:2004ed}. Although such CP-even observables are very sensitive to the CP phases (the observables can change by an order of magnitude and more), CP-odd (T-odd) observables have to be measured for a direct evidence of CP violation. CP-odd observables are, for example, rate asymmetries of cross sections, distributions, and partial decay widths~\cite{otherCPodd}. However, these observables require the presence of absorptive phases, e.g. from loops, and thus usually do not exceed the size of $10\%$.

Larger CP asymmetries in particle decay chains can be obtained with triple products of final particle  momenta~\cite{tripleprods}. They already appear at tree level due to spin correlations. Triple product asymmetries have been intensively studied in the production and decay  of neutralinos~\cite{Bartl:2003tr,Bartl:2003ck,NEUT2,NEUT3,Kittel:2004rp} and charginos~\cite{Kittel:2004rp,CHAR2,CHAR3} at the ILC, also using transversely polarized beams~\cite{Trans}. For recent reviews, see Ref.~\cite{CPreview}.

%-----------------------------------------------------------------------------
%                                       F I G U R E   -1-
%-----------------------------------------------------------------------------
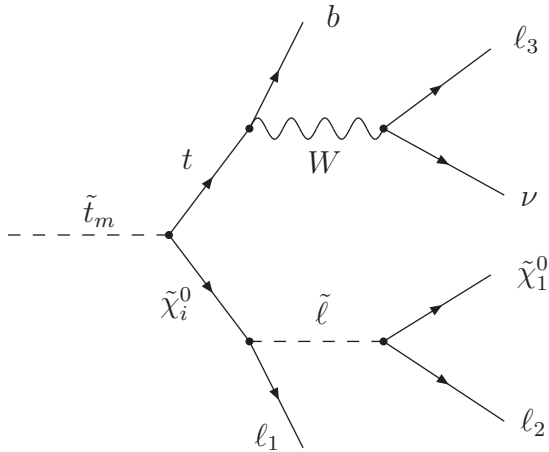
\begin{figure}[t]
\scalebox{1}{
\begin{picture}(10,5)(-3.4,0)
	\DashLine(-10,80)(50,80){5}
	\Vertex(50,80){1.5}
	\ArrowLine(50,80)(80,120)
	\Vertex(80,120){1.5}
	\ArrowLine(80,120)(100,160)
	\Photon(80,120)(130,120){4}{4}
	\Vertex(130,120){1.5}
	\ArrowLine(130,120)(170,150)
	\ArrowLine(130,120)(175,95)
%----------------------------------------------------
	\ArrowLine(50,80)(80,40)
	\Vertex(80,40){1.5}
	\ArrowLine(80,40)(100,0)
	\DashLine(80,40)(130,40){5}
	\Vertex(130,40){1.5}
	\ArrowLine(130,40)(170,65)
	\ArrowLine(130,40)(175,10)
	\put( 0.5,3.0){ $\tilde t_m $}
	\put( 1.51,1.8){ $\tilde\chi_i^0 $}
	\put( 2.75,0.03){ $\ell_1$}
	\put(3.55,1.65){ $\tilde\ell$}
	\put(6.2,2.2){ $\tilde\chi_1^0$}
	\put(6.25,0.22){ $ \ell_2$}
%----------------------------------------------------
	\put( 1.8,3.7){ $t$}
	\put(3.45,3.64){ $W$}
	\put( 3.7,5.58){ $b$}
	\put(6.25, 3.2){ $\nu $}
	\put(6.16,5.3){ $\ell_3$}
\end{picture}}
\caption{Schematic picture of top squark decay.}
\label{Fig:decayStop}
\end{figure}

In this paper, we study CP asymmetries in two-body decays of a stop,
\begin{eqnarray}
\label{eq:decayStop}
	\tilde t_m \to t + \tilde\chi^0_i; \quad m=1,2; \quad i=2,3,4;
\end{eqnarray}
followed by the subsequent two-body decay chains of the neutralino,
\begin{eqnarray}
\label{eq:decayChi}
	\tilde\chi^0_i \to \ell_1         + \tilde\ell_{n}; \quad 
	\tilde\ell_{n}   \to \tilde\chi^0_1 + \ell_2;     \qquad n=L,R,   \qquad  \ell= e,\mu,
\end{eqnarray}
and the top quark
\begin{eqnarray}
\label{eq:decayTop}
	t \to b + W; \quad 
	W \to \nu_\ell + \ell_3; \qquad \ell = e,\mu.
\end{eqnarray}
See Figure~\ref{Fig:decayStop} for a schematic picture of the entire stop decay chain. Since the stop decay, Eq.~(\ref{eq:decayStop}), is a two-body decay of a scalar particle, only the spin-spin correlations of the outgoing fermions contain a CP-sensitive part. Spin-spin correlations are terms in the amplitude squared that include the polarization vectors of both the neutralino and the top. The angular distributions of their decay products are then correlated to each other due to total angular-momentum conservation. Thus for the measurement of the top-neutralino spin-spin correlations, it is vital to include momenta from \emph{both} their decays into the triple products, otherwise the CP-sensitive information in that process will be lost~\cite{Bartl:2004jr}.

%-----------------------------------------------------------------------------
%                                       T A B L E  -1-
%-----------------------------------------------------------------------------
\begin{table}[t]
\renewcommand{\arraystretch}{1.6}
\caption{ 
         Overview of sfermion two-body decays studied in the literature to
         analyze CP phases. Two-body decays are denoted by $a\to bc$ and  
         three-body decays by $a\to bcd$. The column 'CP Phases' lists the phases the process is sensitive to, and 'Max ${\mathcal A}_{\rm CP}$' gives the maximal asymmetries which have been found in the rest frame of the sfermion.
         Asymmetries in parentheses correspond 
         to the decay $ W \to   c  \bar s $.
         \label{tab:OverviesSfermDecays}}
\begin{center}
      \begin{tabular}{|c|c|c|c|c|c|}
\hline
  Process
& Sub Decay 1
& Sub Decay 2
& CP Phases
& Max ${\mathcal A}_{\rm CP}$
& Ref.
\\ \hline
  $\tilde t_m \to t  \tilde\chi_i^0 $
& $ t \to  b  W  $
& $ W \to  \bar\ell  \nu  ( c  \bar s )$
&
&
&
\\ %\hline
&  $ \tilde\chi_i^0 \to \tilde\ell \ell $
& $	\tilde\ell \to  \ell \tilde\chi_1^0 $
& $ \phi_{A_t}, \phi_{\mu}, \phi_{M_1}$
& $	40\%$
& \cite{Bartl:2004jr}
\\ \hline
  $\tilde t_m \to t  \tilde\chi_i^0 $
& $ t \to  b  W  $
&
&
&
&
\\ 
&  $ \tilde\chi_i^0 \to \ell \bar\ell \tilde\chi_1^0$
& 
& $ \phi_{A_t}, \phi_{\mu}, \phi_{M_1}$
& $ 15\%$
& \cite{Ellis:2008hq}
\\ \hline
  $\tilde b_m \to t  \tilde\chi_j^\pm $
& $ t \to  b  W  $
& $ W \to  \bar \ell  \nu( c  \bar s )  $
&
&
&
\\ %\hline
&  $ \tilde\chi_j^\pm \to \ell \tilde\nu $
& 
& 
& $ 15(40)\%$
& 
\\ %\hline
&  $ \tilde\chi_j^\pm \to \tilde\ell \nu $
&  $ \tilde\ell \to  \ell \tilde\chi_1^0 $
& 
& $ 5(10)\% $
& 
\\ %\hline
&  $ \tilde\chi_j^\pm \to W \tilde\chi_1^0 $
& 
& 
&  $ 2(5)\%$
& 
\\ %\hline
&  $ \tilde\chi_j^\pm \to W \tilde\chi_1^0 $
& $	W \to  \ell  \bar\nu (  \bar c   s ) $
& $ \phi_{A_b}, \phi_{\mu}$
& $	2(5)\%$
& \cite{Bartl:2006hh}
\\ \hline
  $\tilde f \to f  \tilde\chi_i^0 $
& $ \tilde\chi_i^0 \to  Z \tilde\chi_1^0 $
& $ Z \to  \ell   \bar\ell $
&
& $	5\%$
&
\\ %\hline
&  
& $ Z \to  c \bar c  $
&  
& $ 15\%$
& 
\\% \hline
&  
& $ Z \to  b \bar b $
&  $\phi_{\mu}, \phi_{M_1}$
& $ 20\%$
& \cite{Bartl:2003ck}
\\ \hline
  $\tilde q \to q  \tilde\chi_i^0 $
& $\tilde\chi_i^0 \to \ell \bar\ell \tilde\chi_1^0$
& 
& $\phi_{\mu}, \phi_{M_1}$
& 	n/a
& \cite{Langacker:2007ur}
\\ \hline
  $\tilde \ell \to \ell \tilde\chi_i^0 $
& $\tilde\chi_i^0 \to \ell \bar\ell \tilde\chi_1^0$
& 
& $\phi_{\mu}, \phi_{M_1}$
& $ 15\% $
& \cite{AguilarSaavedra:2004hu}
\\ \hline
\end{tabular}
\end{center}
\renewcommand{\arraystretch}{1.0}
\end{table}
% -----------------------------------------------------------------------------

Triple product asymmetries have been studied in two-body decays of stops~\cite{Bartl:2004jr} and sbottoms~\cite{Bartl:2006hh}, however in their rest frame only.  At colliders like the LHC the particles are highly boosted, which will generally reduce the triple product asymmetries~\cite{Ellis:2008hq,Langacker:2007ur}. We also include the production of stops at the LHC, and analyze the stop boost distributions as well as their impact on the triple product asymmetries in detail. In our numerical analysis we work within the minimal Supergravity (mSUGRA) framework of SUSY breaking~\cite{mssm}, and explicitly add CP phases for the parameters $A_t, M_1, \mu$  at the electroweak scale, fully taking into account the implications on the EDMs. In addition, we calculate the amplitude squared for the entire stop decay in the spin-density matrix formalism~\cite{Haber:1994pe}.  The compact form of the amplitude squared allows us to identify the optimal CP observables for squark decays. For a recent study of CP-odd observables in stop decays, including their production, but for neutralino three-body decays, see Ref.~\cite{Ellis:2008hq}.  A summary of the literature in sfermion two-body decays is shown in Table~\ref{tab:OverviesSfermDecays}. Three-body decays of stops have been studied in~Ref.~\cite{Bartl:2002hi}.

The paper is organized as follows. In Section~\ref{sec:TripleProducts}, we identify the CP-sensitive parts in the amplitude squared, classify the CP asymmetries in top squark decays and discuss their dependence on the complex stop-top-neutralino couplings. Section~\ref{sec:EDMs} contains a discussion of  the experimental EDM constraints on the SUSY CP phases and asymmetries.  For the LHC, we discuss in Section~\ref{sec:AsymmetryLHC} the SUSY parameter dependence of stop production, boost distribution and branching ratios, and give lower bounds on the required luminosities to observe the asymmetries over their statistical fluctuations.  Section~\ref{sec:AsymmetryLHC} also includes a discussion of angular and volume distributions of the triple products. We summarize and give our conclusions in Section~\ref{sec:Conclusion}. In the Appendix, the stop and neutralino Lagrangians with complex couplings are reviewed, we give the phase space and calculate the stop decay amplitudes in the spin-density matrix formalism. We also show that absorptive phases can be eliminated from the T-odd asymmetries to obtain true CP asymmetries.

%------------------------------------------------------------------------------
\section{CP Asymmetries in Top Squark Decays}
\label{sec:TripleProducts}
%------------------------------------------------------------------------------

In this Section, we identify the CP-sensitive parts in the amplitude squared of the two-body decay of the top squark, see Eqs.~(\ref{eq:decayStop})-(\ref{eq:decayTop}) and Figure~\ref{Fig:decayStop}. In order to probe these parts, we systematically classify T- and CP-odd asymmetries. These can be defined from various triple or epsilon products of the different final state particle three- or four-momenta ${ p}_t, { p}_b , { p}_{\ell_1}, { p}_{\ell_2}, { p}_{\ell_3}$, respectively. Explicit expressions for the squared amplitude, Lagrangians, couplings, and phase space elements are 
summarized in the Appendix. 

%------------------------------------------------------------------------------
\subsection{T-odd Products}
\label{sec:CPterms}
%------------------------------------------------------------------------------

The amplitude squared for the stop decay chain, see Figure~\ref{Fig:decayStop},
can be decomposed into contributions from the top spin correlations, the neutralino 
spin correlations, the top-neutralino spin-spin correlations, and 
an unpolarized part, see Eq.~(\ref{eq:matrixelement2}).
The amplitude squared is a sum of terms which are even and one term which is odd under a T transformation. Assuming CPT invariance, CP asymmetries are proportional to the T-odd term in the amplitude squared. Such a term originates only from the spin-spin correlations.
It is given by the last summand in Eq.~(\ref{CPterm}), which is proportional to 
\begin{equation}
	|T|^2 \supset
	{\rm Im}\{ a^{\tilde t}_{mi} (b^{\tilde t}_{mi})^\ast \}
	[p_{\tilde t},p_t,p_{\ell_3},p_{\ell_1}], \quad m=1,2, \quad i=1,\dots,4.
\label{CPterm2}
\end{equation}
The left and right couplings  $a^{\tilde t}_{mi}$ and $b^{\tilde t}_{mi}$ 
are defined through the $\tilde t_m$--$t$--$\tilde\chi^0_i$ Lagrangian~\cite{Bartl:2004jr},
\begin{eqnarray}
{\scr L}_{t \tilde t\tilde\chi^0}
= g\,\bar t\,(a_{mi}^{\tilde t}\,P_R + b_{mi}^{\tilde t}\,P_L)
            \,\tilde\chi_i^0\,\tilde t_m
+ {\rm h.c.},
\end{eqnarray}
which we give explicitly in Appendix~\ref{Lagrangians and couplings}.
These couplings depend on the mixing in the stop and neutralino sector, and thus also on the CP phases $\phi_{A_t}$, $\phi_{\mu}$ and $\phi_{M_1}$. The imaginary part of the coupling product, ${\rm Im}\{ a^{\tilde t}_{mi} (b^{\tilde t}_{mi})^\ast \}$, in Eq.~(\ref{CPterm2}) is multiplied by a T-odd epsilon product, for which we have used the short hand notation 
\begin{equation}
	[p_{1},p_2,p_{3},p_{4}]\equiv 
	\varepsilon_{\mu\nu\alpha\beta}~
	p_{1}^{\mu}~p_2^{\nu}~p_{3}^{\alpha}~p_{4}^{\beta},
\label{eq:epsilonA}                
\end{equation}
with the convention $\varepsilon_{0123}=1$. Since each of the spatial components of the four-momenta changes sign under a naive time transformation, $t \to -t$, this product is T-odd. 

In order to construct asymmetries that probe this CP-sensitive part, we consider an epsilon product of four four-momenta $p_i$  
\begin{equation}
	{\mathcal E}=
	[p_{1},p_2,p_{3},p_{4}].
\label{eq:epsilon}                
\end{equation}
The $p_i$ can in principle be any four linearly independent momenta in the two-body decay-chain of the top squark, $p_{\tilde t},p_{t}, p_{b} , p_{\ell_1}, p_{\ell_2}, p_{\ell_3}$, see Figure~\ref{Fig:decayStop}. Note however, that momenta from both the decay products of neutralino \emph{and} the top have to be included in the epsilon product to obtain non-vanishing asymmetries~\cite{Bartl:2004jr}. Otherwise the top-neutralino spin-spin correlations are lost. Only they have CP-sensitive terms, since the stop decay, Eq.~(\ref{eq:decayStop}), is a two-body decay of a 
scalar particle\footnote{ 
                   If for example momenta from the top decay are not taken into account, only 
                   CP asymmetries can be obtained, which are sensitive to the CP phases 
                   $\phi_{\mu}$ and $\phi_{M_1}$ alone,  which enter solely from the neutralino 
                   decay. Still in that case, a three-body neutralino decay is 
                   required~\cite{Ellis:2008hq, AguilarSaavedra:2004hu, Langacker:2007ur}, or a 
                   two-body decay chain with an intermediate $Z$-boson~\cite{Bartl:2003ck}. 
                   However, the sfermion then merely serves as a production channel for neutralinos, 
                   whose CP properties are studied through their subsequent 
                   decays~\cite{Ellis:2008hq, AguilarSaavedra:2004hu, Langacker:2007ur,Bartl:2003ck}. 
                   The asymmetries are then only of the order of 10\%, which is typical for 
                   neutralino three-body decays~\cite{NEUT3,Ellis:2008hq, AguilarSaavedra:2004hu, 
                   Langacker:2007ur,Bartl:2003ck}, and also chargino three-body decays~\cite{CHAR3}.
                  }.

%------------------------------------------------------------------------------
\subsection{T-odd Asymmetries}
\label{sec:TOddProducts}
%------------------------------------------------------------------------------

For a given T-odd product ${\mathcal E}$, Eq.~(\ref{eq:epsilon}), we can now define the 
T-odd asymmetry of the partial stop decay width $\Gamma$~\cite{Bartl:2004jr},
\begin{equation}
	{\mathcal A}=
	\frac{\Gamma({\mathcal E}>0)-\Gamma({\mathcal E}<0)}
	     {\Gamma({\mathcal E}>0)+\Gamma({\mathcal E}<0)}
	=
	\frac{\int{\rm Sign}[{\mathcal E}] |T|^2 d{\rm Lips}}
	     {\int |T|^2 d{\rm Lips}},
\label{eq:Toddasym}
\end{equation}
with the amplitude squared $|T|^2$, and the Lorentz invariant phase-space element $d{\rm Lips}$, 
such that $\frac{1}{2m_{\tilde t}}\int |T|^2 d{\rm Lips}=\Gamma$. 
The T-odd asymmetry is also CP-odd, if absorptive phases (from higher order final-state
interactions or finite-width effects) can be neglected~\cite{tripleprods}.
In Appendix~\ref{sec:AbsorptivePhases}, we also construct a  CP asymmetry,
where those contributions are eliminated.

In general the largest asymmetries are obtained by using an epsilon product which matches the 
kinematic dependence of the CP-sensitive terms in the amplitude squared. In the literature, 
this technique is sometimes referred to \emph{optimal observables}~\cite{optimal}.
Thus the largest asymmetry is obtained from the epsilon product
\begin{equation}
	{\mathcal E} =	
	[p_{\tilde t},p_t,p_{\ell_3},p_{\ell_1}].
	\label{eq:Optepsilon}             
\end{equation}
Other combinations of momenta in general lead to smaller asymmetries, see Ref.~\cite{Bartl:2004jr}.

Triple products of three spatial momenta can also be used to define asymmetries~\cite{tripleprods}. 
In the stop rest frame, $p_{\tilde t}^\mu = (m_{\tilde t}, \mathbf{0})$, 
the epsilon product reads~\cite{Bartl:2004jr}
\begin{equation}
	[p_{\tilde t},p_t,p_{\ell_3},p_{\ell_1}] =
	m_{\tilde t} \; \,\mathbf{p}_t\cdot 
        ( \mathbf{p}_{\ell_3 } \times 	\mathbf{p}_{\ell_1}) \equiv 
	m_{\tilde t} \, {\mathcal T}.
\label{tripleterm}
\end{equation}
That triple product has been found to give the largest 
asymmetries in the stop rest frame, and other combinations of momenta for ${\mathcal T}$ lead 
to smaller asymmetries~\cite{Bartl:2004jr}. Note that the asymmetries of an epsilon product ${\mathcal E}$ are by 
construction Lorentz invariant whereas those constructed with  a triple product ${\mathcal T}$ are 
not~\cite{Langacker:2007ur}. The triple product asymmetries will therefore depend on the stop boost, 
$\beta_{\tilde t} = |{\mathbf p}_{\tilde t}|/E_{\tilde t}$, and are generally reduced if not 
evaluated in the stop rest frame~\cite{AguilarSaavedra:2006fy}. We will discuss in detail the impact 
of the stop boost on the asymmetries at the LHC in Section~\ref{sec:AsymmetryLHC}.

%------------------------------------------------------------------------------
\subsection{Phase Dependence}
\label{sec:PhaseDependence}
%------------------------------------------------------------------------------

%
% ------------------------------------------------------------------------------
%                                       T A B L E  -2-
% ------------------------------------------------------------------------------
\begin{table}[t]
\renewcommand{\arraystretch}{1.6}
\caption{
         mSUGRA benchmark scenario, extended with SUSY CP phases at the electroweak scale.
\label{tab:ReferenceScenario}}
%\vspace{1cm}
\begin{center}
\begin{tabular}{ccccccc} 
\hline
  \multicolumn{1}{c}{$ m_0 $} 
& \multicolumn{1}{c}{$ m_{1/2} $}
& \multicolumn{1}{c}{$ A_0 $}  
& \multicolumn{1}{c}{$ \tan\beta $}  
& \multicolumn{1}{c}{$ \phi_{A_t} $}
& \multicolumn{1}{c}{$ \phi_{\mu}$}
& \multicolumn{1}{c}{$ \phi_{M_1} $}
\\\hline
  \multicolumn{1}{c}{$ 70~\rm{GeV} $} 
& \multicolumn{1}{c}{$ 270~\rm{GeV} $}
& \multicolumn{1}{c}{$ 500~\rm{GeV} $}  
& \multicolumn{1}{c}{$ 5 $} 
& \multicolumn{1}{c}{$ \frac{1}{5}\,\pi $}
& \multicolumn{1}{c}{$ 0 $}
& \multicolumn{1}{c}{$ 0 $}
\\\hline
\end{tabular}
%\\[5.0ex]
\end{center}
\renewcommand{\arraystretch}{1.0}
\end{table}
% ------------------------------------------------------------------------------

% ------------------------------------------------------------------------------
%                                       T A B L E  -3-
% ------------------------------------------------------------------------------
\begin{table}[t]
\renewcommand{\arraystretch}{1.6}
\caption{ 
       SUSY parameters and masses at the electroweak scale
       for the benchmark scenario of Table~\ref{tab:ReferenceScenario} ($\ell = e,\mu$).
         \label{tab:ReferenceScenarioDerived}}
\begin{center}
      \begin{tabular}{|c|c|c|c|}
\hline
$     M_1  =   110~{\rm GeV}$ &
$     M_2  =   206~{\rm GeV}$ & 
$     \mu  =   344~{\rm GeV}$ &
$ |A_{t}|  =   329~{\rm GeV}$ 
%${\rm BR}(\tilde\chi_2^0\to \tilde \ell_R \ell) = 66 \%$ \\
\\ \hline
\hline
$   m_{\tilde t_1}  =   472~{\rm GeV}$ &
$   m_{\tilde t_2}  =   612~{\rm GeV}$ &
$   m_{\tilde b_1}  =   560~{\rm GeV}$ &
$   m_{\tilde b_2}  =   581~{\rm GeV}$ 
\\ \hline
$   m_{\chi^0_1}  =   102~{\rm GeV}$ &
$   m_{\chi^0_2}  =   189~{\rm GeV}$ &
$   m_{\chi^0_3}  =   350~{\rm GeV}$ &
$   m_{\chi^0_4}  =   376~{\rm GeV}$ 
\\ \hline
$    m_{\chi^\pm_1}  =   187~{\rm GeV}$ &
$    m_{\chi^\pm_2}  =   376~{\rm GeV}$ &
$  m_{\tilde\ell_R}  =   130~{\rm GeV}$ & 
$  m_{\tilde\ell_L}  =   201~{\rm GeV}$  
\\ \hline
\end{tabular}
\end{center}
\renewcommand{\arraystretch}{1.0}
\end{table}
% ------------------------------------------------------------------------------

In order to analyze the phase dependence of the asymmetry ${\mathcal A}$, 
Eq.~(\ref{eq:Toddasym}), we insert the explicit form of the amplitude squared 
in the spin-density formalism. As shown in Appendix~\ref{Stop decay width}, we obtain
\begin{equation}
	{\mathcal A}= 2 \, \eta \,
	\frac{\int {\rm Sign}({\mathcal E})
	(p_b\cdot p_{\nu_\ell})\,
	[p_{\tilde t},p_t,p_{\ell_3},p_{\ell_1}]~
	d{\rm Lips}  }{(m_{\chi_i^0}^2- m_{\tilde\ell }^2)\int
	(p_t\cdot p_{\ell_3})
	(p_b\cdot p_{\nu_\ell})~
	d{\rm Lips} },
\label{eq:Adependence2}
\end{equation}
with the coupling function
\begin{equation}
	\eta = 
	\frac{{\rm Im}\{ a^{\tilde t}_{mi} (b^{\tilde t}_{mi})^\ast \}}
	{ \frac{1}{2}\left(|a^{\tilde t}_{mi}|^2 + |b^{\tilde t}_{mi}|^2\right)
	\frac{m_{\tilde t}^2 - m_{\chi_i^0}^2 - m_t^2}
	{2 m_{\chi_i^0} m_t}
	- {\rm Re}\{ a^{\tilde t}_{mi} (b^{\tilde t}_{mi})^\ast \}}.
\label{eq:couplingfunct}
\end{equation}
Thus the asymmetry can be separated into a kinematical part, and the effective coupling factor 
$\eta$, that is approximately independent of the particle masses, but governs the complete
phase dependence of the stop-top-neutralino couplings.
%$a^{\tilde t}_{mi}$ and $b^{\tilde t}_{mi}$.

To discuss the phase dependence of  $\eta$ (and ${\mathcal A}$), we choose the 
mSUGRA framework of SUSY breaking~\cite{mssm}. We extend it by adding the CP phases 
$\phi_{A_t}$,  $\phi_\mu$ and $\phi_{M_1}$ at the electroweak scale.  
We define a benchmark scenario in Table~\ref{tab:ReferenceScenario}, 
and give the other low energy parameters and masses in Table~\ref{tab:ReferenceScenarioDerived}.
The scenario is chosen to approximately maximize the signals at the LHC, while still respecting 
current bounds from electric dipole moments, see the detailed discussion in Section~\ref{sec:EDMs}.

\begin{figure}[t!]
\centering
\includegraphics[clip,width=0.495\textwidth]{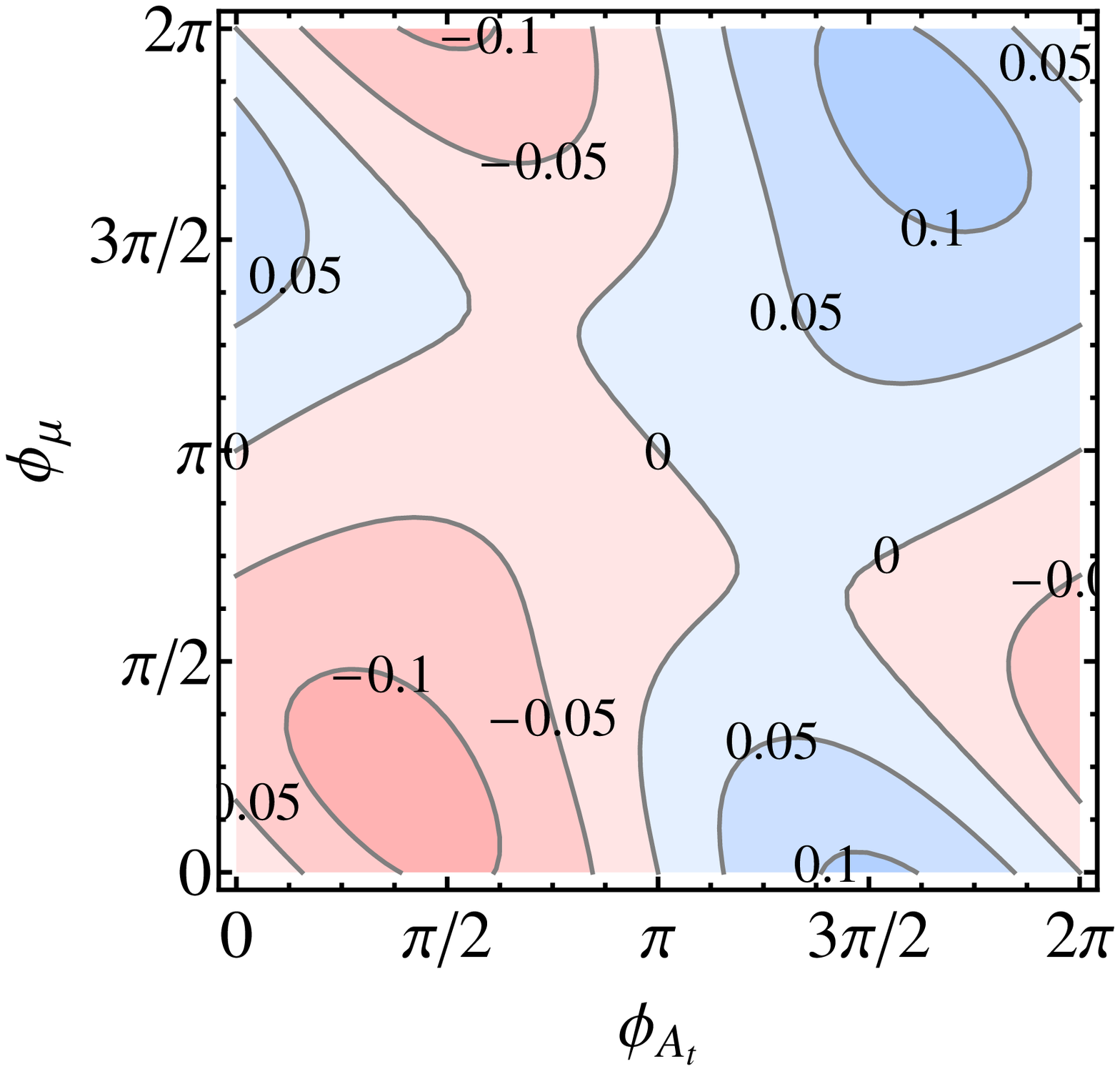}
\includegraphics[clip,width=0.495\textwidth]{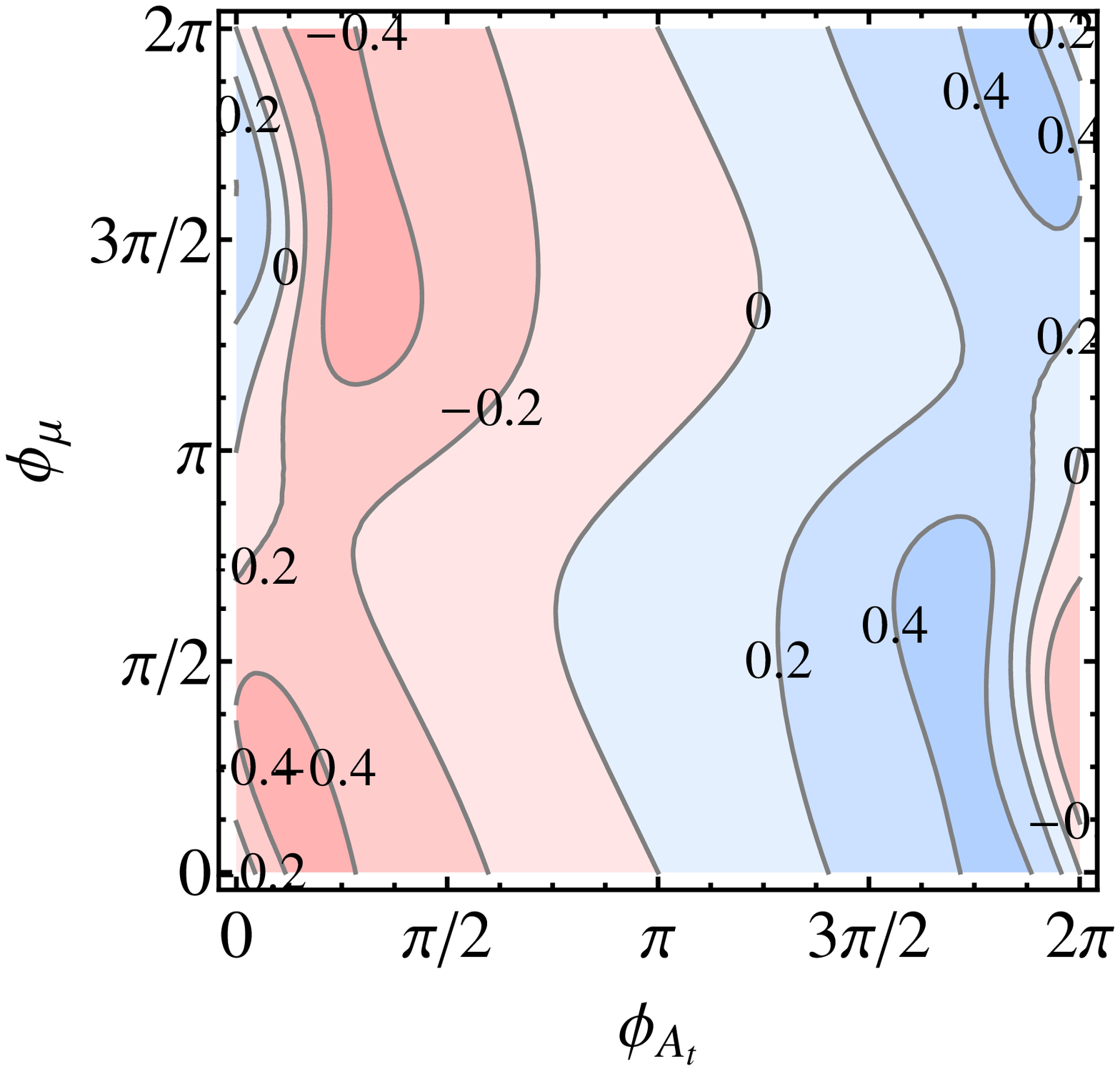}
\caption{Approximated CP-odd coupling factor ${\rm Im}\{ a^{\tilde t}_{12} (b^{\tilde t}_{12})^\ast \}$,  Eq.~(\ref{eq:ApproxPhaseDependence}), (left), and complete effective coupling factor $\eta$, Eq.~(\ref{eq:couplingfunct}), (right) for the decay  $\tilde t_1 \to t \tilde\chi^0_2$, as a function of the CP phases $\phi_{A_t}$ and $\phi_{\mu}$. All other phases are set to zero. The mSUGRA parameters are given in Table~\ref{tab:ReferenceScenario}.}
\label{fig:EtaPhiAtPhiMu}
\end{figure}
Within mSUGRA, the lightest neutralino is essentially a bino whereas the second lightest neutralino is a wino. For the stop decay via the second lightest neutralino, $\tilde t_1 \to t \tilde\chi^0_2$, the asymmetries will primarily depend on $\phi_{A_t}$ and $\phi_{\mu}$, through the influence of stop mixing. We expect a weak dependence on $\phi_{M_1}$, since $\tilde\chi_2^0$ has a negligible bino component. In this limit of the neutralino mixing matrix~\cite{Choi:2004rf}, valid for a large part of the mSUGRA parameter space, the phase dependence of the CP-odd coupling factor ${\rm Im}\{ a^{\tilde t}_{mi} (b^{\tilde t}_{mi})^\ast \}$ for the decay  $\tilde t_1 \to t \tilde\chi^0_2$ can be approximated as
\begin{eqnarray}
\label{eq:ApproxPhaseDependence}
	{\rm Im}\{ a^{\tilde t}_{12} (b^{\tilde t}_{12})^\ast \} 
	&\propto&\,\,\,\,
	    \frac{\cos^2\theta_{\tilde t}}{\tan\beta}\frac{m_t|\mu|}{|\mu|^2-M_2^2}
	    \sin(\phi_\mu)                    \nonumber\\
	&&+\, \frac{\sin 2\theta_{\tilde t}}{2}\left|\frac{m_t M_2}{|\mu|^2-M_2^2}\right|^2
	    \sin(\phi_{A_t})                         \nonumber\\
	&&+\, \frac{\sin 2\theta_{\tilde t}}{\tan\beta}\left|\frac{m_t \sqrt{|\mu| M_2}}{|\mu|^2-M_2^2}\right|^2
	    \sin(\phi_{A_t} + \phi_\mu)        \nonumber\\
	&&+\, \frac{\sin 2\theta_{\tilde t}}{\tan^2\beta}\left|\frac{m_t|\mu|}{|\mu|^2-M_2^2}\right|^2
	    \sin(\phi_{A_t} + 2\phi_\mu),
\end{eqnarray}
with the stop mixing angle $\theta_{\tilde t}$ , defined in Eq.~(\ref{eq:thstop}). 

In Figure~\ref{fig:EtaPhiAtPhiMu}, we show the phase dependence of ${\rm Im}\{ a^{\tilde t}_{12} (b^{\tilde t}_{12})^\ast \}$ in its approximation, Eq.~(\ref{eq:ApproxPhaseDependence}), for the mSUGRA scenario of Table~\ref{tab:ReferenceScenario}. 
For comparison we also show the coupling factor $\eta$,  Eq.~(\ref{eq:couplingfunct}), without approximations. The deviations are due to CP-even terms in the denominator of $\eta$. The CP-even parts are generally functions of the cosine of the phases. Figure~\ref{fig:EtaPhiAtPhiMu} illustrates that such terms have an important impact on the phase dependence of the CP asymmetries.

\begin{figure}[t!]
\centering
\includegraphics[clip,width=0.495\textwidth]{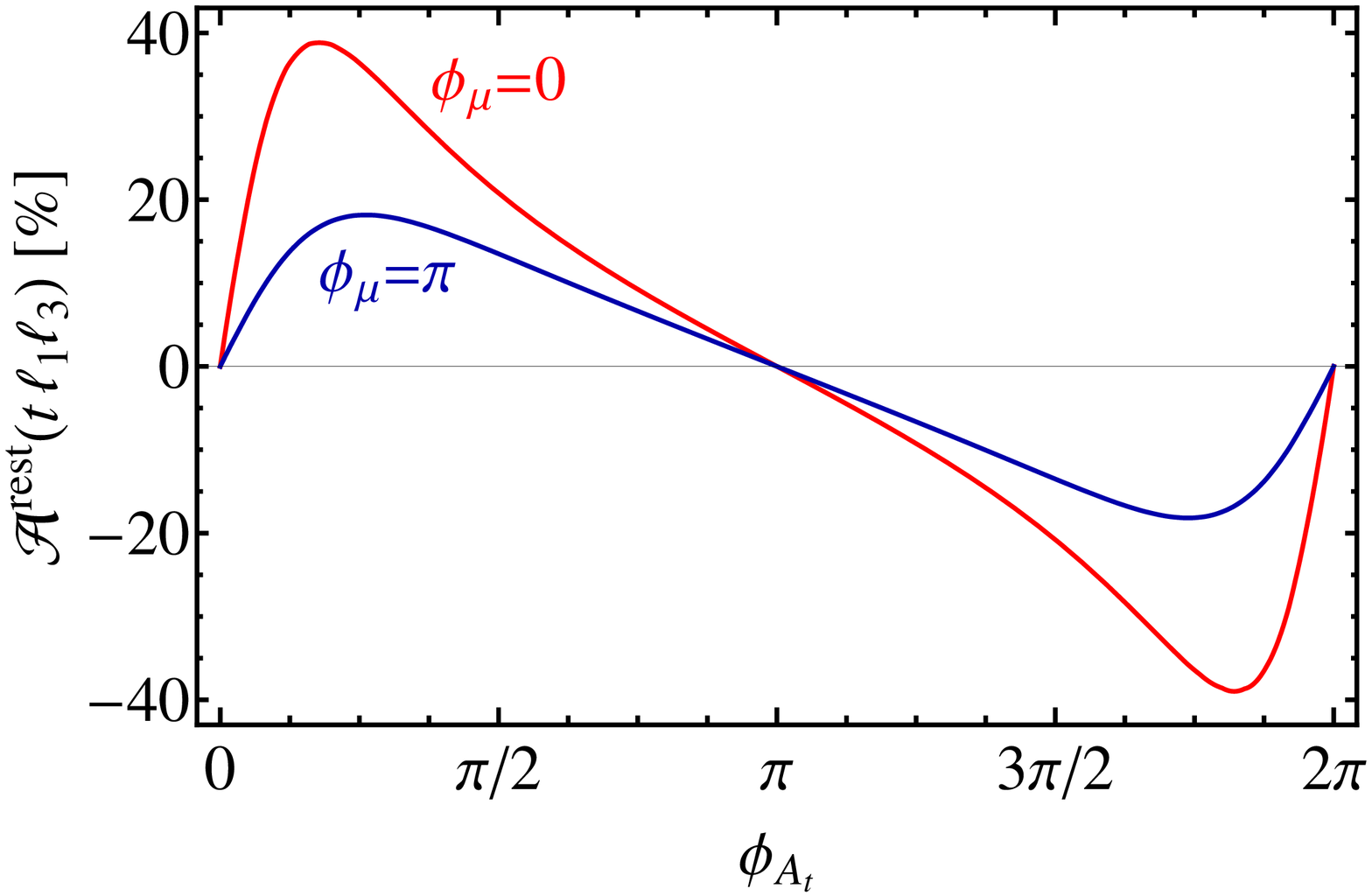}
\includegraphics[clip,width=0.495\textwidth]{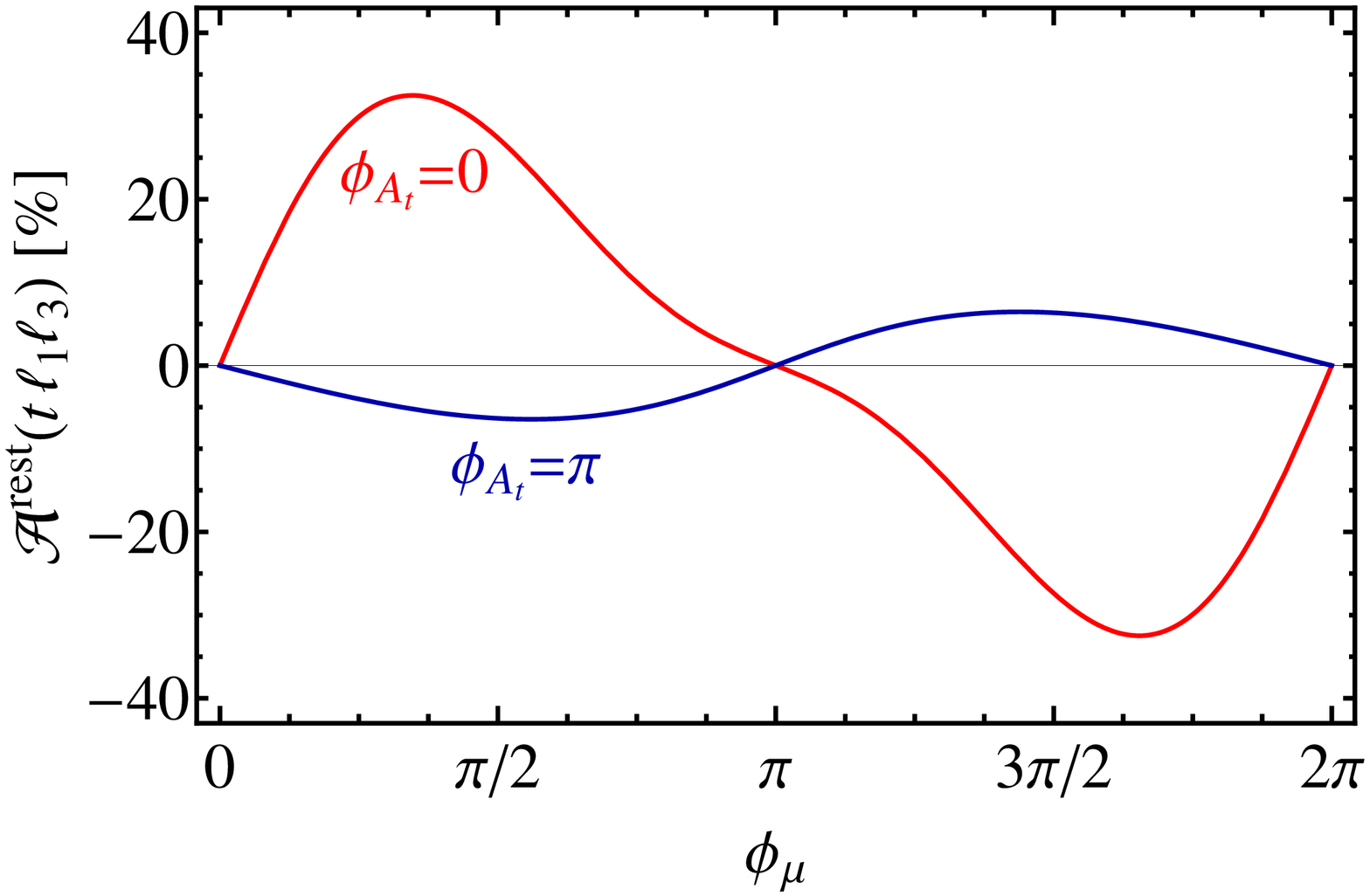}
\caption{
        Asymmetry $\mathcal{A}(t\ell_1\ell_3)$, Eq.~(\ref{eq:Toddasym}), 
        as a function of
        the CP phase $\phi_{A_t}$ with $\phi_\mu = 0,\pi$ (left), 
        and as a function of $\phi_{\mu}$ with $\phi_{A_t} = 0,\pi$ (right),
        for the stop decay $\tilde t_1 \to t \tilde\chi^0_2$  
        and $\tilde\chi^0_2\to \ell_1 \tilde \ell_{R}$, see Figure~\ref{Fig:decayStop},
        in the stop rest frame.
        The mSUGRA parameters are given in Table~\ref{tab:ReferenceScenario}.}
\label{fig:phase1D}
\end{figure}
%\newpage

We further illustrate this effect on the asymmetry  ${\mathcal A}$~(\ref{eq:Adependence2})
for the optimal epsilon product ${\mathcal E}$~(\ref{eq:Optepsilon}),
for $pp\to\tilde t_1\tilde t_1^\ast $ production and the subsequent
decays $\tilde t_1 \to t \tilde\chi^0_2$  and $\tilde\chi^0_2\to \ell \tilde \ell_{R}$ via the second lightest neutralino and a right slepton.
In the stop rest frame, this asymmetry is equivalent to the triple product asymmetry with
${\mathcal T} =\mathbf{p}_t\cdot ( \mathbf{p}_{\ell_3 } \times \mathbf{p}_{\ell_1})$, as given in
Eq.~(\ref{tripleterm}). We use the short hand notation  $\mathcal{A}(t\ell_1\ell_3)$,
and similarly for other asymmetries to indicate the momenta used for the triple product.

In  Figure~\ref{fig:phase1D}, we show $\mathcal{A}(t\ell_1\ell_3)$ as a function of $\phi_{A_t}$ (left panel),
and $\phi_{\mu}$ (right panel). The dependence on $\phi_{M_1}$ is numerically small as expected. 
We clearly see that the maxima of $\mathcal{A}(t\ell_1\ell_3)\approx\pm40\%$ 
are not necessarily obtained for 
maximal CP phases $\phi_\mu$, $\phi_{A_t} =\pi/2,3/2 \pi$.  The reason is that the phase dependence of 
$\mathcal{A}(t\ell_1\ell_3)$ is almost governed by the coupling factor $\eta$, 
see Figure~\ref{fig:EtaPhiAtPhiMu},
right panel. Since the CP-odd (CP-even) numerator (denominator) of $\eta$ has a sine-like (cosine-like) 
dependence on the phases, the maxima of $\eta$, and thus in turn of 
$\mathcal{A}(t\ell_1\ell_3)$, are shifted  away from 
$\phi_{A_t}$, $\phi_\mu=\pi/2,3/2 \pi$ in Figure~\ref{fig:phase1D}. Thus the asymmetry can be sizable for smaller values of the phases, which is favored by EDM constraints. For example, the asymmetry has a maximum at $\phi_\mu\approx 3/8\pi$ for $\phi_{A_t}=0$, and for $\phi_\mu=0$ the asymmetry is maximal at $\phi_{A_t}\approx 1/5\pi$. The positions of the maxima will remain when including the effects of the stop boost, as we will discuss in Section~\ref{sec:StopProduction}. 
This motivates the choice $\phi_{A_t}= 1/5 \pi$,  $\phi_\mu=0$ at the weak scale in our benchmark scenario (Table~\ref{tab:ReferenceScenario}).

%------------------------------------------------------------------------------
\section{Bounds from Electric Dipole Moments}
\label{sec:EDMs}
%------------------------------------------------------------------------------

The CP-violating phases in SUSY models are constrained by limits on the T-odd electric dipole 
moments (EDMs)~\cite{Ellis:2008zy}. The current experimental bounds on the EDMs of 
Thallium~\cite{Regan:2002ta}, neutron~\cite{Baker:2006ts}, and 
Mercury~\cite{Griffith:2009zz} are, respectively,
\begin{eqnarray}
\label{eq:EDMBounds}
	|d_{\textrm{Tl}}| &=& 9 \cdot 10^{-25} \; e\textrm{ cm}, \\
	|d_{\textrm{n}}|  &=& 3 \cdot 10^{-26} \; e\textrm{ cm}, \\
	|d_{\textrm{Hg}}| &=& 3 \cdot 10^{-29} \; e\textrm{ cm}.
\end{eqnarray}
A planned experiment for measuring the deuteron EDM  aims at a sensitivity of~\cite{Semertzidis:2003iq}
\begin{equation}
	\label{eq:EDMFutureBounds}
	|d_{\textrm{d}}| = 1-3 \cdot 10^{-27}  \; e\textrm{ cm}.
\end{equation}
These limits generically restrict the CP phases to be smaller than $\pi/10$, in particular the 
phase $\phi_\mu$~\cite{Choi:2004rf,Barger:2001nu}. However, cancellations between different contributions 
to the EDMs can allow for larger CP phases of the order of $ \pi/4$~\cite{Ellis:2008zy}.

\begin{figure}[t]
\centering
\includegraphics[clip,width=0.7\textwidth]{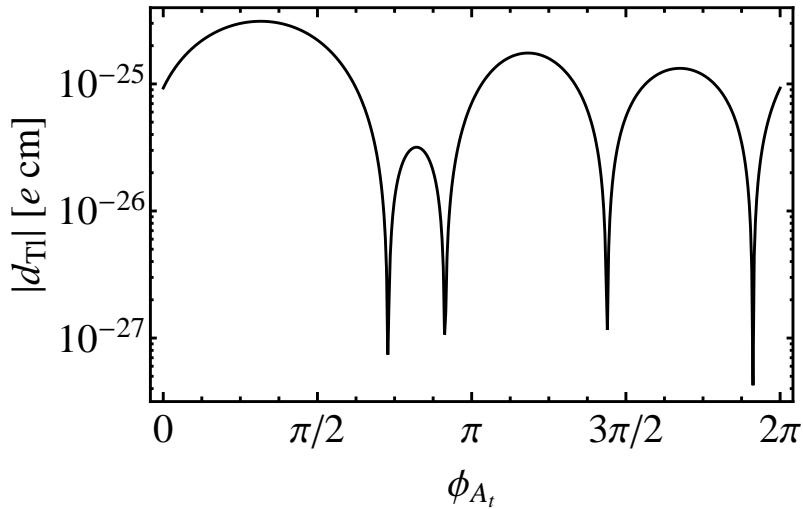}
\caption{Electric dipole moment of Thallium as a function of $\phi_{A_t}$ with 
 $\phi_{A_b}= 40/180~ \pi$, $\phi_{A_\tau}=  1/180 ~\pi$, $\phi_{M_3}= 10/180 ~\pi$ 
and $\phi_{M_1}=\phi_\mu=0$. The mSUGRA parameters are given in Table~\ref{tab:ReferenceScenario}.}
\label{fig:EDMCancellations}
\end{figure}

As an example of such cancellations, Figure~\ref{fig:EDMCancellations} shows the EDM of Thallium 
as a function of $\phi_{A_t}$, for $\phi_{A_b}= 40/180~ \pi$, $\phi_{A_\tau}=  1/180 ~\pi$, 
$\phi_{M_3}= 10/180 ~\pi$ and $\phi_{M_1}=\phi_\mu=0$. The other  mSUGRA parameters are given in 
Table~\ref{tab:ReferenceScenario}.
At four particular values of $\phi_{A_t}$, seen by distinct dips 
in the plot, cancellations reduce $d_{\textrm{Tl}}$ by more than two orders of magnitude. 

\begin{figure}[t]
\centering
\includegraphics[clip,width=0.7\textwidth]{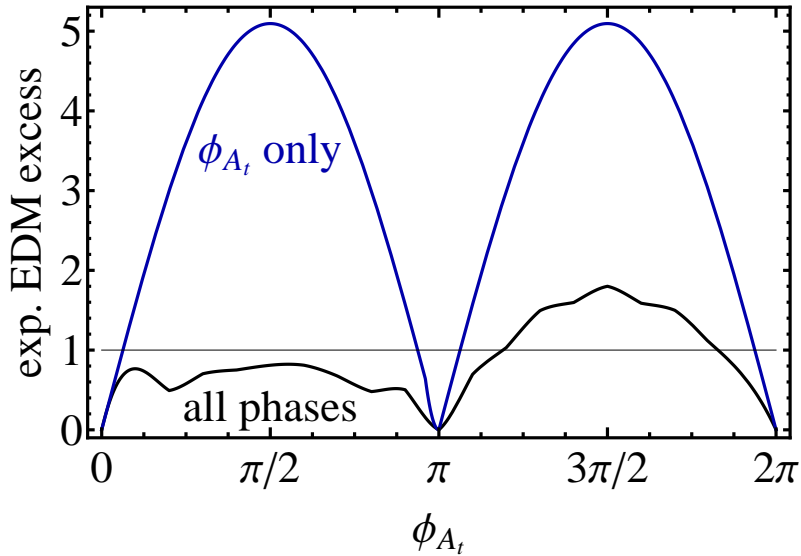}
\caption{Experimental excess in electric dipole moments as a function of $\phi_{A_t}$ for $\phi_\mu=0$. The experimental excess is defined as the maximum $\max_i |d_i(\phi_{A_t},...)|/|d_i^{exp}|$, $i=\textrm{Hg, \textrm{n}, \textrm{Tl}}$, over all current experimental limits. The two curves correspond to the case where:  all phases except $\phi_{A_t}$ are set to zero ('$\phi_{A_t}$ only'),  and the phases $\phi_{A_t}$, $\phi_{A_b}$, $\phi_{A_\tau}$, $\phi_{M_1}$ and $\phi_{M_3}$ are randomly scattered ('all phases'). For a given value of $\phi_{A_t}$, the minimal excess is plotted. The mSUGRA parameters are given in Table~\ref{tab:ReferenceScenario}.}
\label{fig:EDMBounds}
\end{figure}
We now randomly scatter the phases $\phi_{A_b}$, $\phi_{A_\tau}$, $\phi_{M_3}$, $\phi_{M_1}$ and $\phi_{A_t}$, setting the most constrained phase $\phi_\mu=0$. For each scatter point, we calculate the maximal excess of the EDMs over the corresponding experimental limit with the software package CPsuperH~\cite{Lee:2007gn}.  For the set of scatter points, we then show the line of the minima of all the excesses obtained as a function of $\phi_{A_t}$ in Figure~\ref{fig:EDMBounds}. For $\phi_{A_t}<\pi$ the minima are well below the experimental limit, and for $\phi_{A_t}>\pi$ only slightly above. So even large values of $\phi_{A_t}$ are not yet excluded by the current EDM limits. Note that a more exhaustive scan to find narrow cancellation gaps would probably make even smaller experimental EDM excesses possible. For comparison, we show in Figure~\ref{fig:EDMBounds} also the experimental excess in the case that all phases except $\phi_{A_t}$ are set to zero. This demonstrates that a conspiracy 
among the phases can considerably reduce the predicted electric dipole moments. 

We conclude that due to cancellations among different contributions to the EDMs only isolated points 
in the CP phase space can give large CP signals at the LHC.
It is important to search for these signals, since 
the cancellations could be a consequence of a deeper model that correlates the phases. For instance, 
the existing EDM bounds can also be fulfilled by including lepton flavor violating couplings in the 
slepton sector~\cite{Bartl:2003ju}. This is important when considering for example SUSY Seesaw models, 
where CP violation in the slepton sector is connected to the neutrino sector and 
Leptogenesis~\cite{Leptogen,reviewLeptogen}.

The interrelation between the EDMs and the SUSY CP phases show that CP observables have 
to be determined  outside the low energy sector, for example by measurements at the LHC and ILC. 
Asymmetries of triple products generally depend in a different way on the SUSY parameters than the EDMs, 
in particular on the CP phases, giving rise to a potential synergy between experiments at the low and 
the high energy scale. Depending on the experimental results, the EDM  bounds could be either verified, 
or the CP-violating sectors of the model have to be modified. 
CP-odd observables like triple product asymmetries would be the ideal tool 
for such independent measurements.

%------------------------------------------------------------------------------
\section{CP Asymmetries at the LHC}
\label{sec:AsymmetryLHC}
%------------------------------------------------------------------------------

At the LHC, squarks are produced with a distinct boost distribution which reduces the triple product asymmetries in the laboratory (lab) frame.
In contrast to the epsilon product  asymmetries they are not Lorentz invariant.
If the stop rest frame cannot be reconstructed, 
the triple product asymmetries have to be folded with the stop boost distribution.

In this Section we will discuss the dependence of triple product asymmetries on the stop boost. To this end we first review the production cross section of stops pairs, show their boost distribution, and explain how the triple product asymmetries in the lab frame are obtained. After defining their theoretical statistical significance, we study the asymmetries, stop and neutralino branching ratios, and the significances in the mSUGRA parameter space. 

%---------------------------------------------------------------------------------
\subsection{Stop Pair Production and Stop Boost Distribution}
\label{sec:StopProduction}
%--------------------------------------------------------------------------------
%
\begin{figure}[t]
\centering
\includegraphics[clip,width=0.410\textwidth]{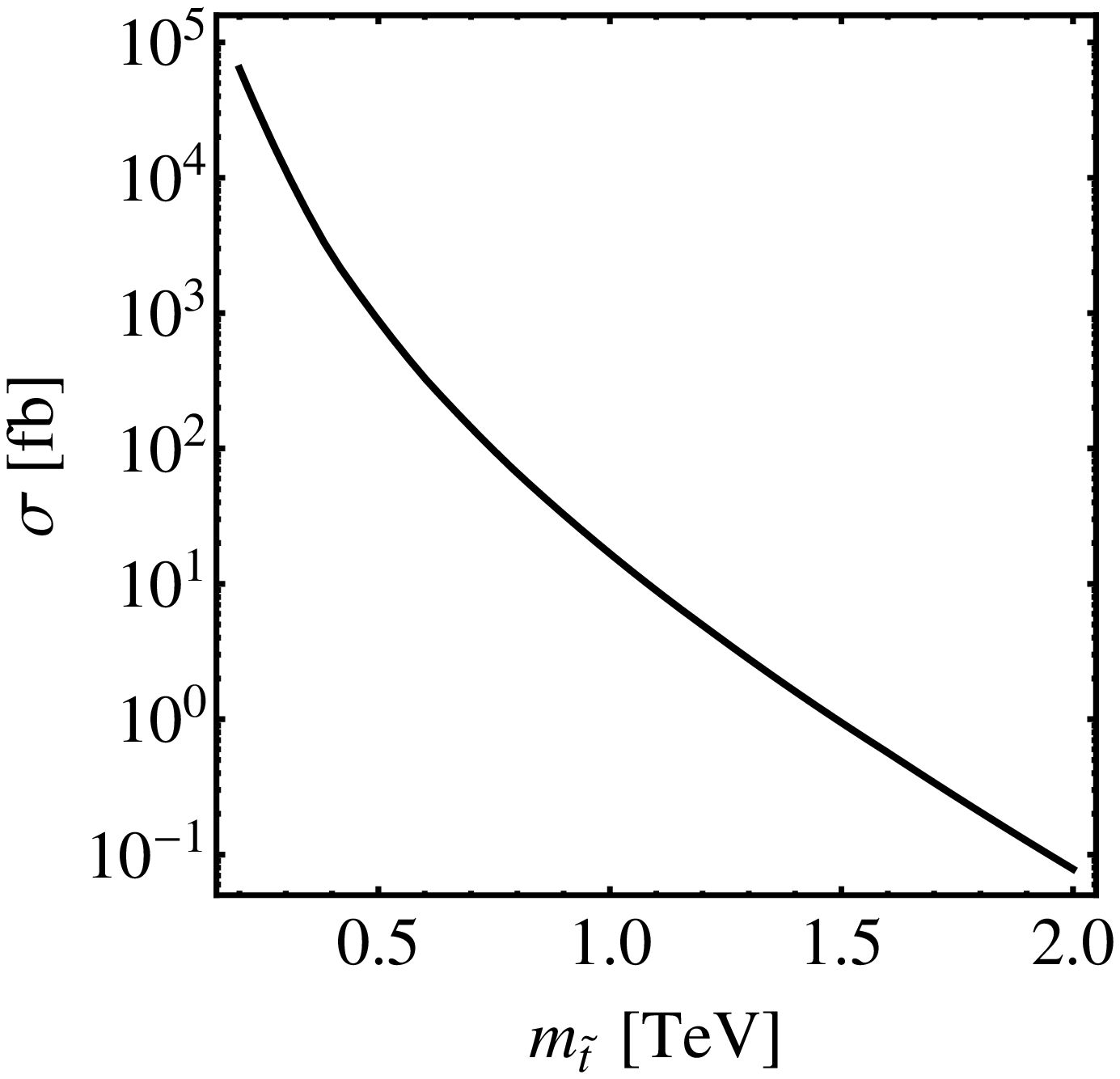}
\includegraphics[clip,width=0.580\textwidth]{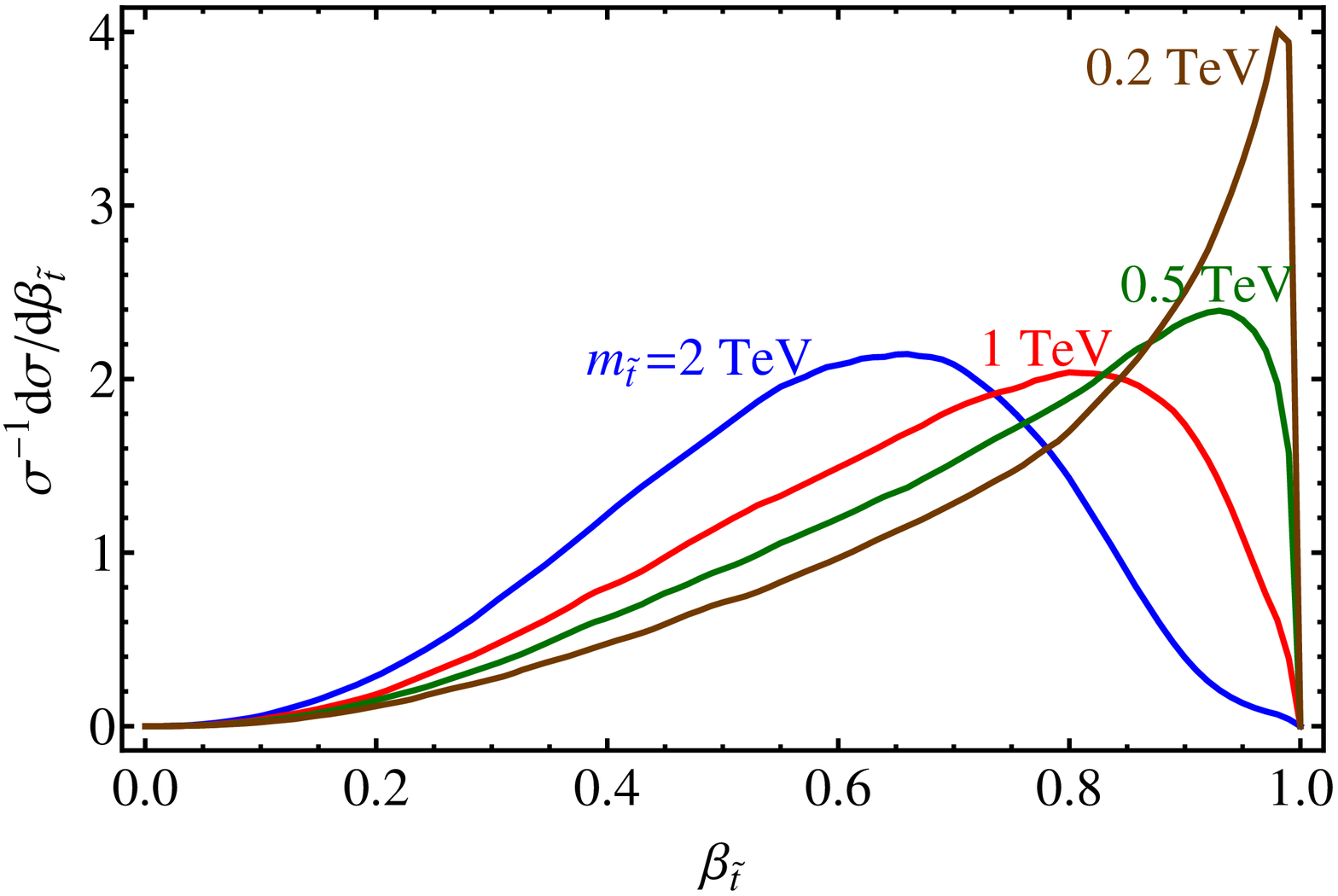}
\caption{Total stop pair production cross section \(\sigma(pp\to\tilde t\tilde t^*)\) at the LHC as 
a function of the stop mass (left). Normalized stop pair production distribution  
\(\frac{1}{\sigma}\frac{d\sigma}{d\beta_{\tilde t}}\) with respect to the 
stop boost factor \(\beta_{\tilde t}\) (right).
The leading order
cross sections have been calculated at $\sqrt{s_{pp}}=14$~TeV using MadGraph~\cite{Stelzer:1994ta}.}
\label{fig:StopPairProduction}
\end{figure}
The production of stop pairs at the LHC~\cite{NLOstop}
\begin{equation}
p + p  \to   \tilde t_m + \tilde t_m^\ast, \quad m =1,2,
\label{eq:stopProd}
\end{equation}
dominantly proceeds via gluon fusion.
As a result, the leading order cross section $\sigma(pp\to\tilde t_m\tilde t_m^\ast)$ 
is independent of any other SUSY model 
parameters than the stop mass~\cite{NLOstop}.
The strong dependence can be seen in Figure~\ref{fig:StopPairProduction}~(left),
where the cross section drops by six orders of magnitude with an increase of the stop mass 
from $0.2$~TeV to $2$~TeV. For the LHC center-of-mass energy $\sqrt{s_{pp}}=14$~TeV,
we have calculated the hadronic cross section at leading order  with the software package 
MadGraph~\cite{Stelzer:1994ta}. Next to leading order (NLO) calculations for stop production 
have been performed in Ref.~\cite{NLOstop}, 
and soft gluon resummation for light-flavored squark-antisquark pairs
at next to NLO  has been performed in Ref.~\cite{Langenfeld:2009eg}. 
Mixed stop pairs can only be generated 
at order $\alpha_s^4$  with a strongly suppressed rate~\cite{NLOstop}.

The produced stops have a distinct distribution in their boost 
\begin{equation}
\beta_{\tilde t}=\frac{|\mathbf{p}_{\tilde t}|}{E_{\tilde t}},
\label{eq:stopBoost}
\end{equation}
along the direction of their momenta. In  Figure~\ref{fig:StopPairProduction}~(right), we
show the normalized boost distribution for several values for the stop mass. 
In our benchmark scenario (Table~\ref{tab:ReferenceScenario}), the lightest stop has a mass of $m_{\tilde t_1}=470$~GeV and its' boost distribution peaks at $\beta_{\tilde t_1}\approx 0.9$. Thus the decaying stop is highly boosted in the laboratory frame, which reduces the triple product asymmetries as 
shown in the next Section.

%
%------------------------------------------------------------------------------
\subsection{Triple Product Asymmetries in the Laboratory Frame}
\label{sec:AsymmetryLab}
%------------------------------------------------------------------------------

%
\begin{figure}[t]
\centering
\includegraphics[clip,width=0.580\textwidth]{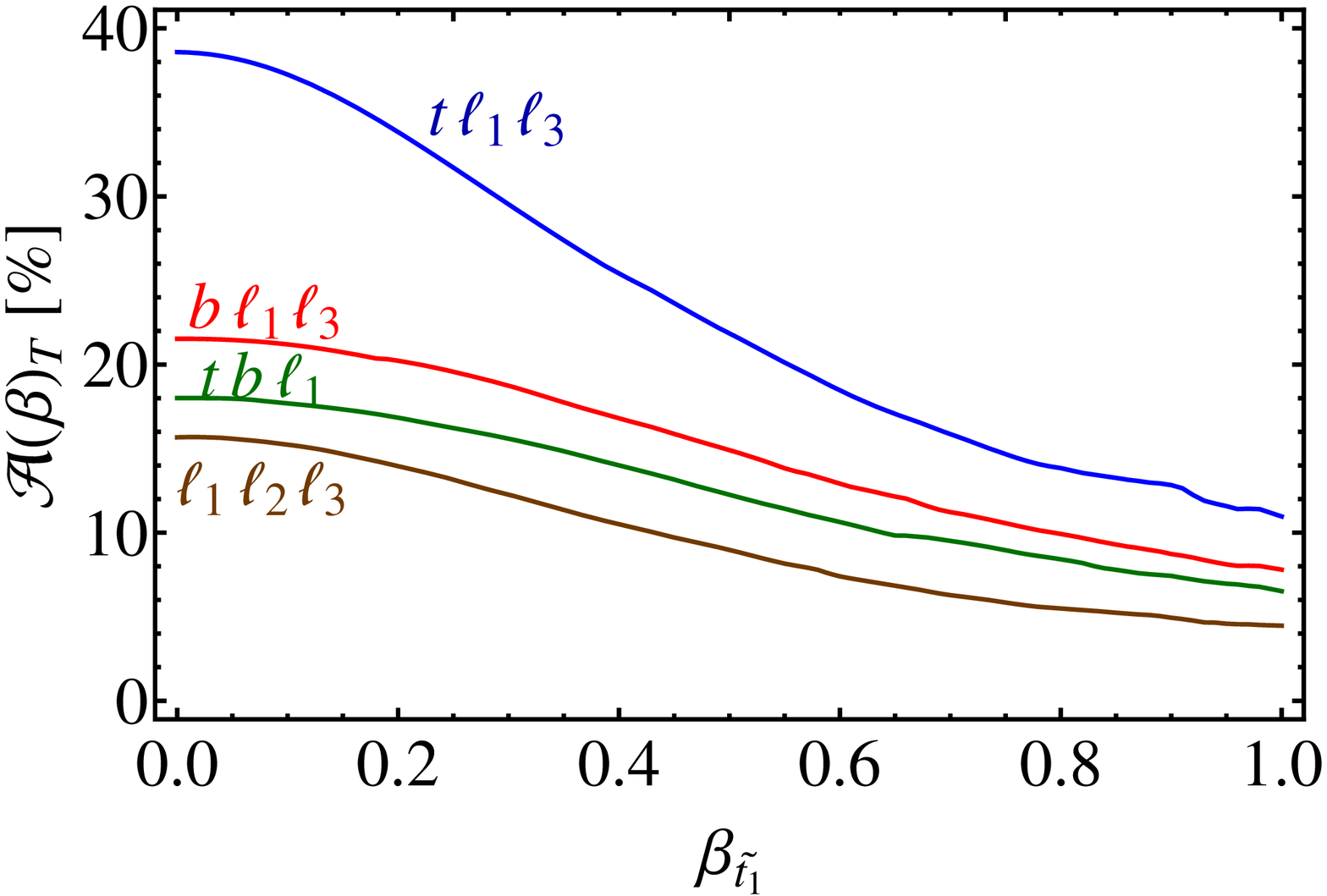}
\includegraphics[clip,width=0.410\textwidth]{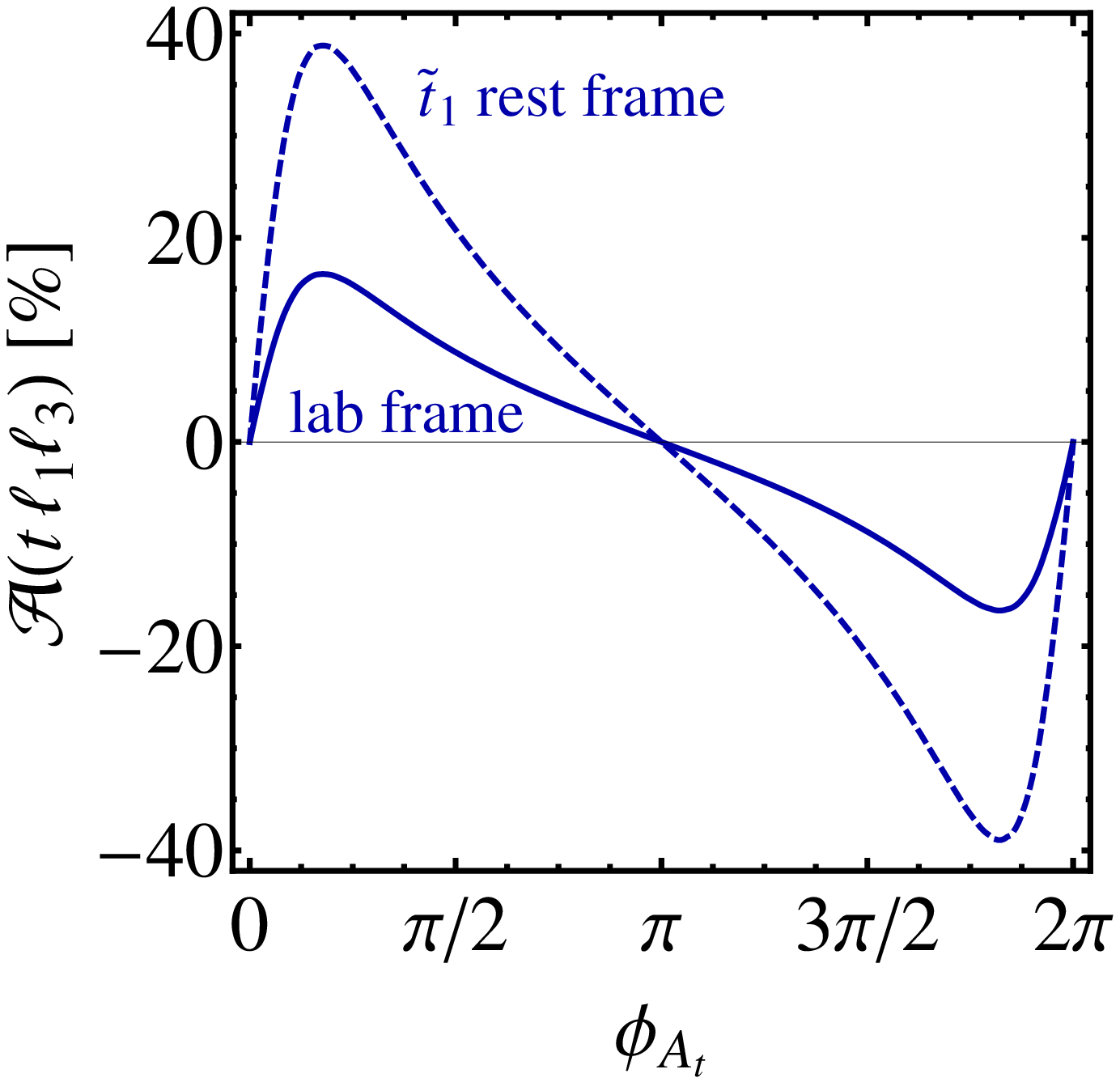}
\caption{Asymmetries \({\mathcal A}(t\ell_1\ell_3)\), \({\mathcal A}(b\ell_1\ell_3)\),
 \({\mathcal A}(tb\ell_1)\), \({\mathcal A}(\ell_1\ell_2\ell_3)\), see Eq.~(\ref{eq:Toddasym}),   for
stop decay ($\tilde t_1 \to t \tilde\chi^0_2$, $\tilde\chi^0_2\to \ell_1 \tilde \ell_{R}$,
 see Figure~\ref{Fig:decayStop}) as a function of the stop boost \(\beta_{\tilde t_1}\) (left). Asymmetry 
\({\mathcal A}(t\ell_1\ell_3)\) in the stop rest frame and the laboratory frame as a function of 
$\phi_{A_t}$ with $\phi_\mu=\phi_{M_1}=0$ (right). The mSUGRA parameters are given in Table~\ref{tab:ReferenceScenario}.}
\label{fig:BoostedAsymmetry}
\end{figure}

The asymmetries ${\mathcal A}$, Eq.~(\ref{eq:Toddasym}),
which are based on epsilon products  ${\mathcal E}= [p_{1},p_2,p_{3},p_{4}]$, Eq.~(\ref{eq:epsilon}),
are by construction Lorentz invariant. The corresponding asymmetries 
with triple products
depend on the stop boost.
In Figure~\ref{fig:BoostedAsymmetry}~(left), we show the boost dependence of asymmetries
for various triple product combinations ${\mathcal T}= (t\ell_1\ell_3)$, $(b\ell_1\ell_3)$, $(t b\ell_1)$, 
and $(\ell_1\ell_2\ell_3)$. The asymmetries are calculated for 
 $pp\to\tilde t_1\tilde t_1^\ast $ production at the LHC with the subsequent
decays $\tilde t_1 \to t \tilde\chi^0_2$  and $\tilde\chi^0_2\to \ell \tilde \ell_{R}$.
The functional dependence of the asymmetries on the stop boost is purely kinematic. 
In the stop rest frame, $\beta_{\tilde t}=0$, they coincide with the corresponding  
epsilon product asymmetries, and then decrease with increasing $\beta_{\tilde t}$. 
Note that the size of the asymmetries strongly depends
on the choice of momenta, and largest values are obtained for the optimal
triple product $(t\ell_1\ell_3)$ as given in Eq.~(\ref{eq:epsilon}).
However this asymmetry also decreases most strongly with $\beta_{\tilde t}$, see  
Figure~\ref{fig:BoostedAsymmetry}~(left).

A triple product asymmetry in the laboratory frame is then obtained by folding the boost dependent 
asymmetry ${\mathcal A}$ with the normalized stop boost distribution,
\begin{equation}
	{\mathcal A}^{\rm lab} =
	\frac{1}{\sigma}
	\int_{0}^{1} 
	\frac{d\sigma}{d\beta_{\tilde t}}
	~{\mathcal A}(\beta_{\tilde t})
	~d\beta_{\tilde t},
\label{eq:boostasymmetry}
\end{equation}
with the production cross section  $\sigma=\sigma(pp\to\tilde t_m\tilde t_m^\ast)$.
This leads to a reduction to roughly $30$-$40$\% of the asymmetry ${\mathcal A}^{\rm lab}$
compared to ${\mathcal A}(\beta_{\tilde t}=0)$.
This can be seen in Figure~\ref{fig:BoostedAsymmetry}~(right), where we show
both asymmetries as a function of the CP phase $\phi_{A_t}$. 
Thus the stop boost effectively reduces the asymmetries, but does not change their 
shape with respect to the phase dependence.
In the following we will focus on the folded asymmetry  ${\mathcal A}(t\ell_1\ell_3)$ in the laboratory frame but similar discussions hold for asymmetries based on the other triple products.

%\newpage
%------------------------------------------------------------------------------
\subsubsection*{Theoretical Statistical Significance}
\label{sec:Significances}
%------------------------------------------------------------------------------

So far we have only discussed the asymmetry of the triple product, but in order to determine the statistical significance of a given asymmetry it is also necessary to take into account the total signal rate. For example, a large  asymmetry does not necessarily result in a large significance if the signal rate is low as this does increase the statistical uncertainty.

Assuming that the fluctuations of the signal rate are binomially distributed with the selection probability $p=1/2(\mathcal{A}+1)$, the significance of an asymmetry is
\begin{eqnarray}
	{\mathcal S} = 
	\frac{|{\mathcal A}|}{\sqrt{1-{\mathcal A}^2}} 
	\sqrt{\sigma{\mathcal L}},
\label{eq:significance}
\end{eqnarray} 
with the integrated LHC luminosity ${\mathcal L}$.
The cross section  is 
\begin{eqnarray}
	\sigma &=& F_N \times
	\sigma(pp\to\tilde t_m\tilde t_m^\ast)  
	\times {\rm Br}(\tilde t_m \to t  \tilde\chi^0_i) 
	\times {\rm Br}(t\to b  W) \times {\rm Br}(W \to \nu_e e)
 \nonumber\\[2mm]
	&& \phantom{F_N }
        \times {\rm Br}(\tilde\chi^0_i\to  e^+\tilde e^-_{n} ) 
	\times {\rm Br}(\tilde e^-_{n} \to \tilde\chi^0_1 e^-),
	\label{eq:crosssection}
\end{eqnarray}
for $m=1,2$, $i=2,3,4$, and $n=L,R$.
The combinatorial factor $F_N$ takes into account the possible $W$ and  neutralino $\tilde\chi^0_i$ 
decays into leptons with different flavors.
We assume that the branching ratios do not depend on the flavor, i.e.,
${\rm Br}(W \to \nu_e e) = {\rm Br}(W \to \nu_\mu \mu)$,
and $
 {\rm Br}(\tilde\chi^0_i\to e^{+}\tilde e^{-}_{n})=
 {\rm Br}(\tilde\chi^0_i\to \mu^{+}\tilde \mu^{-}_{n})$.
The   factor is thus $F_N=8$, if we sum  the lepton flavors $e, \mu$,
and also the slepton charges, $ \tilde\ell^{-}_{n}$ and $ \tilde\ell^{+}_{n}$.

The statistical significance  $\mathcal{S}$ is equal to the number of standard 
deviations that the asymmetry can be determined to be non-zero. For example a value of $\mathcal{S} = 1$ implies a measurement at the $68\%$ confidence level. The minimal required luminosity is then
\begin{eqnarray}
{\mathcal L} &=& \frac{1}{\sigma}\left( \frac{1}{{\mathcal A^2}} -1 \right).
\label{eq:lumi}
\end{eqnarray}
Our definition of the statistical significance $\mathcal{S}$ and the required luminosity $\mathcal{L}$ is purely based on the theoretical signal rate and its asymmetry. Detector efficiency effects and contributions from CP-even backgrounds are neglected, which would reduce the effective asymmetry. The definitions have thus to be regarded as absolute upper bounds on the confidence levels and as absolute lower bounds on the minimal required luminosities, respectively. We will use these terms in the following to exclude mSUGRA parameter regions which cannot be probed at the LHC. In order to give realistic values of the statistical significances and required luminosities to observe a CP signal, a detailed experimental study is necessary.

Also higher order corrections will be important and have to be included in a comprehensive analysis. 
Leading-order (LO) QCD calculations strongly depend on the a priori 
unknown renormalization and factorization scales.  As a result, theoretical predictions of LO QCD 
cross sections and branching ratios are easily uncertain within a factor of $2$, 
see for example Ref.~\cite{NLOstop}.
However, we expect that asymmetries and distributions will have much smaller scale dependencies.  
Since the higher order corrections enter both in the numerator and the denominator, 
they might cancel partially,  and may lead to a significantly reduced scale dependence of the asymmetries.
For instance, the normalized stop boost distribution shown in 
Fig.~\ref{fig:StopPairProduction} (right) only changes
slightly if switching on initial and final state radiations in PYTHIA~\cite{Sjostrand:2006za}. 
Soft gluon radiation could in principle change the spin structure of the stop decay process, 
see Figure~\ref{Fig:decayStop}.
However, only the top-neutralino spin-spin correlations are CP-sensitive,
and thus they could only be altered by gluon radiation of the top alone.

Finally, our asymmetries probe CP-violating phases of the stop-top-neutralino 
couplings, which only belong to the electroweak sector.
However, we expect that also the effect of electroweak corrections 
to our observables and  asymmetries is rather small. 
Although electroweak corrections to  neutralino masses, for example, 
can be $10\%$ at one-loop level~\cite{Oller:2005xg},
and neutralino branching ratios for two-body decays may receive  
electroweak CP-even one-loop corrections of up to $16\%$ in some cases~\cite{Drees:2006um}, 
we expect that asymmetries are less sensitive to these electroweak corrections, since they again enter 
both in the numerator and denominator. 

%\newpage
%------------------------------------------------------------------------------
\subsection{Phase Dependence}
\label{sec:PhaseDependenceN}
%------------------------------------------------------------------------------

%
\begin{figure}[t]
\centering
\includegraphics[clip,width=0.495\textwidth]{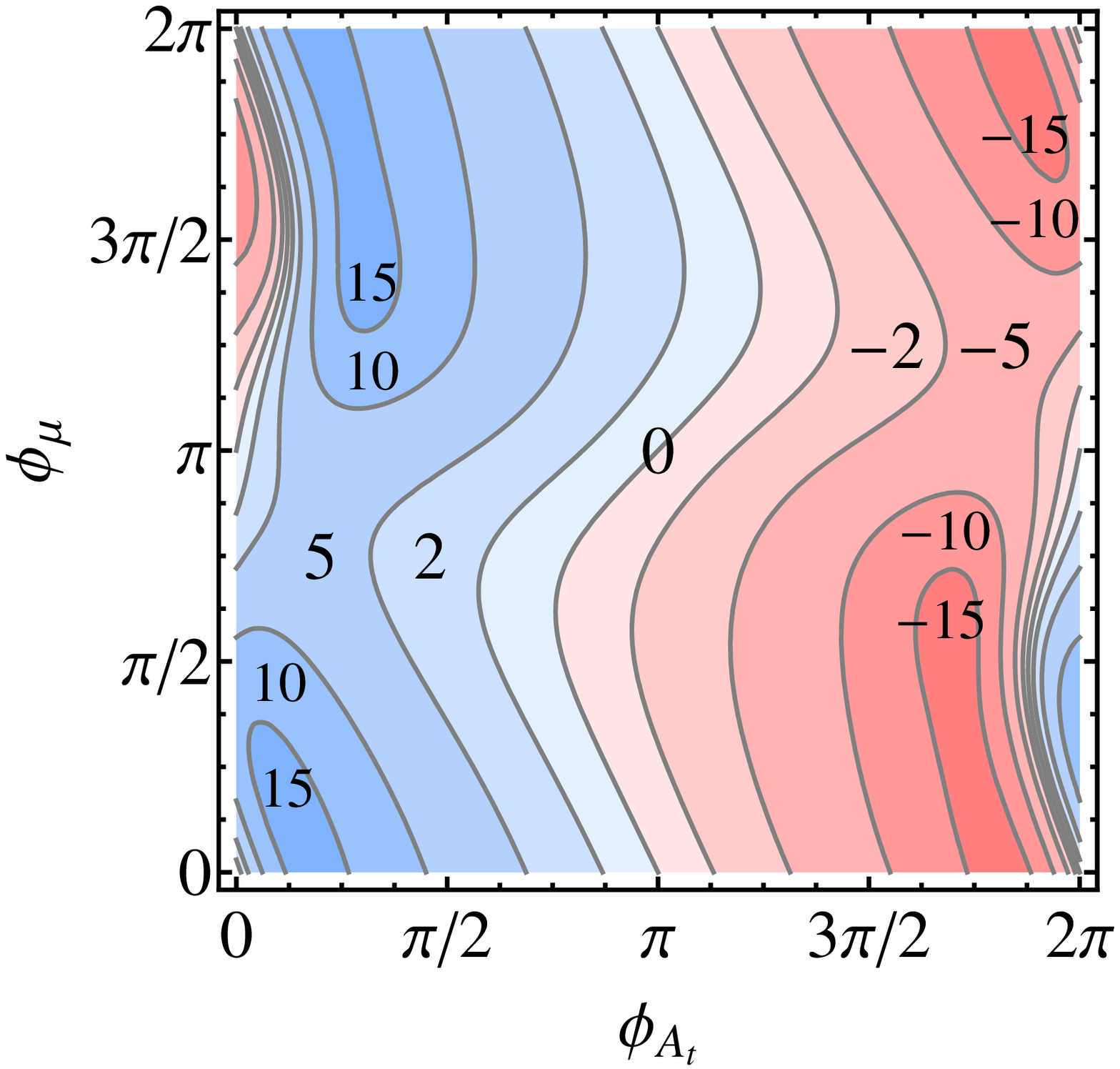}
\includegraphics[clip,width=0.495\textwidth]{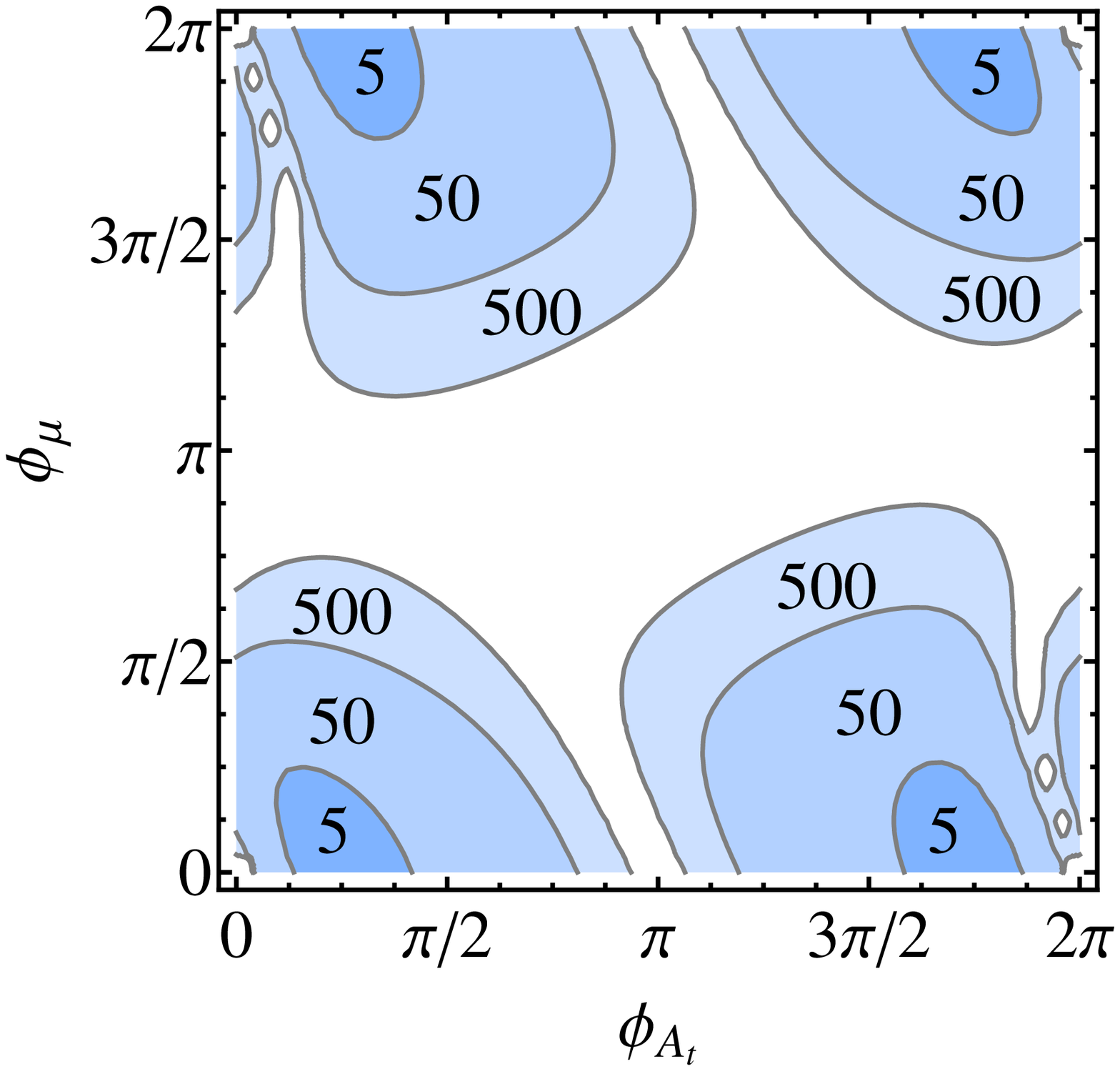}
\caption{Triple product asymmetry $\mathcal{A}^{\rm lab}(t\ell_1\ell_3)$, 
         see Eq.~(\ref{eq:boostasymmetry}), in percent (left), and minimal bound on the 
         luminosity ${\mathcal L}$, Eq.~(\ref{eq:lumi}), in fb$^{-1}$ to detect such an asymmetry 
         at  \(1\sigma\) above fluctuations (right), 
         as functions of the CP phases $\phi_{A_t}$, $\phi_{\mu}$, 
         with $\phi_{M_1}=0$, for stop production at the LHC 
         with $\sqrt{s_{pp}}=14$~TeV and decay  $\tilde t_1 \to t \tilde\chi^0_2$, 
         $\tilde\chi^0_2\to \ell_1 \tilde \ell_{R}$ see Figure~\ref{Fig:decayStop}.
         The mSUGRA parameters are given in Table~\ref{tab:ReferenceScenario}.}
\label{fig:PhaseDependence}
\end{figure}

We  analyze the $\phi_{A_t}$ and $\phi_\mu$ dependence of the triple product asymmetry 
${\mathcal A}^{\rm lab}$ in the lab frame, as defined in Eq.~(\ref{eq:boostasymmetry}).
For stop $pp\to\tilde t_1\tilde t_1^\ast $ production and subsequent
decays $\tilde t_1 \to t \tilde\chi^0_2$  and $\tilde\chi^0_2\to \ell \tilde \ell_{R}$,
we show in Figure~\ref{fig:PhaseDependence}~(left) contour lines of
${\mathcal A}(t \ell_1\ell_3)$ in the  $\phi_{A_t}$--$\phi_\mu$ plane
for our reference scenario (Table~\ref{tab:ReferenceScenario}). 
The maximal asymmetry in the lab frame is  about $15\%$,
whereas the dependence on $\phi_{M_1}$ is weak, due to the negligible bino admixture of 
$\tilde\chi_2^0$ in mSUGRA scenarios.
Note that the phase dependence of the asymmetry Figure~\ref{fig:PhaseDependence}~(left) closely 
resembles that of the effective coupling factor $\eta$, Eq.~(\ref{eq:couplingfunct2}),
shown in Figure~\ref{fig:EtaPhiAtPhiMu}~(right) for the same scenario.
Thus the asymmetry directly probes the CP-violating structure of the stop-top-neutralino couplings.
The kinematical part of the asymmetry is largely phase independent.

To determine the potential observability of the asymmetry, we calculate the minimal required luminosity 
${\mathcal L}$, Eq.~(\ref{eq:lumi}), to observe a $1\sigma$ deviation of the asymmetry above the 
statistical fluctuations. Contour lines of ${\mathcal L}$ in the $\phi_{A_t}$--$\phi_\mu$ plane
are shown in Figure~\ref{fig:PhaseDependence}~(right).
A large part of the parameter space can be probed with ${\mathcal L}=500$~fb$^{-1}$. 
Even small phases close to the CP-conserving points $\phi_{A_t}$, $\phi_{\mu}=0,\pi$ 
can be probed with ${\mathcal L}=50$~fb$^{-1}$.
The significance also depends on the stop and neutralino branching ratios,
see Eq.~(\ref{eq:crosssection}), which depend differently on the phases $\phi_{A_t}$, $\phi_\mu$. 
Thus the contour lines of the luminosity do not directly follow those of the asymmetry. 

%\newpage
%------------------------------------------------------------------------------
\subsection{mSUGRA Parameter Space}
\label{sec:mSUGRAScans}
%------------------------------------------------------------------------------

So far we have analyzed the asymmetries within our mSUGRA scenario 
(Table~\ref{tab:ReferenceScenario}). 
We now extend our study to the entire mSUGRA parameter space, and study
the dependence of the asymmetries and  the event rates on the parameters
 $ m_0$, $m_{1/2}$, $\tan\beta$, and $A_0$. In such an mSUGRA framework, the  
production of light stops will dominate, $pp\to\tilde t_1\tilde t_1^\ast $, followed
by the subsequent decays $\tilde t_1 \to t \tilde\chi^0_2$,
and $\tilde\chi^0_2\to \ell \tilde \ell_{R}$.

%------------------------------------------------------------------------------
\subsubsection*{Dependence on $ m_0$ and $m_{1/2}$}
\label{sec:m12mo}
%------------------------------------------------------------------------------
\begin{figure}[t]
\centering
\includegraphics[clip,width=0.32\textwidth]{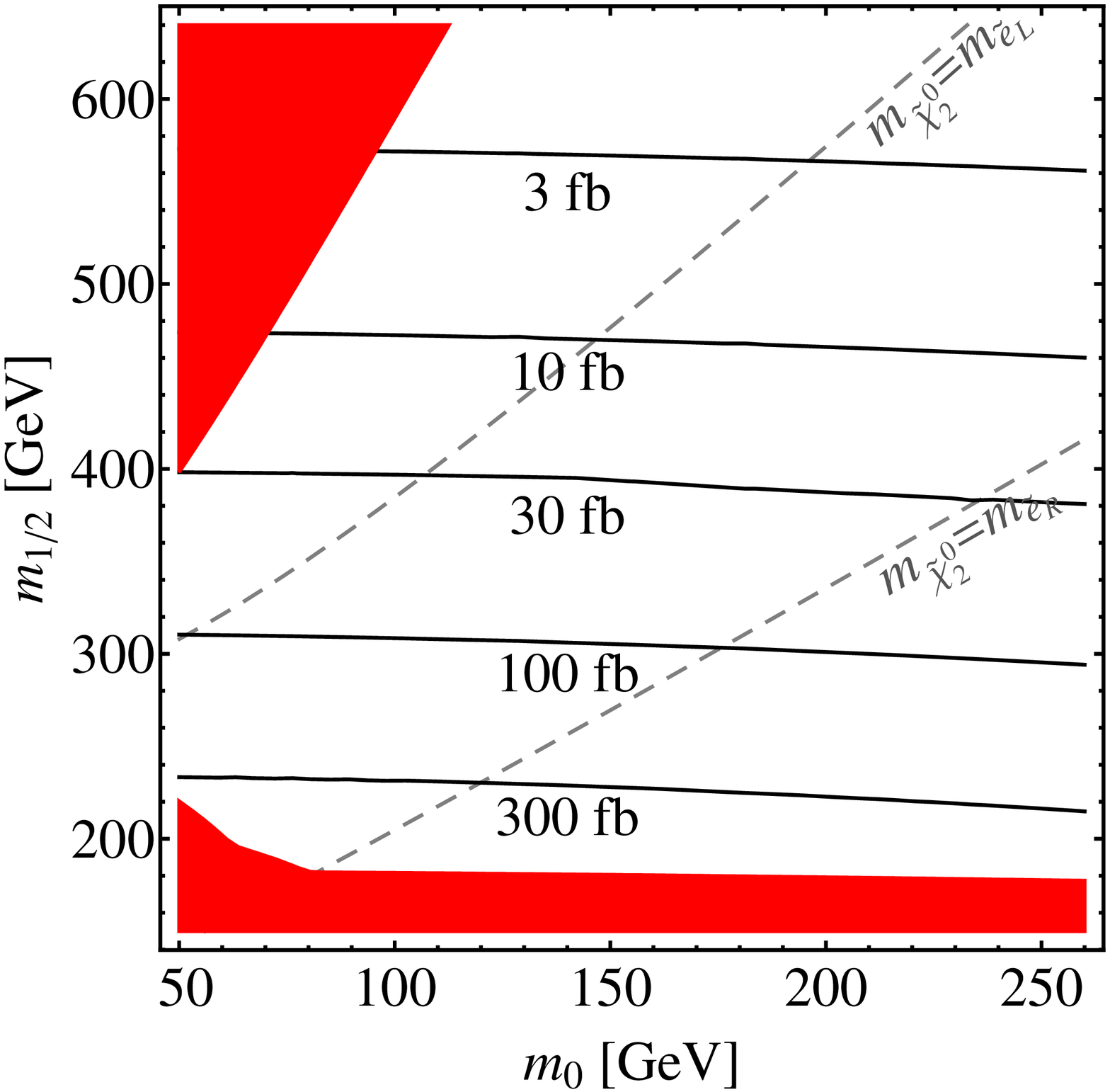}
\includegraphics[clip,width=0.32\textwidth]{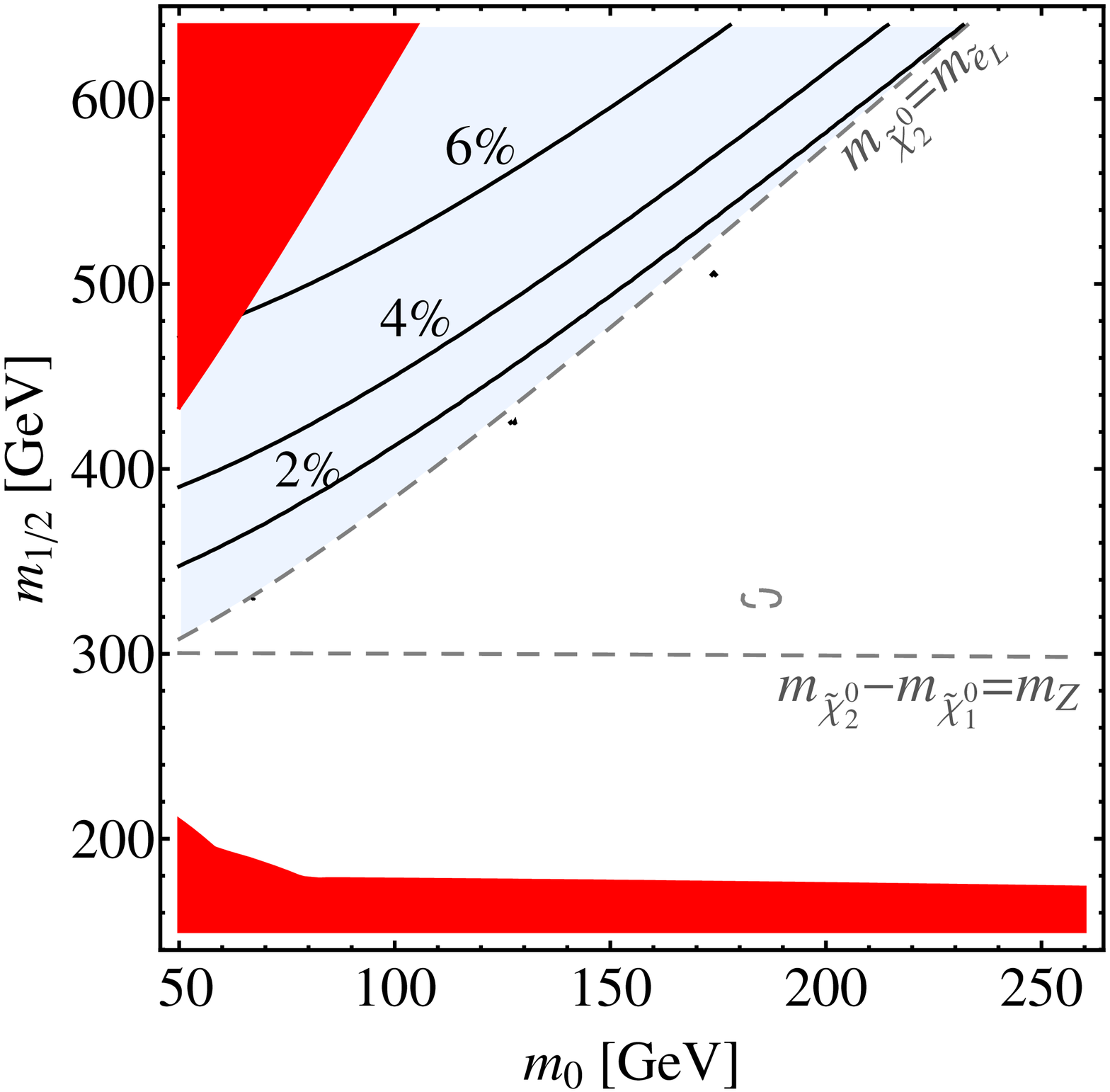}
\includegraphics[clip,width=0.32\textwidth]{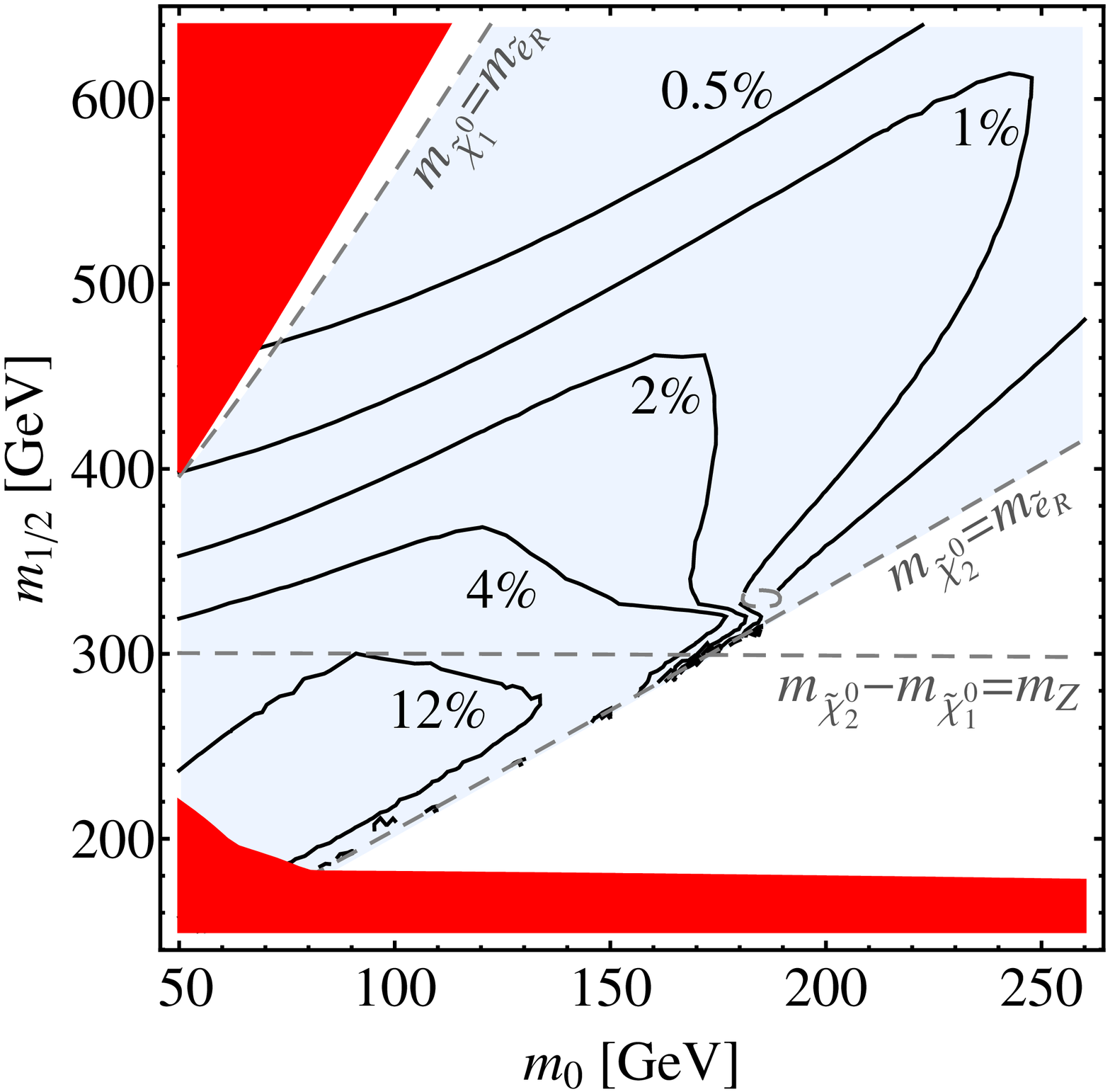}
\caption{Cross section $\sigma(pp\to\tilde t_1\tilde t_1^\ast)\times {\rm Br}(\tilde t_1\to t\tilde\chi_2^0)$ (left), neutralino branching ratio into left slepton ${\rm Br}(\tilde\chi_2^0\to e^{+}\tilde e_L^{-})$ (middle), and  $\tilde\chi_2^0$ branching ratio into right slepton ${\rm Br}(\tilde\chi_2^0\to e^{+}\tilde e_R^{-})$ (right), as a function of the mSUGRA parameters \(m_0\) and \(m_{1/2}\). The other mSUGRA parameters and the CP phases are given in Table~\ref{tab:ReferenceScenario}. The dark red area is excluded by direct SUSY searches.}
\label{fig:Brs}
\end{figure}

The asymmetries in the stop rest frame and in the LHC lab frame are 
rather independent of the masses of the stop and its decay products.
Thus we find an almost constant asymmetry  ${\mathcal A}^{\rm lab}(t\ell_1\ell_3)\approx 15\%$ 
within the kinematical allowed region of the $ m_0$--$m_{1/2}$ parameter plane.
However the stop production rate and the branching ratios 
of the neutralinos sensitively depend on $m_0$ and $m_{1/2}$.

In Figure~\ref{fig:Brs}~(left), we show contour lines of the  leading order (LO)  cross section 
$\sigma(pp\to\tilde t_1\tilde t_1^\ast)\times {\rm Br}(\tilde t_1\to t\tilde\chi_2^0)$ 
in the $m_0$--$m_{1/2}$ plane.  
While the  stop decay branching ratio  ${\rm Br}(\tilde t_1\to t\tilde\chi_2^0)\approx 0.1 $ 
is almost constant, the stop production cross section %$\sigma(pp\to\tilde t_1\tilde t_1^\ast)$ 
strongly decreases with increasing stop mass, see Figure~\ref{fig:StopPairProduction}~(left).
Therefore the cross section also rapidly drops with increasing $m_{1/2}$, which clearly  can be 
seen in Figure~\ref{fig:Brs}~(left).
The dashed lines $m_{\chi_2^0} = m_{\tilde e_{R,L}}$ 
%and $m_{\chi_2^0} =m_{\tilde e_L}$
indicate the neutralino decay thresholds into sleptons.
We show their branching ratios  ${\rm Br}(\tilde\chi_2^0\to e^{+}\tilde e_L^{-})$ and 
${\rm Br}(\tilde\chi_2^0\to e^{+}\tilde e_R^{-})$ in Figure~\ref{fig:Brs}~(middle) and (right),
respectively. The branching ratio  into right selectrons
reaches up to $12\%$ for small values of $m_0$ and $m_{1/2}$. 
The decay fraction is considerably reduced for $m_{1/2}>300$~GeV, where decays into the $Z$ and 
the lightest neutral Higgs boson open, see the dashed
line $m_{\chi_2^0} - m_{\chi_1^0} = m_Z$ in  Figure~\ref{fig:Brs}.
The significance for observing an asymmetry for the decay $\tilde\chi^0_2\to \ell \tilde \ell_{L}$
is suppressed to that for the decay $\tilde\chi^0_2\to \ell \tilde \ell_{R}$
for most of the $m_0$--$m_{1/2}$ parameter space. 
This is because the branching ratio 
${\rm Br}(\tilde\chi_2^0\to\ell \tilde \ell_{L})$ is larger for higher values of $m_{1/2}$,
where however the stop production cross section is suppressed. 
We thus find  that both $m_0$ and $m_{1/2}$ should be as small as possible 
to obtain the highest process rates.

\begin{figure}[t]
\centering
\includegraphics[clip,width=0.495\textwidth]{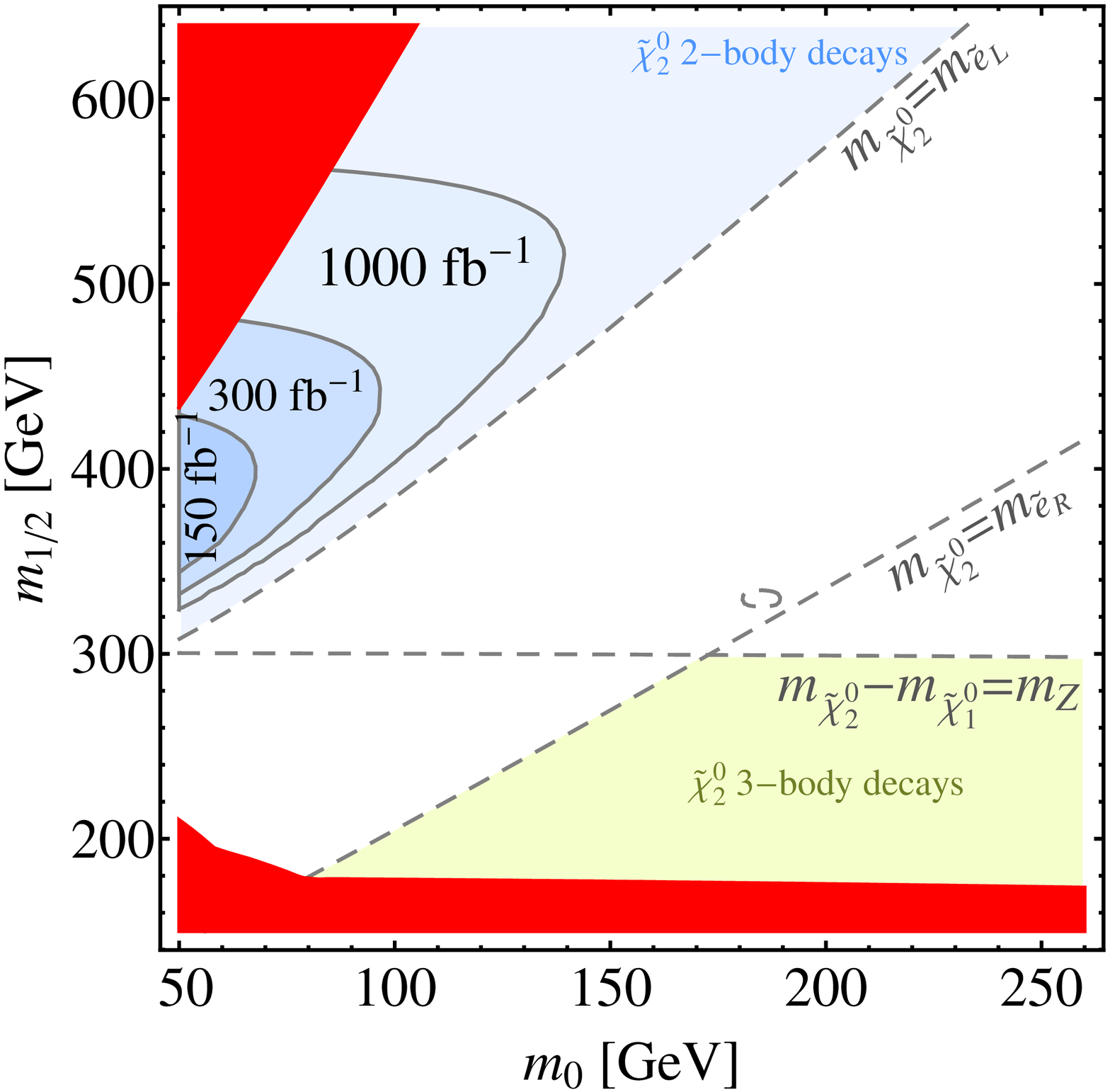}
\includegraphics[clip,width=0.495\textwidth]{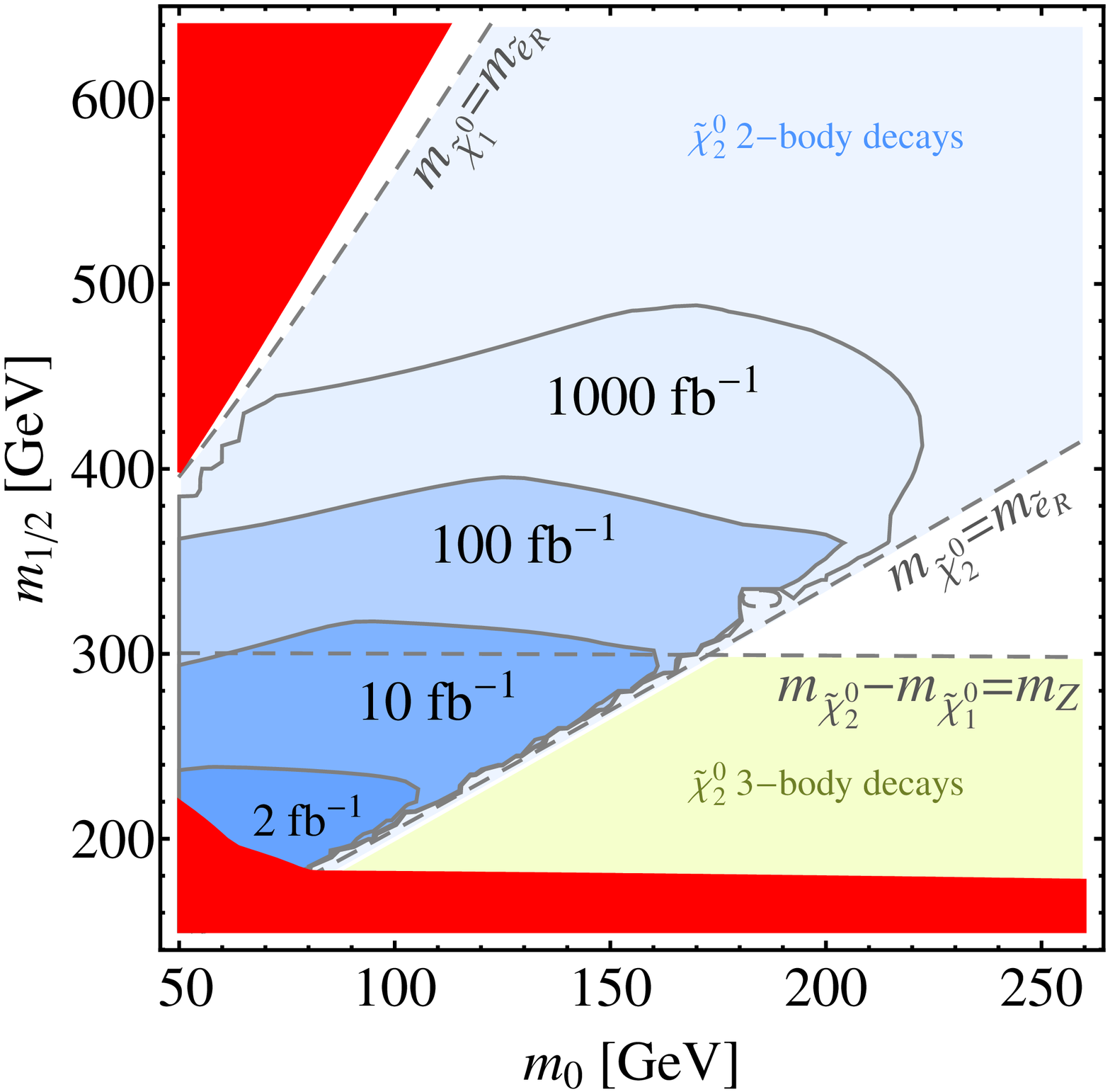}
\caption{Contour lines in the $m_0$--$m_{1/2}$ plane of the 
         minimal required luminosity  ${\mathcal L}$, Eq.~(\ref{eq:lumi}), to observe
         the asymmetry $\mathcal{A}^{\rm lab}(t\ell_1\ell_3)$, Eq.~(\ref{eq:boostasymmetry}), 
         at $1\sigma$ above fluctuations 
         for stop production at the LHC with $\sqrt{s_{pp}}=14$~TeV 
         and decay $\tilde t_1 \to t \tilde\chi^0_2$, 
         via a left slepton $\tilde\chi^0_2\to \ell_1 \tilde \ell_{L}$ (left) 
         and right slepton $\tilde\chi^0_2\to \ell_1 \tilde \ell_{R}$ (right),
         see Figure~\ref{Fig:decayStop}. 
         The other mSUGRA parameters and the CP phases are given in 
         Table~\ref{tab:ReferenceScenario}. 
         The dark red area is excluded by direct SUSY searches.}
\label{fig:ScanM0M12}
\end{figure}
These findings are confirmed when we calculate the minimal 
luminosity ${\mathcal L}$, Eq.~(\ref{eq:lumi}), which is required to observe the 
asymmetry at $1\sigma$  for stop production $pp\to\tilde t_1\tilde t_1^\ast $ at the LHC, with the subsequent decays
$\tilde t_1 \to t \tilde\chi^0_2$ and $\tilde\chi^0_2\to \ell \tilde \ell_{L(R)}$,
see Figure~\ref{fig:ScanM0M12} (left) and (right), respectively.
The mSUGRA parameters and the CP phases are given in Table~\ref{tab:ReferenceScenario}. 
For right sleptons, the required integrated luminosity can be as low as $2$~fb$^{-1}$ 
for  $m_0\approx 50-100$~GeV and $m_{1/2}\approx 200$~GeV. 
Asymmetries involving left sleptons require at least $\mathcal{L}\approx 150$~fb$^{-1}$ for
$m_0\approx 50$~GeV and $m_{1/2}\approx 340-420$~GeV. In the overlap between the allowed 
kinematical regions for decays into left and right sleptons, the rates can be comparable, 
though. As the two asymmetries have the same magnitude but opposite signs, 
$\mathcal{A}^{\tilde \ell_L}(t\ell_1\ell_3)= - \mathcal{A}^{\tilde \ell_R}(t\ell_1\ell_3)$, 
this would lead to a reduction of the total asymmetry in this intermediate region.
In Figure~\ref{fig:ScanM0M12}, we also indicate the $m_0$--$m_{1/2}$ parameter region 
where three-body decays $\tilde\chi^0_2\to\ell^+\ell^-\tilde\chi^0_1$ 
have been used to study CP asymmetries~\cite{Ellis:2008hq}. 

%\newpage
%------------------------------------------------------------------------------
\subsubsection*{Dependence on $ A_0$ and $\tan\beta$}
\label{sec:A0tanb}
%------------------------------------------------------------------------------

%
\begin{figure}[t]
\centering
\includegraphics[clip,width=0.495\textwidth]{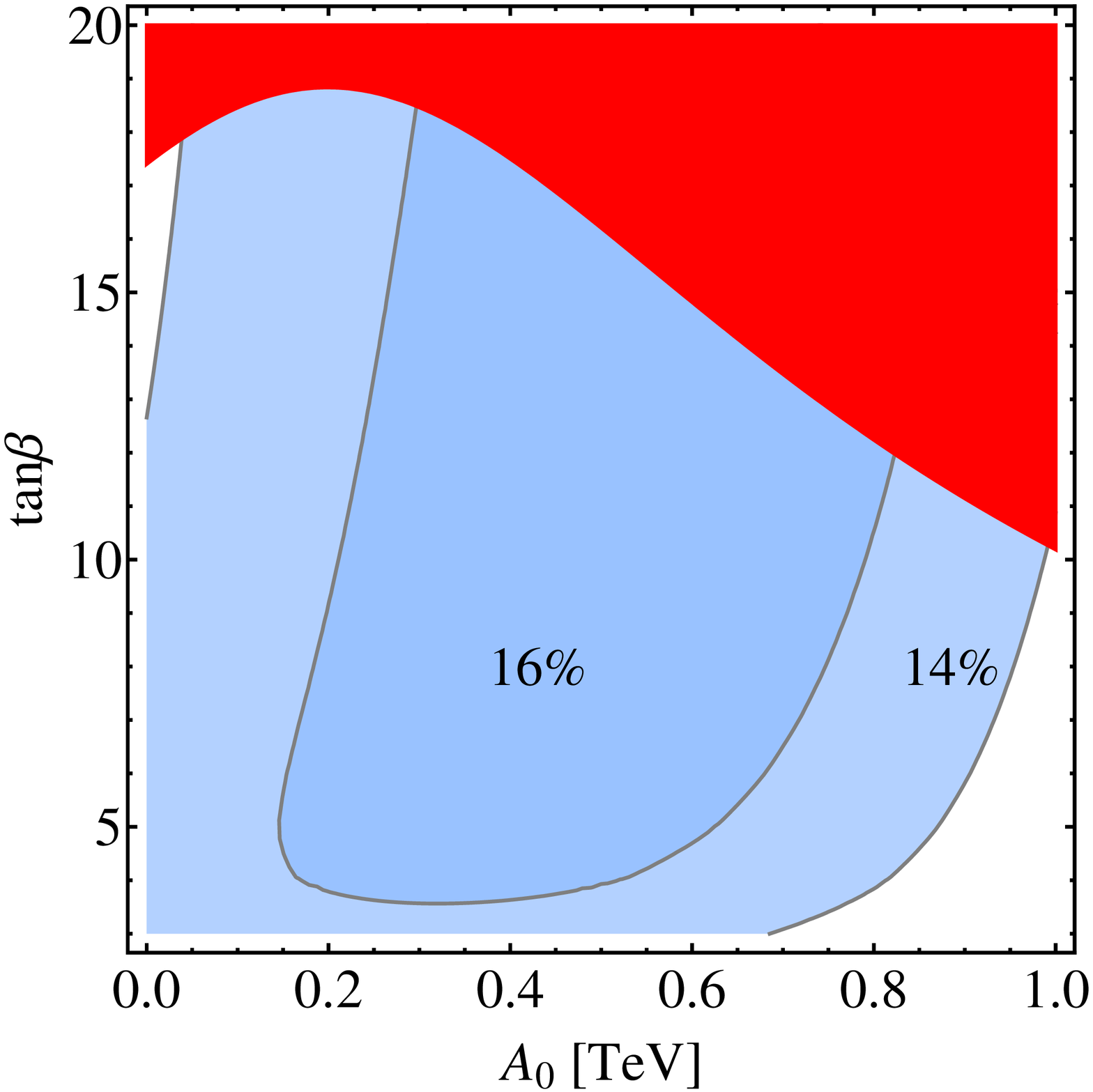}
\includegraphics[clip,width=0.495\textwidth]{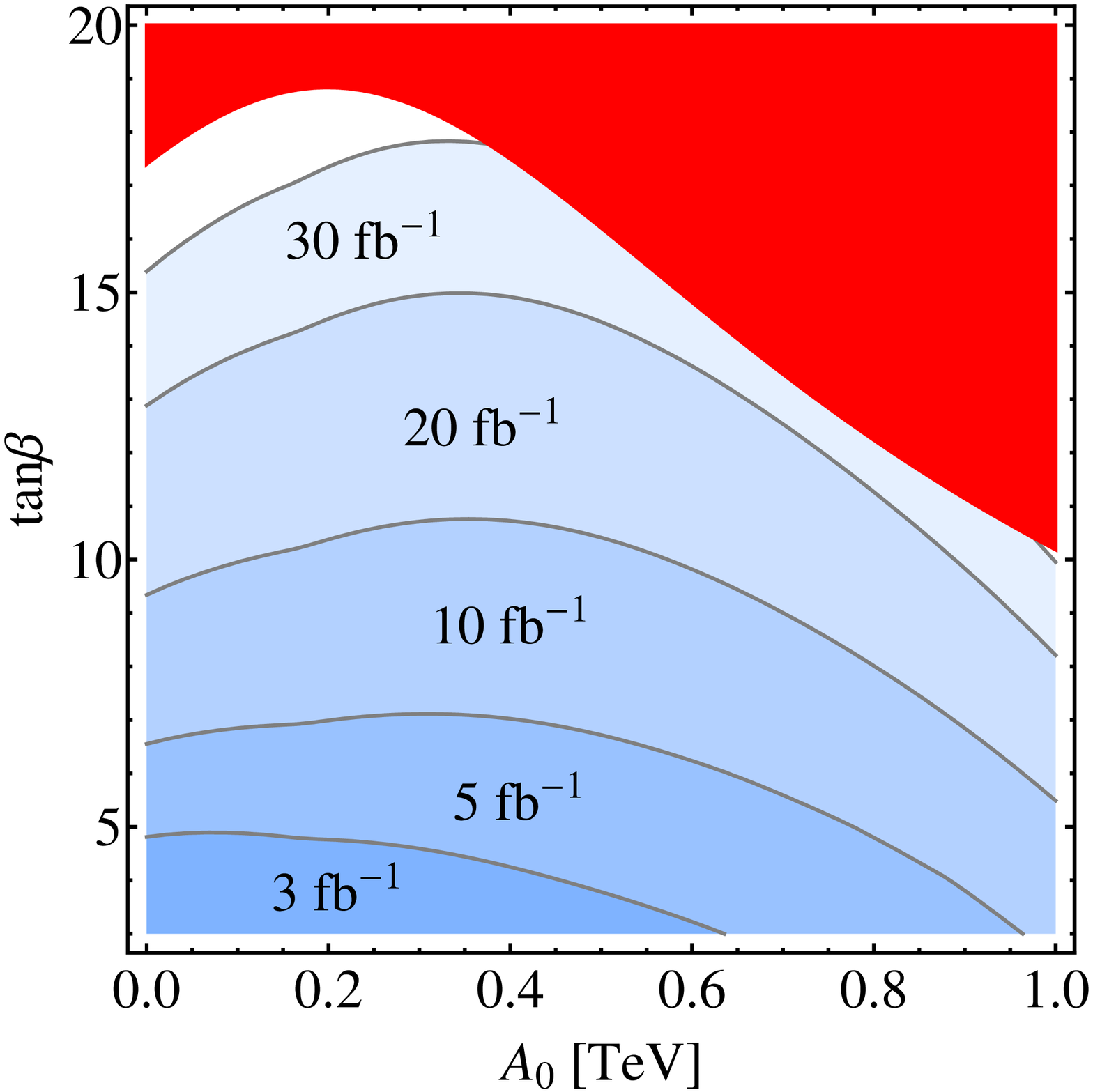}
\caption{Contour lines in the $A_0$--$\tan\beta$ plane of 
         the asymmetry ${\mathcal A}^{\rm lab}(t\ell_1\ell_3)$, Eq.~(\ref{eq:boostasymmetry}),
         in the laboratory frame (left), and the minimal required luminosity 
         ${\mathcal L}$, Eq.~(\ref{eq:lumi}), to observe 
         it at $1\sigma$ above fluctuations (right), 
         for stop production at the LHC 
         with $\sqrt{s_{pp}}=14$~TeV and decay  $\tilde t_1 \to t \tilde\chi^0_2$, 
         $\tilde\chi^0_2\to \ell_1 \tilde \ell_{R}$, see Figure~\ref{Fig:decayStop}.
         The other mSUGRA parameters and CP phases are given in 
         Table~\ref{tab:ReferenceScenario}. The dark red area is excluded by 
         direct SUSY searches.}
\label{fig:ScanA0TanBeta}
\end{figure}
In Figure~\ref{fig:ScanA0TanBeta}, we show contour lines in the $A_0$--$\tan\beta$ plane 
of the asymmetry \({\mathcal A}(t\ell_1\ell_3)\) 
for stop production, $pp\to\tilde t_1\tilde t_1^\ast $, 
(and decay $\tilde t_1 \to t \tilde\chi^0_2$,  $\tilde\chi^0_2\to \ell \tilde \ell_{R}$,
see Figure~\ref{Fig:decayStop})
 and the corresponding minimal luminosity to observe it at $1\sigma$ above fluctuations.
All other parameters are given as in the reference scenario, Table~\ref{tab:ReferenceScenario}. 
The asymmetry is essentially independent of $A_0$ and $\tan\beta$ over the region of interest, slightly peaking at 16.5\% at $A_0\approx 500$~GeV and $\tan\beta\approx 10-15$. For small values of $\tan\beta\lsim 2$, a reduced stop mixing leads to smaller asymmetries. On the other hand, an increasing value of $A_0$ also enhances the stau neutralino branching ratios, $\tilde\chi_2^0\to \tau\tilde\tau$. As a result of this, the dependence of the required luminosity on $A_0$ and $\tan\beta$ is dominated by the rate, and the lowest luminosities are achieved for small values of $A_0$ and $\tan\beta$, see Figure~\ref{fig:ScanA0TanBeta} (right).

%\newpage
%------------------------------------------------------------------------------
\subsection{Angular and Volume Distributions of Triple Products}
\label{sec:AsymmetryDistributions}
%------------------------------------------------------------------------------
%

A measurement of an asymmetry requires the reconstruction 
of the spatial momenta of the three particles  which form the triple product.
It has to be determined whether the three momenta form a left- or right-handed trihedron. 
This can certainly only be answered by a detailed experimental simulation for the LHC.
However to gain some insight, we briefly discuss the distributions of the stop decay width with respect to the volume and opening angles of the trihedron spanned by the momenta.

\begin{figure}[t]
\centering
\includegraphics[clip,width=0.495\textwidth]{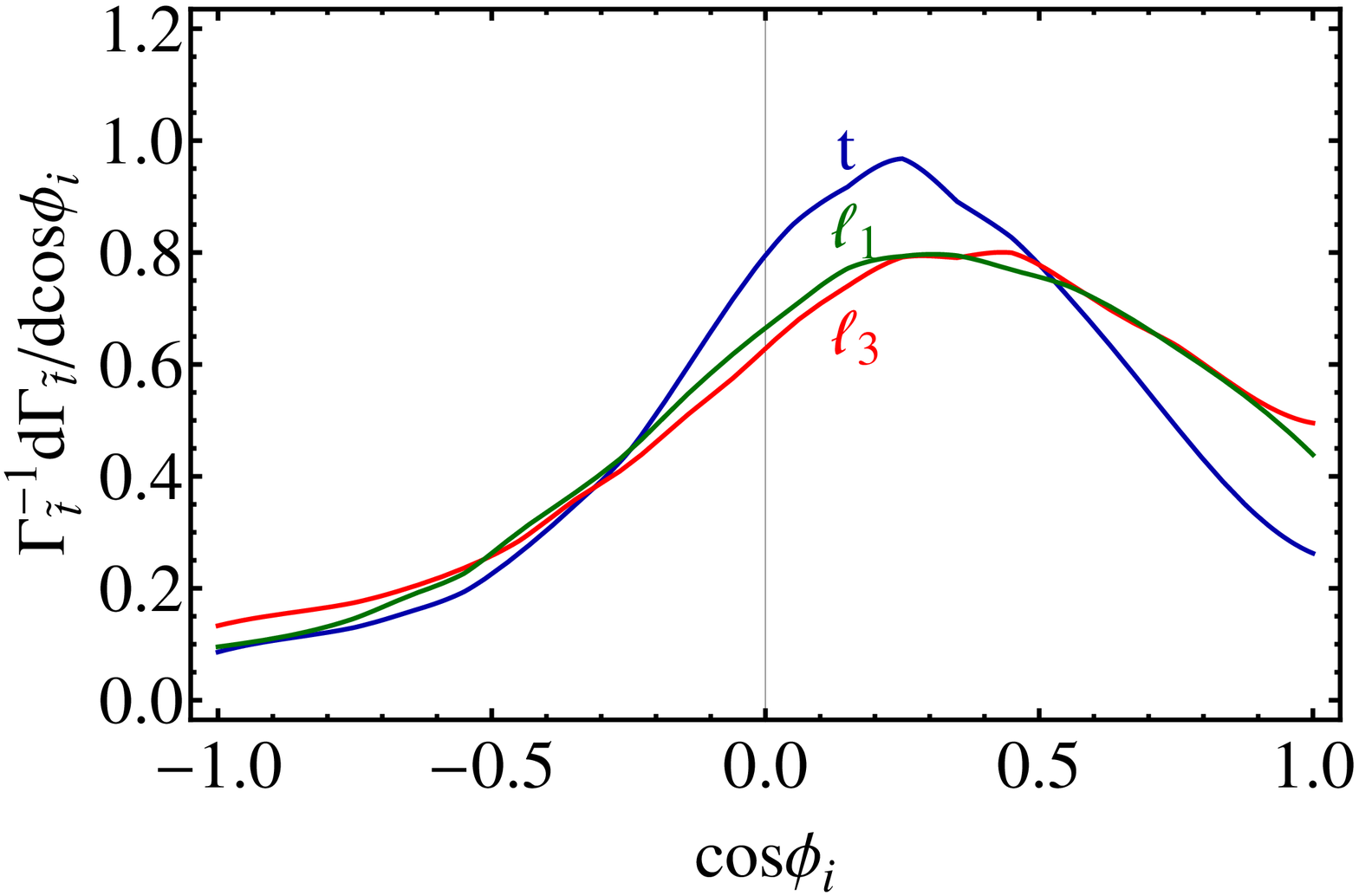}
\includegraphics[clip,width=0.495\textwidth]{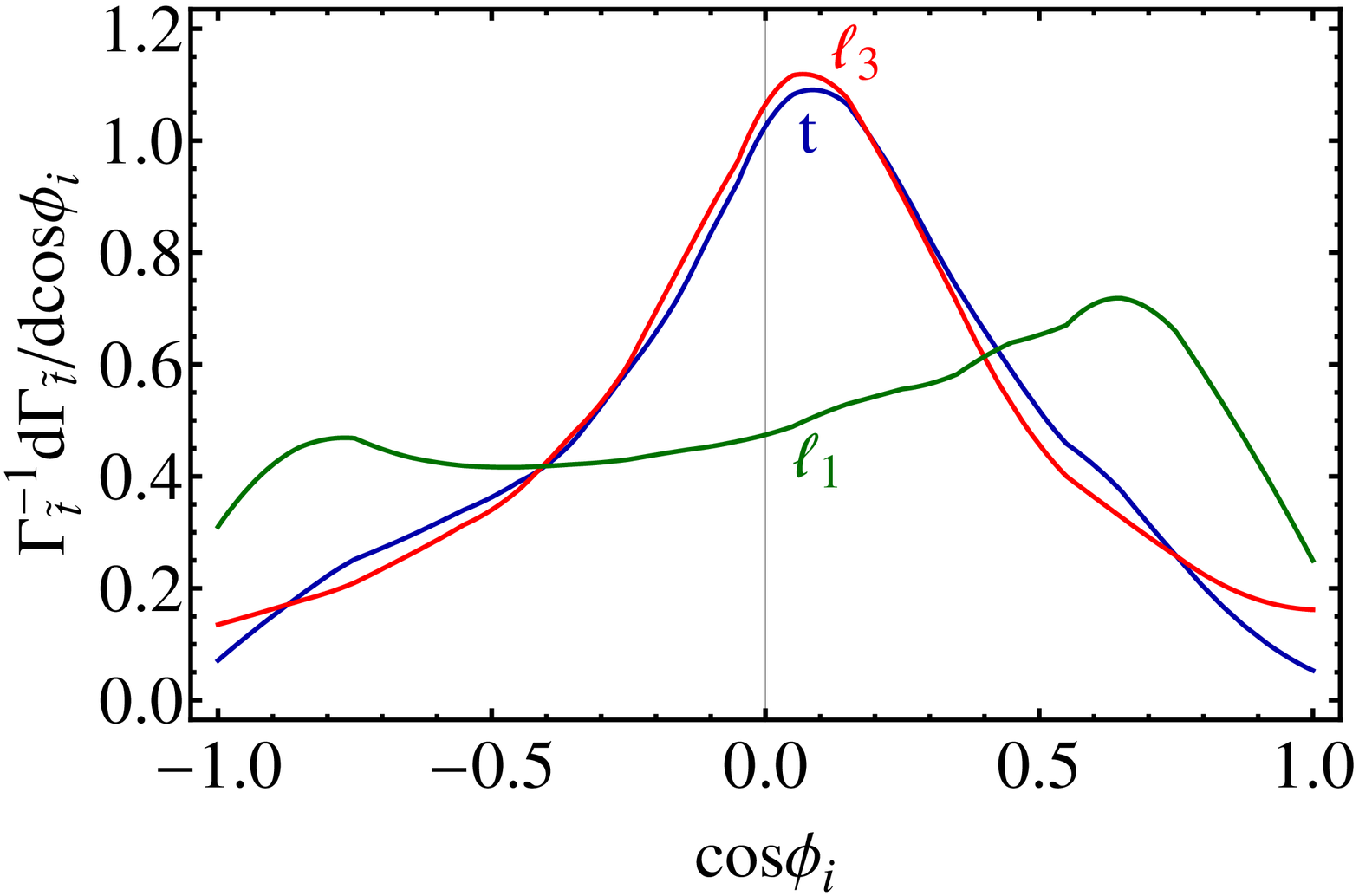}
\caption{Normalized distributions of the stop decay width $\Gamma$ with respect to the cosine of the three 
opening angles $\phi_i$ ($i=t,\ell_1,\ell_3$) of the parallelepiped ($\mathbf{p}_t,\mathbf{p}_{\ell_1},\mathbf{p}_{\ell_3}$), e.g.  \(\phi_t  =\varangle(\mathbf{p}_t,\mathbf{p}_{\ell_1}\times\mathbf{p}_{\ell_3})\). Results are shown for stop boost factors \(\beta_{\tilde t_1}=0\) (left), and \(\beta_{\tilde t_1}=0.8\) (right). The mSUGRA parameters and CP phases are given in Table~\ref{tab:ReferenceScenario}.}
\label{fig:AngularDistribution}
\end{figure}
\begin{figure}[ht]
\centering
\includegraphics[clip,width=0.495\textwidth]{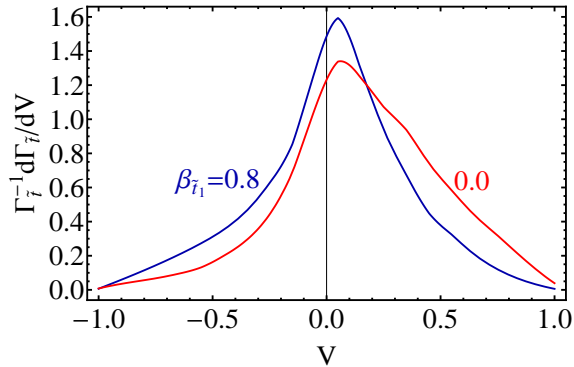}
\caption{Normalized distributions of the stop decay width $\Gamma$ with respect to the
 normalized volume  \(V=\mathbf{p}_t\cdot(\mathbf{p}_{\ell_1}\times\mathbf{p}_{\ell_3})/(|\mathbf{p}_t|| \mathbf{p}_{\ell_1}||\mathbf{p}_{\ell_3}|)\) of the three particle momenta
which form the parallelepiped. Results are shown for stop boost factors \(\beta_{\tilde t_1}=0\) and \(\beta_{\tilde t_1}=0.8\). The mSUGRA parameters and CP phases are given in Table~\ref{tab:ReferenceScenario}.}
\label{fig:VolumeDistribution}
\end{figure}

In Figure~\ref{fig:AngularDistribution}, we show the distribution of the stop decay width, 
\(d\Gamma/d\cos\phi_i\), with respect to one of the three opening angles 
of the trihedron formed by the momenta $\mathbf{p}_t,\mathbf{p}_{\ell_1},\mathbf{p}_{\ell_3}$, 
where \(\phi_i\) (\(i=t,\ell_1,\ell_3\)) is the opening angle between one particle momentum and  the plane formed by the other two, e.g., 
\(\phi_t =\varangle(\mathbf{p}_t,\mathbf{p}_{\ell_1}\times\mathbf{p}_{\ell_3})\). In Figure~\ref{fig:AngularDistribution}~(left), the distribution is shown in the stop rest frame, 
and in Figure~\ref{fig:AngularDistribution} (right) for a boost factor 
\(\beta_{\tilde t_1}=0.8\). The  parallelepiped collapses to a plane for \(\cos\phi_i=0\). For angles close to this configuration it will be  difficult to reconstruct the handedness of the trihedron. 
For larger stop boost factors, the distributions become even
more symmetric and the peaks move closer to \(\cos\phi_i=0\). Only for
\(\phi_{\ell_1}=\varangle(\mathbf{p}_{\ell_1},\mathbf{p}_{\ell_3}\times\mathbf{p}_t)\), Figure~\ref{fig:AngularDistribution}~(right), double peaks are formed at \(\phi_{\ell_1}\approx \pm 45^\circ\), 
which should simplify the determination of the handedness in a boosted frame.

In Figure~\ref{fig:VolumeDistribution}, we show the normalized distribution \(d\Gamma/dV\) with respect 
to the  normalized volume 
\(V=\mathbf{\hat p}_t\cdot(\mathbf{\hat p}_{\ell_1}\times\mathbf{\hat p}_{\ell_3})\)
of the unit vectors \(\mathbf{\hat p}=\mathbf{p}/|\mathbf{p}|\) for a stop boost 
factor  $\beta_{\tilde t_1}=0$ and $\beta_{\tilde t_1}=0.8$. This succinctly
demonstrates that the event configurations mainly have the form of a collapsed parallelepiped, in particular for large stop boosts.

%------------------------------------------------------------------------------
\section{Summary and Conclusions}
\label{sec:Conclusion}
%------------------------------------------------------------------------------

We have analyzed observables of CP violation in the two-body decays of a light stop 
\begin{eqnarray}
	\tilde t_1     &\to&  t + \tilde\chi^0_2. \nonumber
\end{eqnarray}
The CP-sensitive parts appear only in the top neutralino spin-spin correlations, which can be probed by the subsequent decays 
\begin{eqnarray}
	 t              &\to& b + W, \;  \quad W \to  \nu_\ell + \ell_3\nonumber\\
	\tilde\chi^0_2 &\to& \ell_1 + \tilde\ell_{R},   \quad
	\tilde\ell_{R} \to \tilde\chi^0_1 + \ell_2, \qquad \ell=e,\mu.
	\nonumber
\end{eqnarray}
This specific decay chain of the light stop was found to give the largest signal of CP violation over most of the  mSUGRA parameter space.

Due to angular momentum conservation, the decay distributions of the final state momenta are correlated to each other. Asymmetries of triple products of three spatial momenta, as well as epsilon products of four space-time momenta are ideal CP observables. They are sensitive to the CP phases of the trilinear coupling parameter $A_t$, the higgsino mass parameter $\mu$, and the gaugino mass parameter $M_1$.

We have analyzed the asymmetries and event rates in an mSUGRA framework, with CP-violating phases explicitly added at the weak scale. For a strong stop mixing, the asymmetries are rather constant in the mSUGRA space, and reach up to $40\%$ in the rest frame of the stop for $\phi_{ A_t}=1/5 \pi$ and $\phi_\mu=0$. The influence of $\phi_{M_1}$ is negligible for a wino-like neutralino. 

The asymmetries of triple products are not Lorentz invariant and therefore frame-dependent. Since the stops are produced highly boosted  at the LHC, the asymmetries are reduced by a factor of about three when evaluated in the laboratory frame, compared to the stop rest frame. Luminosities of at least $10 \textrm{ fb}^{-1}$ for small values of $m_{1/2}\lsim 300$~GeV, $m_0\lsim 150$~GeV and $\tan\beta\approx 5$ are required to observe a CP signal of $1\sigma$ above statistical fluctuations at the LHC.

%The above stop decay chain is a dominant mode within the light mSUGRA framework. The lightest stop production cross section $pp \to \tilde t_1\tilde t_1^\ast$ is about an order of magnitude larger than that for heavy  $\tilde t_2\tilde t_2^*$ production. The channel into heavier neutralinos $\tilde t_1 \to t \tilde\chi^0_{3,4}$ is closed for $m_{1/2}\lsim 400$~GeV and the branching ratio ${\rm Br}(\tilde\chi_{3}^0\to \tilde\ell\ell)$ is suppressed. However for $m_{1/2}\gsim 400$~GeV, the branching ratio into left sleptons  $(\tilde\chi_{2}^0\to \tilde\ell\ell_L)$ is larger than that into right sleptons.

There are already stringent constraints on SUSY CP phases from electric dipole moments, but large phases and CP asymmetries at the LHC are possible when taking into account cancellations between EDM contributions from different phases. Since the asymmetries of T-odd products depend differently than EDMs on the SUSY parameters, in particular on the CP phases, asymmetries would be an ideal tool for independent measurements at the LHC and ILC.

Clearly, the measurability of the asymmetries and thus of the CP phases can only
be addressed properly in a detailed experimental analysis, which should take into account background processes, detector simulations and event reconstruction efficiencies. We want to underline the need for such a thorough analysis, to explore the potential of the LHC to probe SUSY CP violation.

\newpage

%------------------------------------------------------------------------------
\subsection*{Acknowledgments}
We thank  J.~A.~Aguilar-Saavedra, F.~del \'Aguila, N.~F.~Castro, S.~Heinemeyer,  
T.~Kernreiter, U.~Nierste,  A.~Pilaftsis, and J.~Reuter for very helpful 
comments and discussions. 
OK would like to thank the School of Physics and Astronomy of the University of Manchester 
for kind hospitality, and J.~S.~Kim, aka Zong, for assistance in using  MadGraph~\cite{Stelzer:1994ta}.
FFD would like to thank the Department of Physics and Astronomy of the University of Granada for kind hospitality. This work is supported by MICINN project FPA.2006-05294.

\clearpage
%------------------------------------------------------------------------------
\begin{appendix}
\noindent{\Large\bf Appendix}
\setcounter{equation}{0}
\renewcommand{\thesubsection}{\Alph{section}.\arabic{subsection}}
\renewcommand{\theequation}{\Alph{section}.\arabic{equation}}
\setcounter{equation}{0}

%------------------------------------------------------------------------------
\section{Stop Mixing}
\label{sec:StopMixing}

The masses and couplings of the stops follow from their mass matrix~\cite{Bartl:2004jr}
\begin{eqnarray}
	{\mathcal{L}}_M^{\tilde t} &=&
	-(\tilde t_L^{\ast},\, \tilde t_R^{\ast})
	\left(\begin{array}{ccc}
		m_{\tilde t_{L}}^2 & 
		e^{-i\phi_{\tilde t}}m_t |\Lambda_{\tilde t}|\\[5mm]
		e^{i\phi_{\tilde t}} m_t |\Lambda_{\tilde t}| & 
		m_{\tilde t_R}^2
	\end{array}\right)
	\left(\begin{array}{ccc}
		\tilde t_L\\[5mm]
		\tilde t_R 
	\end{array}\right),
\label{eq:mm}
\end{eqnarray}
with 
\begin{eqnarray}
	m_{\tilde t_L}^2 & = & 
	M_{\tilde Q}^2+
	\left(\frac{1}{2}-\frac{2}{3}\sin^2\theta_w \right)
	m_Z^2\cos2\beta  +m_t^2, 
\label{eq:ml} \\[2mm]
	m_{\tilde t_R}^2 & = & 
	M_{\tilde U}^2+\frac{2}{3}m_Z^2\sin^2\theta_w\cos2\beta + m_t^2,
\label{eq:mr}
\end{eqnarray}
with the soft SUSY-breaking parameters $M_{\tilde Q}$, $M_{\tilde U}$, the ratio $\tan\beta=\frac{v_2}{v_1}$ of the vacuum expectation values of the two neutral Higgs fields, the weak mixing angle $\theta_w$, the mass $m_Z$ of the $Z$ boson, and the mass $m_t$ of the top quark. The CP phase of the stop sector is
\begin{eqnarray}
	\phi_{\tilde t}& = & \arg\lbrack \Lambda_{\tilde t}\rbrack,\\ [2mm]
	\Lambda_{\tilde t}& = & A_t-\mu^\ast\cot\beta,
\label{eq:mlr}
\end{eqnarray}
with the complex trilinear scalar coupling parameter $A_t$, and the higgsino mass parameter $\mu$. The stop mass eigenstates 
\begin{eqnarray}
	\left(\begin{array}{ccc}
		\tilde t_1 \\[1mm]
		\tilde t_2 
	\end{array}\right)
	&=&
	\mathcal{R}^{\tilde t}
	\left(\begin{array}{ccc}
		\tilde t_L \\[1mm]
		\tilde t_R 
	\end{array}\right),
\label{eq:stopmasseigen}
\end{eqnarray}
are given by the diagonalization matrix~\cite{Bartl:2004jr}
\begin{equation}
	\mathcal{R}^{\tilde t} = 
	\left( \begin{array}{ccc}
		e^{i\phi_{\tilde t}}\cos\theta_{\tilde t} &
		\sin\theta_{\tilde t}\\[5mm]
		-\sin\theta_{\tilde t} &
		e^{-i\phi_{\tilde t}}\cos\theta_{\tilde t}
	\end{array}\right),
\label{eq:rstop}
\end{equation}
with the stop mixing angle $\theta_{\tilde t}$
\begin{equation}
	\cos\theta_{\tilde t}=
	\frac{-m_t |\Lambda_{\tilde t}|}{\sqrt{m_t^2 |\Lambda_{\tilde t}|^2+
	(m_{\tilde t_1}^2-m_{\tilde t_{L}}^2)^2}},\quad
	\sin\theta_{\tilde t}=\frac{m_{\tilde t_{L}}^2-m_{\tilde t_1}^2}
	{\sqrt{m_t^2 |\Lambda_{\tilde t}|^2+(m_{\tilde t_1}^2 
	-m_{\tilde t_{L}}^2)^2}}.
\label{eq:thstop}
\end{equation}
The mass eigenvalues are
\begin{equation}
	m_{\tilde t_{\,1,2}}^2 = 
	\frac{1}{2}\left[(m_{\tilde t_{L}}^2+m_{\tilde t_{R}}^2)\mp 
	\sqrt{(m_{\tilde t_{L}}^2 - m_{\tilde t_{R}}^2)^2
	+4m_t^2 |\Lambda_{\tilde t}|^2}\;\right].
\label{eq:m12}
\end{equation}
Note that for $|A_t|\gg |\mu| \cot\beta$ we have $\phi_{\tilde t}\approx\phi_{A_t}$.

%------------------------------------------------------------------------------
\section{Neutralino Mixing}
\label{sec:NeutralinoMixing}

The complex symmetric mass matrix of the neutralinos in the photino, zino, higgsino basis ($\tilde{\gamma},\tilde{Z}, \tilde{H}^0_a, \tilde{H}^0_b$), is given by~\cite{Bartl:1986hp}
\begin{equation}
	{\mathcal M}_{\chi^0} =
	\left(\begin{array}{cccc}
		M_2 \, s^2_w + M_1 \, c^2_w & 
		(M_2-M_1) \, s_w c_w & 0 & 0 \\
		(M_2-M_1) \, s_w c_w & 
		M_2 \, c^2_w + M_1 \, s^2_w & m_Z  & 0 \\
		0 & m_Z &  \mu \, s_{2\beta} & -\mu \, c_{2\beta} \\
		0 &  0  & -\mu \, c_{2\beta} & -\mu \, s_{2\beta} 
	\end{array}\right),
\label{eq:neutmass}
\end{equation}
with the short hand notation for the angles $s_w = \sin\theta_w$, $c_w = \cos\theta_w$, and $s_{2\beta} = \sin(2\beta)$, $c_{2\beta} = \cos(2\beta)$, and the $SU(2)$ gaugino mass parameter $M_2$. The phases of the complex parameters $M_1=|M_1|e^{i\phi_{M_1} }$ and $\mu=|\mu|e^{i\phi_\mu }  $ can lead to CP-violating effects in the neutralino system. We diagonalize the neutralino mass matrix with a complex, unitary $4\times 4$ matrix $N$~\cite{mssm},
\begin{equation}
	N^* \cdot {\mathcal M}_{\chi^0} \cdot N^{\dagger} =
	{\rm diag}(m_{\chi^0_1},\dots,m_{\chi^0_4}),
\label{eq:neutn}
\end{equation}
with the real neutralino masses $ 0 < m_{\chi^0_1} < m_{\chi^0_2} < m_{\chi^0_3} < m_{\chi^0_4}$. This can be achieved by diagonalizing the hermitian matrix ${\mathcal M}_{\chi^0}^\dagger \cdot {\mathcal M}_{\chi^0}$ with a unitary matrix $U$
\begin{equation}
	U \cdot 
	({\mathcal M}_{\chi^0}^\dagger \cdot {\mathcal M}_{\chi^0}) \cdot
	U^\dagger =
	{\rm diag}(m^2_{\chi^0_1},\dots,m^2_{\chi^0_4}).
\end{equation}
The neutralino mixing matrix $N$ can then be expressed as
\begin{equation}
	N = {\rm diag}(e^{i\phi_1/2},\dots,e^{i\phi_4/2}) \cdot U,
\end{equation}
with the phases $\phi_i$ of the elements of the diagonal matrix $U^* \cdot {\mathcal M}_{\chi^0} \cdot U^{\dagger}$,
\begin{equation}
	\phi_i = {\rm arg}(U^* \cdot {\mathcal M}_{\chi^0} \cdot U^{\dagger})_{ii}, 
	\quad i=1,\dots,4.
\end{equation}

%------------------------------------------------------------------------------
\section{Lagrangians and Couplings}
\label{Lagrangians and couplings}

The interaction Lagrangian for stop decay
$\tilde t_m \to t\tilde\chi_i^0 $
is given by~\cite{Bartl:2004jr}
\begin{eqnarray}
{\scr L}_{t \tilde t\tilde\chi^0}
= g\,\bar t\,(a_{mi}^{\tilde t}\,P_R + b_{mi}^{\tilde t}\,P_L)
            \,\tilde\chi_i^0\,\tilde t_m
+ {\rm h.c.} ,
\end{eqnarray}
with $P_{L, R}=(1\mp \gamma_5)/2$, and the weak coupling constant
$g=e/\sin\theta_w$, $e>0$.
The couplings are
\begin{equation}
a_{mi}^{\tilde t} = \sum^2_{n=1}\, (\mathcal{R}^{\tilde t}_{mn})^{\ast}\,
                    {\cal A}_{in}^t, \qquad
b_{mi}^{\tilde t} = \sum^2_{n=1}\, (\mathcal{R}^{\tilde t}_{mn})^{\ast}\,
                    {\cal B}_{in}^{t} ,
\label{eq:abstop}
\end{equation}
with the stop diagonalization matrix 
$\mathcal{R}^{\tilde t}$, see Eq.~(\ref{eq:rstop}),
and
\begin{equation}
{\cal A}_i^t =
  \left(\begin{array}{ccc}
   f_{ti}^L \\[2mm]
   h_{ti}^R \end{array}\right), \qquad
  {\cal B}_i^t =
 \left(\begin{array}{ccc}
    h_{ti}^L\\[2mm]
    f_{ti}^R\end{array}\right),
\label{eq:AB}
\end{equation}
with, in the photino, zino, higgsino basis
($\tilde{\gamma},\tilde{Z}, \tilde{H}^0_a, \tilde{H}^0_b$),
\begin{eqnarray}
f_{t i}^L &=& -\sqrt{2}\bigg[\frac{1}{\cos
        \theta_w}\left(\frac{1}{2}-\frac{2}{3}\sin^2\theta_w\right)N_{i2}+
        \frac{2}{3}\sin \theta_w N_{i1}\bigg],
\label{eq:flt}\\[2mm]
 f_{t i}^R &=& -\frac{2\sqrt{2}}{3} \sin \theta_w
               \left(\tan\theta_w N_{i2}^*-N_{i1}^*\right),
\label{eq:frt}\\[2mm]
 h_{t i}^L &=& (h_{t i}^R)^{\ast} = 
            Y_t( N_{i3}^\ast\sin\beta - N_{i4}^\ast\cos\beta),
\label{eq:hrlt}\\[2mm]
      Y_t  &=& \frac{m_t}{\sqrt{2}\,m_W \sin\beta}, 
\label{eq:yt}
\end{eqnarray}
and $m_W$ the mass of the $W$ boson.

\medskip

The interaction Lagrangian for neu\-tra\-lino decay
$\tilde\chi_i^0 \to \tilde\ell_{R,L}^\pm \ell^\mp$, followed by
$\tilde\ell_{R,L}^\pm \to \tilde\chi_1^0 \ell^\pm$ with $\ell = e,\mu$
is given by~\cite{Bartl:1986hp}
\begin{eqnarray}
      {\scr L}_{\ell \tilde \ell \tilde\chi^0} & = & 
             g \bar\ell f_{\ell i}^L  P_R \tilde\chi_i^0  \tilde \ell_L 
     +       g \bar\ell f_{\ell i}^R  P_L \tilde\chi_i^0  \tilde \ell_R
     +       \mbox{h.c.},
\label{eq:slechie}
\end{eqnarray}
with the couplings~\cite{Bartl:1986hp} 
\begin{eqnarray}
f_{\ell i}^L &=& \sqrt{2}\bigg[\frac{1}{\cos
        \theta_w}\left(\frac{1}{2}-\sin^2\theta_w\right)N_{i2}+
         \sin \theta_w N_{i1}\bigg],
\label{eq:fl}\\[2mm]
f_{\ell i}^R &=& \sqrt{2} \sin \theta_w
        \left(\tan\theta_w N_{i2}^*-N_{i1}^*\right).
\label{eq:fr}
\end{eqnarray}
The Lagrangian for $\ell = \tau$ with complex $A_\tau$ 
are given, for example, in Ref~\cite{Bartl:2002uy}.

%---------------------------------------------------------------------
\section{Kinematics and Phase Space}
\label{Phase space}
%---------------------------------------------------------------------

For the squark decay
$\tilde t_m \to t\tilde\chi_i^0 $,
we choose a coordinate frame in the laboratory (lab) frame
such that the momentum of the top squark $\tilde t_m$
points in the $z$-direction 
\begin{eqnarray}
  p_{\tilde t}^{\mu} &=& (E_{\tilde t},\, 0,\, 0,\, |{\mathbf p}_{\tilde t}| ), \\[2mm]
  p_t^{\mu} &=& 
   (E_t,\, |{\mathbf p}_t|\sin\theta_t,\, 0,\, |{\mathbf p}_t|\cos\theta_t). 
\label{eq:momenta1}
\end{eqnarray}
The decay angle 
$\theta_t =\varangle ({\mathbf p}_{\tilde t},{\mathbf p}_t)$
of the top quark is constrained by
$\sin\theta^{\rm max}_t= |{\mathbf p}_{\tilde t}^\prime | /
                         |{\mathbf p}_{\tilde t}|$
for $|{\mathbf p}_{\tilde t}|>|{\mathbf p}_{\tilde t}^\prime|=
\lambda^{\frac{1}{2}}(m^2_{\tilde t},m^2_t,m^2_{\chi^0_i})/2m_t$,
with the triangle function $\lambda(a,b,c)=a^2+b^2+c^2-2(a b + a c + b c)$.
In this case there are two solutions~\cite{Kittel:2004rp,Byckling} 
\begin{eqnarray}
| {\mathbf p}^{\pm}_t|&=& \frac{
(m^2_{\tilde t}+m^2_t-m^2_{\chi^0_i}) |{\mathbf p}_{\tilde t}|\cos\theta_t\pm
E_{\tilde t}\sqrt{\lambda(m^2_{\tilde t},m^2_t,m^2_{\chi^0_i})-
         4|{\mathbf p}_{\tilde t}|^2~m^2_t~\sin^2\theta_t}}
        {2|{\mathbf p}_{\tilde t}|^2 \sin^2\theta_t
         +2 m^2_{\tilde t}}.\nonumber \\
&&
\label{maxpt}
\end{eqnarray}
For $|{\mathbf p}_{\tilde t}|<|{\mathbf p}_{\tilde t}^\prime|$
the angle $\theta_t$ is unbounded, and only the physical solution 
$ |{\mathbf p}^+_t|$ is left.

The momenta of the subsequent decays of the neutralino
$\tilde\chi_i^0  \to \ell_1 \tilde\ell$; 
$\tilde\ell\to \tilde\chi_1^0 \ell_2$~(\ref{eq:decayChi}),
and those of the top quark
$t \to b W$, $W \to \ell_3 \nu_\ell$~(\ref{eq:decayTop}), 
can be parametrized by
\begin{eqnarray}
  p_b^{\mu} &=& E_b(1,\, \sin \theta_b \cos \phi_b,
                      \, \sin \theta_b \sin \phi_b,
                      \, \cos \theta_b), \\[2mm]
  p_{\ell_1}^{\mu} &=& E_{\ell_1}(1,\, \sin \theta_1 \cos \phi_1,
                      \, \sin \theta_1 \sin \phi_1,
                      \, \cos \theta_1), \\[2mm]
  p_{\ell_2}^{\mu} &=& E_{\ell_2}(1,\, \sin \theta_2 \cos \phi_2,
                      \, \sin \theta_2 \sin \phi_2,
                      \, \cos \theta_2),\\[2mm]
  p_{\ell_3}^{\mu} &=& E_{\ell_3}(1,\, \sin \theta_3 \cos \phi_3,
                      \, \sin \theta_3 \sin \phi_3,
                      \, \cos \theta_3),
\label{eq:momenta2}
\end{eqnarray}
with the energies~\cite{Kittel:2004rp}
\begin{eqnarray}
E_{\ell_1} = \frac{ m^2_{\chi_i^0}-m^2_{\tilde\ell} }
            {2(E_{\chi_i^0}-|{\mathbf p}_{\chi_i^0}|\cos\theta_{D_1}  )},\quad
E_{\ell_2} = \frac{ m^2_{\tilde\ell}- m^2_{\chi_1^0} }
            {2(E_{\tilde\ell}-|{\mathbf p}_{\tilde\ell}|\cos\theta_{D_2}  )},\\[2mm]
       E_b = \frac{ m^2_t-m_W^2 }
            {2(E_t-|{\mathbf p}_t|\cos\theta_{D_b}  )}, \quad
E_{\ell_3} = \frac{ m^2_W }
             {2(E_W-|{\mathbf p}_W|\cos\theta_{D_3}  )}, 
\label{eq:energies}
\end{eqnarray}
and the decay angles,
$\theta_{D_1} =\varangle ({\mathbf p}_{\chi_i^0},{\mathbf p}_{\ell_1})$,
$\theta_{D_2} =\varangle ({\mathbf p}_{\tilde\ell},{\mathbf p}_{\ell_2})$,
$\theta_{D_3} =\varangle ({\mathbf p}_W,{\mathbf p}_{\ell_3})$, and
$\theta_{D_b} =\varangle ({\mathbf p}_{t},{\mathbf p}_b)$,
for example,
\begin{eqnarray}
\cos\theta_{D_1} &=& 
   \frac{({\mathbf p}_{\tilde t}-{\mathbf p}_{t})\cdot\hat{\mathbf p}_{\ell_1}}
       { | {\mathbf p}_{\tilde t}-{\mathbf p}_{t} | },
\label{eq:decayangle}
\end{eqnarray}
with the unit momentum vector 
$\hat{\mathbf p} = {\mathbf p}/|{\mathbf p}|$.
The other momenta 
${\mathbf p}_{\chi_i^0}$, ${\mathbf p}_{\tilde\ell}$,
and ${\mathbf p}_W$,
follow from momentum conservation.

%%The top quark and neutralino spin vectors can be chosen by, for example,
%%\begin{eqnarray}
%%s^{1,\,\mu}_t = \left(0,\frac{{\bf s}^2_t\times{\bf s}^3_t}
%%        {|{\bf s}^2_t\times{\bf s}^3_t|}\right),\;
%%s^{2,\,\mu}_t = \left(0, \frac{{\bf p}_{\tilde t}\times{\bf p}_t}
%%        {|{\bf p}_{\tilde t}\times{\bf p}_t|}\right),\;
%%s^{3,\,\mu}_t = \frac{1}{m_t} \left(|{\bf p}_t|, 
%%        \frac{E_t}{|{\bf p}_t|}{\bf p}_t \right),\qquad 
%%\label{eq:spinvectorst}\\[3mm]
%%s^{1,\,\mu}_{\chi_i^0}=\left(0,\frac{{\bf s}^2_{\chi_i^0}\times{\bf s}^3_{\chi_i^0}}
%%        {|{\bf s}^2_{\chi_i^0}\times{\bf s}^3_{\chi_i^0}|}\right),
%%s^{2,\,\mu}_{\chi_i^0}=\left(0, \frac{{\bf p}_{\tilde t}\times{\bf p}_t}
%%        {|{\bf p}_{\tilde t}\times{\bf p}_t|}\right),
%%s^{3,\,\mu}_{\chi_i^0}=\frac{1}{m_{\chi_i^0}} \left(|{\bf p}_{\chi_i^0}|, 
%%        \frac{E_{\chi_i^0}}{|{\bf p}_{\chi_i^0}|}{\bf p}_{\chi_i^0} \right).
%%\nonumber \\
%%\label{eq:spinvectorsneut}
%%\end{eqnarray}
%%They form an orthonormal set
%%\begin{eqnarray}
%%&&s^a_t\cdot s^b_t=-\delta^{ab}, \quad
%%s^a_t\cdot \hat p_t=0, \qquad
%%s^a_{\chi_i}\cdot s^b_{\chi_i}=-\delta^{ab}, \quad
%%s^a_{\chi_i}\cdot \hat p_{\chi_i}=0,
%%\end{eqnarray}
%%with the notation
%%$\hat p^{\mu} = p^{\mu}/m$. 

The Lorentz invariant phase-space element
for the squark decay chain, see Eqs.~(\ref{eq:decayStop})-(\ref{eq:decayTop}),
can be decomposed into two-body  phase-space elements~\cite{Kittel:2004rp,Byckling}
\begin{eqnarray}
 d{\rm Lips}(s_{\tilde t}\,;
p_{\nu_\ell},p_{\chi_1^0},p_b,
p_{\ell_1},p_{\ell_2},p_{\ell_3})=
\frac{1}{(2\pi)^4}~
\sum_{\pm}~d{\rm Lips}(s_{\tilde t}\,;p_t,p_{\chi_i^0})~
\nonumber \\  \times
d s_{\chi_i^0}~d{\rm Lips}(s_{\chi_i^0};p_{\ell_1},p_{\tilde\ell})~
d s_{\tilde\ell}~d{\rm Lips}(s_{\tilde\ell};p_{\chi_1^0},p_{\ell_2})~
\nonumber \\[2mm]  \times
d s_{t}~d{\rm Lips}(s_{t};p_{b},p_{W})~
d s_{W}~d{\rm Lips}(s_{W};p_{\ell_3},p_{\nu_\ell}),
\label{eq:phasespace}
 \end{eqnarray}
where we have to sum the two solutions 
$| {\mathbf p}^{\pm}_t|$ of the top quark momentum,
see Eq.~(\ref{maxpt}),
if the decay angle $\theta_t$ is constrained.
The different factors are
 \begin{eqnarray}
d{\rm Lips}(s_{\tilde t}\,;p_t,p_{\chi_i^0})&=&
\frac{1}{8\pi}~
\frac{|{\mathbf p}_t|^2}{|E_t~|{\mathbf p}_{\tilde t}|\cos\theta_t-
        E_{\tilde t}~|{\mathbf  p_t}||}~\sin\theta_t~ d\theta_t,\\
d{\rm Lips}(s_{\chi_i^0};p_{\ell_1},p_{\tilde\ell})&=&
        \frac{1}{2(2\pi)^2}~
        \frac{|{\mathbf p}_{\ell_1}|^2}{m_{\chi_i^0}^2-m_{\tilde\ell}^2}
        ~d\Omega_1,\\
        d{\rm Lips}(s_{\tilde\ell} ;p_{\chi_1^0},p_{\ell_2})&=&
\frac{1}{2(2\pi)^2}~
        \frac{|{\mathbf p}_{\ell_2}|^2}{m_{\tilde\ell}^2-m_{\chi_1^0}^2}
        ~d\Omega_2,\\
        d{\rm Lips}(s_{t};p_{b},p_{W})&=&
\frac{1}{2(2\pi)^2}~
        \frac{|{\mathbf p}_b|^2}{m_t^2-m_W^2}
        ~d\Omega_b,\\
        d{\rm Lips}(s_W;p_{\ell_3},p_{\nu_\ell})&=&
\frac{1}{2(2\pi)^2}~\frac{|{\mathbf p}_{\ell_3}|^2}{m_W^2}
        ~d\Omega_3,
\end{eqnarray}
with $s_j=p^2_j$ and 
$ d\Omega_j=\sin\theta_j~ d\theta_j~ d\phi_j$.
We use the narrow width approximation 
 \begin{eqnarray}
\int|\Delta(j)|^2 ~ d s_j &=& 
\frac{\pi}{m_j\Gamma_j},
\label{narrowwidth}
\end{eqnarray}
for the propagators
\begin{eqnarray}
     \Delta(j) &=& \frac{i}{s_j -m_j^2 +im_j\Gamma_j},
\label{eq:propagators}
\end{eqnarray}
%with mass $m_j$ and width $\Gamma_j$ of particle $j$.
which is justified for $\Gamma_j/m_j\ll1$,
which holds in our case for particle widths
$\Gamma_j\lsim {\mathcal O}(1 {\rm GeV}) $,
and masses
$m_j\approx {\mathcal O}(100 {\rm GeV}) $.
Note, however, that the naive
${\mathcal O}(\Gamma/m)$-expectation of the error can easily receive
large off-shell corrections of an order of magnitude and more,
in particular at threshold, or due to interferences
with other resonant or non-resonant processes.
For a recent discussion of these issues, see, for example, 
Ref.~\cite{narrowwidth}.

%\newpage
%--------------------------------------------------------------------------------------
\section{Density Matrix Formalism}
  \label{Density matrix formalism}
%--------------------------------------------------------------------------------------
%

The amplitude squared for the entire stop decay chain, 
Eqs.~(\ref{eq:decayStop})-(\ref{eq:decayTop}),
has been calculated in Ref.~\cite{Bartl:2004jr},
by using the spin formalism of Kawasaki, Shirafuji 
and Tsai~\cite{Kawasaki:1973hf}.
We calculate the amplitude squared 
in the spin-density matrix formalism~\cite{Haber:1994pe,gudi},
which allows a separation of the amplitude  squared 
into contributions from spin correlations,  spin-spin correlations and
from the unpolarized part.
In that way, the CP-sensitive parts of the amplitude squared 
can be easily separated and identified. 
The stop decay amplitude squared in this formalism, however
for a subsequent three-body decay of the neutralino, was calculated 
in Ref.~\cite{Ellis:2008hq}.

\medskip

In the  spin-density matrix formalism of Ref.~\cite{Haber:1994pe},
the amplitude squared of the stop decay chain, 
Eqs.~(\ref{eq:decayStop})-(\ref{eq:decayTop}), can be written as
%For the calculation of the amplitude squared of the 
%stop decay chain, Eqs.~(\ref{eq:decayStop})-(\ref{eq:decayChi}),
%we use the spin-density matrix formalism of~\cite{Haber:1994pe} 
\begin{eqnarray}
        |T|^2=|\Delta(t)|^2~|\Delta(\tilde\chi^0_i)|^2
             ~|\Delta(W)|^2~|\Delta(\tilde\ell)|^2
 \times \nonumber \\ [3mm]
        \sum_{\lambda_i,\lambda^\prime_i,\lambda_t,\lambda^\prime_t,
              \lambda_k\lambda^\prime_k}~
        \rho_{D  }(\tilde t \,)_{\lambda_i\lambda^\prime_i}^{\lambda_t\lambda^\prime_t}~
        \rho_{D_1}(\tilde\chi^0_i)^{\lambda^\prime_i\lambda_i}~
        D_2(\tilde\ell)~
         \rho_{D_3}(t)_{\lambda^\prime_t\lambda_t}^{\lambda^\prime_k\lambda_k}~
         \rho_{D_4}(W)_{\lambda_k\lambda^\prime_k}.
\label{eq:matrixelement}
\end{eqnarray}
The amplitude squared is composed of the propagators $\Delta(j)$, Eq.~(\ref{eq:propagators}),
of particle $j = t$, $\tilde\chi^0_i$, $W$ or $\tilde\ell$, 
and the un-normalized spin density matrices
$\rho_{D  }(\tilde t \,)$, $\rho_{D_1}(\tilde\chi^0_i)$, 
$ \rho_{D_3}(t)$, and $\rho_{D_4}(W)$,
with the helicity indices $\lambda_i,\lambda^\prime_i$ 
of the neutralino, the helicity indices 
$\lambda_t,\lambda^\prime_t$ of the top quark, and those
of the $W$ boson, $\lambda_k,\lambda^\prime_{k}$.
The decay factor  for the selectron is denoted by $D_2(\tilde\ell)$.
%decay~(\ref{eq:decayChi}).

The density matrices can be expanded in terms of the Pauli matrices
\begin{eqnarray} 
\rho_{D  }(\tilde t \,)_{\lambda_i\lambda^\prime_i}^{\lambda_t\lambda^\prime_t}
 &=&
       \delta_{\lambda_t\lambda^\prime_t}~\delta_{\lambda_i\lambda^\prime_i}~D
      +\delta_{\lambda_i\lambda^\prime_i}~
       \sigma_{\lambda_t\lambda^\prime_t}^a~\Sigma_D^a
      +\delta_{\lambda_t\lambda^\prime_t}~
       \sigma_{\lambda_i\lambda^\prime_i}^b~\Sigma_D^b
        \nonumber \\[2mm]
     && 
      +\sigma_{\lambda_t\lambda^\prime_t}^a~
       \sigma_{\lambda_i\lambda^\prime_i}^b~
       \Sigma_D^{ab},
\label{eq:rhoD}\\[2mm]
\rho_{D_1}(\tilde\chi^0_i)^{\lambda^\prime_i\lambda_i}&=&
          \delta_{\lambda_i\lambda^\prime_i}~D_1+
        \sigma^b_{\lambda_i\lambda^\prime_i}~\Sigma^b_{D_1}, 
\label{eq:rhoD1}\\[2mm]
\rho_{D_3}(t)_{\lambda^\prime_t\lambda_t}^{\lambda^\prime_k\lambda_k} &=&
        \left[\delta_{\lambda^\prime_t\lambda_t}~D_3^{\mu\nu}+
            \sigma^a_{\lambda^\prime_t\lambda_t}~
                \Sigma^{a\,\mu\nu}_{D_3}\right] \varepsilon^{\lambda_k\ast}_{\mu}
                \varepsilon^{\lambda^\prime_k}_{\nu},
\label{eq:rhoD3}\\[2mm]
\rho_{D_4}(W)^{\lambda^\prime_k\lambda_k}&=&
        D_4^{\rho\sigma}~\varepsilon_{\rho}^{\lambda_k}~
          \varepsilon^{\lambda^\prime_k\ast}_{\sigma},
\label{eq:rhoD4}
\end{eqnarray}
with an implicit sum over $a,b=1,2,3$.

The polarization vectors 
$\varepsilon^{\lambda_k}_{\mu}$ of the $W$ boson
fulfil the completeness relation
\begin{eqnarray}
\sum_{\lambda_k} \varepsilon^{\lambda_k\ast}_{\mu}
\varepsilon^{\lambda_k}_{\nu}= -g_{\mu\nu}+
   \frac{p_{W, \mu}~p_{W, \nu}}{m_W^2},
\label{eq:Wcompleteness}
\end{eqnarray}
with $p^\mu_W\varepsilon^{\lambda_k}_{\mu}=0$.
Similarly the  spin four-vectors $s^a_{t},$ $a=1,2,3,$ for the top quark $t$,
and $s^b_{\chi_i^0},$ $b=1,2,3,$  for the neutralino $\tilde\chi^0_i$, 
also fulfil completeness relations
\begin{eqnarray}
\sum_{a}s_t^{a,\,\mu} s_t^{a,\,\nu} = -g^{\mu\nu}
      + \frac{p_t^\mu p_t^\nu}{m_t^2},
\qquad
\sum_{b}s_{\chi_i^0}^{b,\,\mu} s_{\chi_i^0}^{b,\,\nu} = -g^{\mu\nu}
      + \frac{p_{\chi_i^0}^\mu p_{\chi_i^0}^\nu}{m_{\chi_i^0}^2},
\label{eq:completeness}
\end{eqnarray}
and they form an orthonormal set
\begin{eqnarray}
&&s^a_t\cdot s^b_t=-\delta^{ab}, \quad
s^a_t\cdot \hat p_t=0, \qquad
s^a_{\chi_i^0}\cdot s^b_{\chi_i^0}=-\delta^{ab}, \quad
s^a_{\chi_i^0}\cdot \hat p_{\chi_i^0}=0,
\end{eqnarray}
with the notation
$\hat p^{\mu} = p^{\mu}/m$.

The expansion coefficients of the 
matrices~(\ref{eq:rhoD})-(\ref{eq:rhoD3}) are 
%
%
%%\newpage
%
%
\begin{eqnarray} 
D &=& 
%%      \frac{g^2}{4}\left(|a^{\tilde t}_{mi}|^2+|b^{\tilde t}_{mi}|^2\right) 
%%      (m_{\tilde t}^2 - m_{\chi_i^0}^2) 
      \frac{g^2}{2}\left(|a^{\tilde t}_{mi}|^2+|b^{\tilde t}_{mi}|^2\right) 
      (p_t\cdot p_{\chi_i^0} )
      -g^2 {\rm Re}\{ a^{\tilde t}_{mi} (b^{\tilde t}_{mi})^\ast \}m_t m_{\chi_i^0},
\label{eq:D} \\ [2mm]
\Sigma^a_{D} &=& \,^{\;\,-}_{(+)}
        \frac{g^2}{2}\left(|a^{\tilde t}_{mi}|^2-|b^{\tilde t}_{mi}|^2\right)
        m_t(p_{\chi_i^0}\cdot s^a_{t}),
\label{eq:SigmaaD} \\ [2mm]
\Sigma^b_{D} &=& \,^{\;\,-}_{(+)}
        \frac{g^2}{2}\left(|a^{\tilde t}_{mi}|^2-|b^{\tilde t}_{mi}|^2\right)
        m_{\chi_i^0}(p_t\cdot s^b_{{\chi_i^0}}),
\label{eq:SigmabD} \\ [2mm]
\Sigma^{ab}_{D} &=& 
         \frac{g^2}{2}\left(|a^{\tilde t}_{mi}|^2+|b^{\tilde t}_{mi}|^2\right)
         (s^a_{t}\cdot s^b_{{\chi_i^0}})m_t m_{\chi_i^0}
        \nonumber \\ [2mm] &&
         + g^2{\rm Re}\{ a^{\tilde t}_{mi} (b^{\tilde t}_{mi})^\ast \}
\left[  (s^a_{t}\cdot   p_{\chi_i^0}) (s^b_{\chi_i^0}\cdot p_t)
       -(s^a_{t}\cdot s^b_{\chi_i^0}) (p_{\chi_i^0}  \cdot p_t)
\right]
        \nonumber \\ [2mm] &&
        -g^2 {\rm Im}\{ a^{\tilde t}_{mi} (b^{\tilde t}_{mi})^\ast \}
%          \epsilon_{\mu\nu\alpha\beta}~
%        s^{a,\mu}_{t}~ p_t^{\nu}~ s^{b,\alpha}_{\chi_i^0}~p_{\chi_i^0}^{\beta};
         [s^{a}_{t},~ p_t,~ s^{b}_{\chi_i^0},~p_{\chi_i^0}],
%        \quad (\epsilon_{0123}=1),
\label{eq:sigmaabD} \\ [2mm]
D_1 &=& \frac{g^2}{2} |f^{R}_{\ell i}|^2 (m_{\chi_i^0}^2 -m_{\tilde\ell}^2 ),
\label{eq:D1} \\ [2mm]
\Sigma^b_{D_1} &=& \,^{\;\,+}_{(-)} g^2 |f^{R}_{\ell i}|^2 
m_{\chi_i^0} (s^b_{\chi_i^0} \cdot p_{\ell_1}),
\label{eq:SigmabD1} \\ [2mm]
{D_3}^{\mu\nu}&=& 
   \frac{g^2}{2}\left[
      p^{\mu}_b p^{\nu}_t +   p^{\nu}_b p^{\mu}_t
        -( p_b \cdot p_t) g^{\mu\nu}
               \right]
        \,^{\;\,+}_{(-)}\frac{g^2}{2}i
     \varepsilon^{\mu\alpha\nu\beta}p_{t, \, \alpha}~p_{b,\,\beta},
%      \quad (\epsilon_{0123}=1),
\label{eq:D3} \\ [2mm]
\Sigma^{a\, \mu\nu}_{D_3} &=& 
   \,^{\;\,-}_{(+)}
    \frac{g^2}{2}m_t \left\{ \left[
      p^{\mu}_b s^{a,\, \nu}_t +  p^{\nu}_b s^{a,\,\mu}_t
        -( p_b \cdot s^a_t) g^{\mu\nu}
               \right]
        \,^{\;\,+}_{(-)}i
     \varepsilon^{\mu\alpha\nu\beta}s_{t, \, \alpha}^a~p_{b,\,\beta}
                \right\},\quad
%      \quad (\epsilon_{0123}=1),
\label{eq:SigmaD3}\\ [2mm]
D_4^{\rho\sigma}&=& 
     g^2\left[p^{\rho}_{\ell_3} p^{\sigma}_{\nu_\ell} 
             +p^{\sigma}_{\ell_3} p^{\rho}_{\nu_\ell}
            -( p_{\ell_3} \cdot p_{\nu_\ell}) g^{\rho\sigma}
         \right]
         \,^{\;\,+}_{(-)} g^2
     i \varepsilon^{\rho\alpha\sigma\beta}p_{\ell_3,\,\alpha}~p_{\nu_\ell,\,\beta},
\label{eq:D4}\\ [2mm]
        D_2(\tilde\ell) &=& g^2 |f^{R}_{\ell 1}|^2 
        ( m_{\tilde\ell}^2-m_{\chi_1^0}^2 ),
\label{eq:D2}
\end{eqnarray}
with the couplings as defined in Appendix~\ref{Lagrangians and couplings},
and the short hand notation
\begin{equation}
     [p_{1},p_2,p_{3},p_{4}]	\equiv 
	\varepsilon_{\mu\nu\alpha\beta}~
	p_{1}^{\mu}~p_2^{\nu}~p_{3}^{\alpha}~p_{4}^{\beta};
      \quad  \varepsilon_{0123}=1.
\label{eq:epsilonB}      
\end{equation}
The coefficients for squark decay,
Eqs.~(\ref{eq:D})-(\ref{eq:sigmaabD}),
agree with those given in Ref.~\cite{Ellis:2008hq}.
The signs in parentheses hold
for the charge conjugated processes, that is
$\tilde t_m^\ast \to \bar t \tilde\chi^0_i$
in Eqs.~(\ref{eq:SigmaaD}) and (\ref{eq:SigmabD}),
$\tilde\chi^0_i \to \ell_1^- \tilde \ell_R^+$
in Eq.~(\ref{eq:SigmabD1}),
$\bar t \to\bar b  W^-$
in Eqs.~(\ref{eq:D3}) and (\ref{eq:SigmaD3}),
and finally $W^- \to \bar\nu_\ell \ell_3^-$
in Eq.~(\ref{eq:D4}).

Inserting the density matrices, Eqs.~(\ref{eq:rhoD})-(\ref{eq:rhoD4})
into Eq.~(\ref{eq:matrixelement}), we obtain 
\begin{eqnarray}
    |T|^2=4~|\Delta(t)|^2~|\Delta(\tilde\chi^0_i)|^2
           ~|\Delta(W)|^2~|\Delta(\tilde\ell)|^2
 \times \nonumber \\ [3mm]
\left[
 D~D_1~D_3^{\rho\sigma}
+ D_1~\Sigma^a_{D}~\Sigma^{a \,\rho\sigma}_{D_3}
+ \Sigma^b_{D}~\Sigma^{b}_{D_1}~D_3^{\rho\sigma}
+  \Sigma^{ab}_{D}~\Sigma^{b}_{D_1}~\Sigma_{D_3}^{a\,\rho\sigma}
\right]
D_{4\,\rho\sigma}~D_2,
\label{eq:matrixelement2}
\end{eqnarray}
with an implicit sum over $a,b$.
The amplitude squared $|T|^2$ is now composed into an
unpolarized part (first summand), into the
spin correlations of the top (second summand),
those of the neutralino (third summand),
and into the spin-spin correlations of top and neutralino
(fourth summand), in Eq.~(\ref{eq:matrixelement2}).

With the completeness relation for the $W$ polarization
vectors~(\ref{eq:Wcompleteness}), we find
\begin{eqnarray}
D_3^{\rho\sigma}~D_{4\,\rho\sigma}&=& 
%%%\,^{\;\,+}_{(-)} 
2g^4(p_t\cdot p_{\ell_3})(p_b\cdot p_{\nu_\ell}),
\label{eq:d3d4}\\[2mm]
\Sigma_{D3}^{a\,\rho\sigma}~D_{4\,\rho\sigma}&=& 
\,^{\;\,-}_{(+)} 2m_tg^4(s^a_t\cdot p_{\ell_3})(p_b\cdot p_{\nu_\ell}),
\label{eq:sig3ad4}
\end{eqnarray}
and the sign in parenthesis for the charge conjugated decay, 
$\bar t \to\bar b  W^-$, $W^- \to \bar\nu_\ell \ell_3^-$.
By also using the completeness relations for the top and
neutralino spin vectors, Eq.~(\ref{eq:completeness}),
the products in Eq.~(\ref{eq:matrixelement2}) can be written as
\begin{eqnarray}
 \Sigma^a_{D}~\Sigma^{a \,\rho\sigma}_{D_3}~D_{4\,\rho\sigma}
&=&
  g^6\left(|a^{\tilde t}_{mi}|^2 - |b^{\tilde t}_{mi}|^2\right)
   (p_b\cdot p_{\nu_\ell})  \nonumber\\ 
   &&\times\left[
      (p_{\chi_i^0}\cdot p_t)(p_{\ell_3}\cdot p_t)
     -m_t^2(p_{\chi_i^0}\cdot p_{\ell_3})
  \right],  \label{eq:product1}\\
\Sigma^b_{D}~\Sigma^{b}_{D_1} 
&=&  \,^{\;\,+}_{(-)}
  \frac{g^4}{2}\left(|a^{\tilde t}_{mi}|^2 - |b^{\tilde t}_{mi}|^2\right)
   |f^{R}_{\ell i}|^2 \nonumber\\ 
  &&\times\left[
      m_{\chi_i^0}^2(p_t\cdot p_{\ell_1})
      -(p_{\chi_i^0}\cdot p_t)(p_{\ell_1}\cdot p_{\chi_i^0})
  \right], \label{eq:product2}
\\
\Sigma^{ab}_{D}~\Sigma^{b}_{D_1}~\Sigma_{D_3}^{a\,\rho\sigma}~D_{4\,\rho\sigma}
&=& \,^{\;\,+}_{(-)}
g^8\left(|a^{\tilde t}_{mi}|^2 + |b^{\tilde t}_{mi}|^2\right)
|f^{R}_{\ell i}|^2
\left[
    (p_{\chi_i^0}\cdot p_{\ell_1})(p_{\chi_i^0}\cdot p_{\ell_3}) m_t^2
\right.
\nonumber \\ [2mm]
&&
   +(p_t\cdot p_{\ell_1})(p_t\cdot p_{\ell_3}) m_{\chi_i^0}^2
   -(p_{\ell_1}\cdot p_{\ell_3}) m_{\chi_i^0}^2 m_t^2
 \nonumber \\ [2mm]
&&
\left.
  -(p_{\chi_i^0}\cdot p_t)(p_t\cdot p_{\ell_3})(p_{\chi_i^0}\cdot p_{\ell_1})
  \right]
(p_b\cdot p_{\nu_\ell})
 \nonumber \\ [2mm]
&&
 \,^{\;\,+}_{(-)} 2g^8{\rm Re}\{ a^{\tilde t}_{mi} (b^{\tilde t}_{mi})^\ast \}
  |f^{R}_{\ell i}|^2  m_{\chi_i^0} m_t(p_b\cdot p_{\nu_\ell})
%\left[
%\right.
\nonumber \\ [2mm]
&&
%\left.
\times
\left[
     (p_{\chi_i^0}\cdot p_t)(p_{\ell_1}\cdot p_{\ell_3})
    -(p_{\chi_i^0}\cdot p_{\ell_3})(p_t\cdot p_{\ell_1})
 \right]
\nonumber \\ [2mm]
&&
  \,^{\;\,+}_{(-)}2g^8{\rm Im}\{ a^{\tilde t}_{mi} (b^{\tilde t}_{mi})^\ast \}
   |f^{R}_{\ell i}|^2 m_{\chi_i^0} m_t(p_b\cdot p_{\nu_\ell})
%          \varepsilon_{\mu\nu\alpha\beta}~
%        p_{\ell_3}^{\mu}~p_t^{\nu}~p_{\ell_1}^{\alpha}~p_{\chi_i^0}^{\beta}.
        [p_{\ell_3},p_t,p_{\ell_1},p_{\chi_i^0}]. 
\nonumber \\ 
\label{CPterm}
\end{eqnarray}
Note that 
%$\varepsilon_{\mu\nu\alpha\beta}~
%   p_{\ell_3}^{\mu}~p_t^{\nu}~p_{\ell_1}^{\alpha}~p_{\chi_i^0}^{\beta} =
%\varepsilon_{\mu\nu\alpha\beta}~
%        p_{\tilde t}^{\mu}~p_t^{\nu}~p_{\ell_3}^{\alpha}~p_{\ell_1}^{\beta}$,
$ [p_{\ell_3}, p_t, p_{\ell_1}, p_{\chi_i^0}] =
  [p_{\tilde t}, p_t, p_{\ell_3}, p_{\ell_1}]$
due to momentum conservation, 
with the short hand notation Eq.~(\ref{eq:epsilonB}).
The signs in parentheses in Eqs.~(\ref{eq:product2}) and (\ref{CPterm})
hold for the charge conjugated process
$\bar t \to\bar b  W^-$, $W^- \to \bar\nu_\ell \ell_3^-$.
In order to obtain the terms for the decay into a positively charged slepton,
$\tilde\chi^0_i \to \ell_1^- \tilde \ell_R^+$, followed by
$\tilde \ell_R^+\to \ell_2^+\tilde\chi^0_1$,
one has to reverse the signs of Eqs.~(\ref{eq:product2}) and (\ref{CPterm}).
This is due to the sign change of the term $\Sigma^b_{D_1} $, see Eq.~(\ref{eq:SigmabD1}).

For the decay into a left slepton $\tilde\chi^0_i \to \ell_1^+ \tilde \ell_L^-$,
Eqs.~(\ref{eq:D1}), (\ref{eq:SigmabD1}), and (\ref{eq:D2}) read
\begin{eqnarray}
D_1 &=& \frac{g^2}{2} |f^{L}_{\ell i}|^2 (m_{\chi_i^0}^2 -m_{\tilde\ell}^2 ),
\label{eq:D1selL} \\ [2mm]
\Sigma^b_{D_1} &=& \,^{\;\,-}_{(+)} g^2 |f^{L}_{\ell i}|^2 
m_{\chi_i^0} (s^b_{\chi_i^0} \cdot p_{\ell_1}),
\label{eq:SigmabD1selL} \\ [2mm]
        D_2(\tilde\ell) &=& g^2 |f^{L}_{\ell 1}|^2 
        ( m_{\tilde\ell}^2-m_{\chi_1^0}^2 ),
\label{eq:D2selL}
\end{eqnarray}
respectively.
The expressions for Eqs.~(\ref{eq:product2}) and  (\ref{CPterm})
have to be changed accordingly.
The sign in parenthesis in Eq.~(\ref{eq:SigmabD1selL}) holds
for the charge conjugated process
$\tilde\chi^0_i \to \ell_1^- \tilde \ell_L^+$.

%lala
%--------------------------------------------------------------------------------------
\section{Stop Decay Widths and Asymmetry}
  \label{Stop decay width}
%--------------------------------------------------------------------------------------
The partial decay width for the stop decay $\tilde t_m \to t \tilde\chi_i^0$
in the stop rest frame is~\cite{Bartl:2003he}
\begin{equation}
\Gamma(\tilde t_m \to t \tilde\chi_i^0)=
\frac{\sqrt{\lambda (m^2_{\tilde t},m^2_t,m^2_{\chi^0_i})}\; }
{4 \pi m^3_{\tilde t}}D,
\label{eq:widtht}
\end{equation}
with the decay function $D$ given in Eqs.~(\ref{eq:D}).
For the decay $\tilde t_m \to b \tilde\chi_j^+$ we have~\cite{Bartl:2003he}
\begin{equation}
\Gamma(\tilde t_m \to b \tilde\chi_j^+)=
\frac{ (m^2_{\tilde t} -m^2_{\chi^+_j})^2 }
{16 \pi m^3_{\tilde t}}g^2
\left(|k^{\tilde t}_{mj}|^2 + |l^{\tilde t}_{mj}|^2\right),
\label{eq:widthb}
\end{equation}
whith the approximation $m_b= 0$. The
 stop-bottom-chargino couplings are~\cite{Bartl:2003he}
\begin{eqnarray}
k^{\tilde t}_{mj} &=&
 {\mathcal R}^{\tilde t \ast}_{m1}\, U_{j2}^\ast \,Y_b, \qquad
l^{\tilde t}_{mj} \:=\:
-{\mathcal R}^{\tilde t \ast}_{m1}\, V_{j1}
+{\mathcal R}^{\tilde t \ast}_{m2}\, V_{j2} \, Y_t,
\label{eq:stcharcouplings}
\end{eqnarray}
with the stop diagonalization matrix ${\mathcal R}^{\tilde t}$,  Eq.~(\ref{eq:rstop}),
the diagonalization matrices $U$, $V$ for the chargino matrix~\cite{mssm}, 
\begin{equation} 
    U^{\ast} \cdot {\mathcal M}_{\chi^\pm} \cdot V^{\dagger} = 
{\rm diag}(m_{\chi^\pm_{1}},m_{\chi^\pm_{2}}),
\label{diagchar}   
\end{equation} 
and the Yukawa couplings $Y_t$,  Eq.~(\ref{eq:yt}), and
\begin{equation}
%      Y_t  = \frac{m_t}{\sqrt{2}\,m_W \sin\beta}, \quad 
      Y_b  = \frac{m_b}{\sqrt{2}\,m_W \cos\beta}.
\label{eq:yb}
\end{equation}

%	N^* \cdot {\mathcal M}_{\chi^0} \cdot N^{\dagger} =
%	{\rm diag}(m_{\chi^0_1},\dots,m_{\chi^0_4}),

The stop decay width for the entire decay chain,
Eqs.~(\ref{eq:decayStop})-(\ref{eq:decayTop}), is  given by 
\begin{equation}
\Gamma(\tilde t \to \nu_\ell\,\tilde\chi_1^0\, b\,\ell_1\ell_2\ell_3)=
        \frac{1}{2 m_{\tilde t}}\int|T|^2
      d{\rm Lips}(s_{\tilde t}\,;
       p_{\nu_\ell},p_{\chi_1^0},p_b,
       p_{\ell_1},p_{\ell_2},p_{\ell_3}),
\label{eq:width}
\end{equation}
with the phase-space element $d{\rm Lips}$ given in 
Eq.~(\ref{eq:phasespace}).

We obtain an explicit expression for the asymmetry,
if we insert the amplitude squared $|T|^2$~(\ref{eq:matrixelement2})
into Eq.~(\ref{eq:Toddasym}),
\begin{equation}
{\mathcal A}=
  \frac{\int {\rm Sign}({\mathcal E})\,
 \Sigma^{ab}_{D}~\Sigma^{b}_{D_1}~\Sigma_{D_3}^{a\,\rho\sigma}
~D_{4\,\rho\sigma}~D_2~
  d{\rm Lips}  }
{\int D~D_1~D_3^{\rho\sigma}~D_{4\,\rho\sigma}~D_2~
d{\rm Lips} },
\label{eq:Adependence}
\end{equation}
where we have already used the narrow width approximation of the propagators,
see Eq.~(\ref{narrowwidth}).
In the numerator, only the spin-spin terms of the
amplitude squared remain, since only they contain 
the epsilon product ${\mathcal E}$, see Eq.~(\ref{eq:epsilon}).
The other terms vanish due to the phase space integration
over the sign of the epsilon product, ${\rm Sign}({\mathcal E})$.
In the denominator, all spin and spin-spin correlation
terms vanish, and only the spin-independent parts contribute.
Inserting now the explicit expressions of the terms
Eqs.~(\ref{eq:D})-(\ref{CPterm})
into the formula for the asymmetry, Eq.~(\ref{eq:Adependence}), 
we find 
%the form of the asymmetry as given in Eqs.~(\ref{eq:Adependence})
\begin{equation}
{\mathcal A}= 2 \, \eta \,
\frac{\int {\rm Sign}({\mathcal E})
                (p_b\cdot p_{\nu_\ell})\,
        [p_{\tilde t},p_t,p_{\ell_3},p_{\ell_1}]~
        d{\rm Lips}  }{(m_{\chi_i^0}^2- m_{\tilde\ell }^2)\int
 (p_t\cdot p_{\ell_3}) (p_b\cdot p_{\nu_\ell})~
d{\rm Lips} },
\label{eq:Adependence3}
\end{equation}
with the coupling function
\begin{equation}
\eta = \frac{{\rm Im}\{ a^{\tilde t}_{mi} (b^{\tilde t}_{mi})^\ast \}}
      { \frac{1}{2}\left(|a^{\tilde t}_{mi}|^2 + |b^{\tilde t}_{mi}|^2\right)
     \frac{m_{\tilde t}^2 - m_{\chi_i^0}^2 - m_t^2}
    {2 m_{\chi_i^0} m_t}
     - {\rm Re}\{ a^{\tilde t}_{mi} (b^{\tilde t}_{mi})^\ast \}}.
\label{eq:couplingfunct2}
\end{equation}

%------------------------------------------------------------------------------
\section{Contributions from Absorptive Phases}
\label{sec:AbsorptivePhases}
%------------------------------------------------------------------------------

The T-odd asymmetries ${\mathcal A}$, as defined in Eq.~(\ref{eq:Toddasym}), are not only sensitive to the CP phases in the stop-neutralino system. Absorptive phases could also enter from final-state interactions or finite-width effects. Although such absorptive contributions are a higher order effect, and thus expected to be small, they do not signal CP violation. However, they can be eliminated in the CP asymmetry 
\begin{eqnarray}
	 {\mathcal A}_{\rm CP} = 
	\frac{1}{2}({\mathcal A} -\bar{\mathcal A}),
\label{eq:CPasymmetry}
\end{eqnarray}
where ${\mathcal A}$ is the asymmetry for the stop decay,
\begin{eqnarray}
	\tilde t_m \to  t + \tilde\chi^0_i; 
	\quad {\rm with} \quad &&
	t \to  b + W^+; \quad 
	W^+ \to \nu_\ell + \ell_3^+; 
\label{eq:decayW+}\\[2mm]
	&&
	\tilde\chi^0_i \to \ell_1^+ + \tilde \ell_R^-; \quad 
	\tilde \ell_R^- \to \tilde\chi^0_1 + \ell_2^-;
\label{eq:decaySel-}
\end{eqnarray}
and $\bar{\mathcal A}$ the corresponding asymmetry for the decay of the charge conjugated stop,
\begin{eqnarray}
	\tilde t_m^* \to  \bar t + \tilde\chi^0_i; 
	\quad {\rm with} \quad &&
	\bar t \to  \bar b + W^-; \quad 
	W^- \to \bar \nu_\ell + \ell_3^-, 
\label{eq:decayW-}
\end{eqnarray}
and the same neutralino $\tilde\chi^0_i$ decay as in Eq.~(\ref{eq:decaySel-}). In forming the CP asymmetry in Eq.~(\ref{eq:CPasymmetry}), the absorptive phases cancel, since they do not change sign under charge conjugation. The truly CP-violating parts of ${\mathcal A}$ however add up, since they change sign for the charge conjugated top decay chain. See the last term of Eq.~(\ref{CPterm}), which is the CP-violating part of the amplitude squared.

There is one subtlety in the decay of the neutralino. It decays to equal parts into a negatively or positively charged slepton, respectively, leading to the decay chains
\begin{eqnarray}
	\tilde\chi^0_i  \to \ell_1^+ + \tilde \ell_R^-; \quad 
	\tilde \ell_R^- \to \tilde\chi^0_1 + \ell_2^-; \\[2mm]
	\tilde\chi^0_i  \to \ell_1^- + \tilde \ell_R^+; \quad 
	\tilde \ell_R^+ \to \tilde\chi^0_1 + \ell_2^+.
\label{eq:decaySel}
\end{eqnarray}
The two corresponding asymmetries for these two decay modes again differ in sign. This is due to a sign change in the spin correlation term of the neutralino $\Sigma^b_{D_1} $, see Eq.~(\ref{eq:SigmabD1}). One would have in principle to distinguish the near and the far leptons, $\ell_1$ and $\ell_2$ respectively, not to sum up the asymmetries to zero~\cite{Bartl:2004jr}. But instead an asymmetry can be  considered of a triple product which includes both leptons from the slepton decay, for example, 
\begin{eqnarray}
	{\mathcal T} =
	(\mathbf{p}_{\ell_1} \times \mathbf{p}_{\ell_2 }) \cdot\mathbf{p}_{\ell_3}
	\equiv
	[ \mathbf{p}_{\ell_1}, \mathbf{p}_{\ell_2 },  \mathbf{p}_{\ell_3}].
\label{eq:difftriple}
\end{eqnarray}
Then a distinction of the near and far leptons ${\ell_1}$, $ {\ell_2}$ from the slepton decay is not needed, since there is an additional sign change due to:
\begin{eqnarray}
	[\mathbf{p}_{\ell_1},\mathbf{p}_{\ell_2 },\mathbf{p}_{\ell_3}] =
	-[ \mathbf{p}_{\ell_2}, \mathbf{p}_{\ell_1 },  \mathbf{p}_{\ell_3}].
\label{eq:xxxtriple}
\end{eqnarray}
Thus one only would have to distinguish the charges of the two leptons by considering the triple product~\cite{private}
\begin{eqnarray}
	{\mathcal T} =
	[\mathbf{p}_{ \ell^+}, \mathbf{p}_{\ell^- },  \mathbf{p}_{\ell_3^+}],
\end{eqnarray}
thereby constructing the CP asymmetry
\begin{eqnarray}
	{\mathcal A}_{\rm CP} = \frac{1}{2}
	{\mathcal A}
	\big(
		[\mathbf{p}_{\ell^+},\mathbf{p}_{\ell^-},\mathbf{p}_{\ell_3^+}]
	\big)
	-\frac{1}{2} \bar{\mathcal A}
	\big(
		[\mathbf{p}_{\ell^+}, \mathbf{p}_{\ell^-}, \mathbf{p}_{\ell_3^-} ]
	\big),
\label{eq:CPasymmetry2}
\end{eqnarray}
which is no longer sensitive to absorptive phases.

\end{appendix}

%-----------------------------------------------------------------------------

%-----------------------------------------------------------------------------

% ----------------------------------------------------------------------------

\begin{thebibliography}{99}
%-----------------------------------------------------------------------------

%%%--------MSSM basics 
%\cite{Haber:1984rc}
\bibitem{mssm}
  H.~E.~Haber and G.~L.~Kane,
%  ``The Search For Supersymmetry: Probing Physics Beyond The Standard Model,''
  Phys.\ Rept.\  {\bf 117}, 75 (1985);\\
  %%CITATION = PRPLC,117,75;%%
%
%%\cite{Nilles:1983ge}
%\bibitem{Nilles:1983ge}
  H.~P.~Nilles,
  %``Supersymmetry, Supergravity And Particle Physics,''
  Phys.\ Rept.\  {\bf 110} (1984) 1;\\
  %%CITATION = PRPLC,110,1;%%
%
%\cite{Drees:2004jm}
%\bibitem{Drees:2004jm}
  M.~Drees, R.~Godbole and P.~Roy,
  \emph{Theory and phenomenology of sparticles}, 
%: An account of four-dimensional N=1
%  supersymmetry in high energy physics},
  World Scientific, Singapur (2004);\\
%\href{http://www.slac.stanford.edu/spires/find/hep/www?irn=6240364}{SPIRES entry}
%
%\cite{Baer:2006rs}
%\bibitem{Baer:2006rs}
  H.~Baer and X.~Tata,
  \emph{Weak scale supersymmetry: From superfields to scattering events},
%\href{http://www.slac.stanford.edu/spires/find/hep/www?irn=6927297}{SPIRES entry}
  Cambridge, UK: Univ. Pr. (2006).


%%%%-------------- dark matter papers

%\cite{Goldberg:1983nd}
\bibitem{Goldberg:1983nd}
  H.~Goldberg,
  %``Constraint On The Photino Mass From Cosmology,''
  Phys.\ Rev.\ Lett.\  {\bf 50} (1983) 1419; \\
  %%CITATION = PRLTA,50,1419;%%
%
%\cite{Ellis:1983ew}
%\bibitem{Ellis:1983ew}
  J.~R.~Ellis, J.~S.~Hagelin, D.~V.~Nanopoulos, K.~A.~Olive and M.~Srednicki,
  %``Supersymmetric Relics From The Big Bang,''
  Nucl.\ Phys.\ B {\bf 238} (1984) 453;\\
  %%CITATION = NUPHA,B238,453;%%
%\cite{Pagels:1981ke}
%\bibitem{Pagels:1981ke}
  H.~Pagels and J.~R.~Primack,
  %``Supersymmetry, Cosmology And New Tev Physics,''
  Phys.\ Rev.\ Lett.\  {\bf 48}, 223 (1982).
  %%CITATION = PRLTA,48,223;%%

%%%---------------Relic Density--------------------------
\bibitem{RelicCP}
%\cite{Belanger:2006qa}
For an analysis of the neutralino relic density with CP phases, see, for example,\\
%\bibitem{Belanger:2006qa}
  G.~Belanger, F.~Boudjema, S.~Kraml, A.~Pukhov and A.~Semenov,
  %``Relic density of neutralino dark matter in the MSSM with CP violation,''
  Phys.\ Rev.\ D {\bf 73}, 115007 (2006)
  [arXiv:hep-ph/0604150];
  %%CITATION = HEP-PH 0604150;%%
%
%\cite{Belanger:2006pc}
%\bibitem{Belanger:2006pc}
%  G.~Belanger, F.~Boudjema, S.~Kraml, A.~Pukhov and A.~Semenov,
  %``Neutralino dark matter in the MSSM with CP violation,''
  AIP Conf.\ Proc.\  {\bf 878}, 46 (2006)
  [arXiv:hep-ph/0610110];\\
  %%CITATION = APCPC,878,46;%%
%
%\cite{Choi:2006hi}
%\bibitem{Choi:2006hi}
  S.~Y.~Choi and Y.~G.~Kim,
  %``Heavy Higgs resonances for the neutralino relic density in the Higgs
  %decoupling limit of the CP-noninvariant minimal supersymmetric standard
  %model,''
  Phys.\ Lett.\ B {\bf 637}, 27 (2006)
  [arXiv:hep-ph/0602109];\\
  %%CITATION = HEP-PH 0602109;%%
%
%\cite{Lee:2007ai}
%\bibitem{Lee:2007ai}
  J.~S.~Lee and S.~Scopel,
  %``Lightest Higgs boson and relic neutralino in the MSSM with CP violation,''
  Phys.\ Rev.\  D {\bf 75}, 075001 (2007)
  [arXiv:hep-ph/0701221];\\
  %%CITATION = PHRVA,D75,075001;%%
%
%\cite{Belanger:2008yc}
%\bibitem{Belanger:2008yc}
  G.~Belanger, O.~Kittel, S.~Kraml, H.~U.~Martyn and A.~Pukhov,
  %``Neutralino relic density from ILC measurements in the CPV MSSM,''
  Phys.\ Rev.\  D {\bf 78}, 015011 (2008)
  [arXiv:0803.2584 [hep-ph]],
  %%CITATION = PHRVA,D78,015011;%%
%\cite{Belanger:2009dz}
%\bibitem{Belanger:2009dz}
%  G.~Belanger, O.~Kittel, S.~Kraml, H.~U.~Martyn and A.~Pukhov,
  %``Neutralino Relic Density in the CPVMSSM at the ILC,''
  arXiv:0901.4838 [hep-ph].
  %%CITATION = ARXIV:0901.4838;%%


%%%%-------------- CP phases in MSSM
%\cite{Haber:1997if}
\bibitem{Haber:1997if}
  See, for example, \\
   H.~E.~Haber,
  %``The status of the minimal supersymmetric standard model and beyond,''
  Nucl.\ Phys.\ Proc.\ Suppl.\  {\bf 62}, 469 (1998)
  [arXiv:hep-ph/9709450]; \\
  %%CITATION = NUPHZ,62,469;%%
%\cite{Ibrahim:2002ry}
%\bibitem{Ibrahim:2002ry}
  T.~Ibrahim and P.~Nath,
  %``Phases and CP violation in SUSY,''
  arXiv:hep-ph/0210251;
  %%CITATION = HEP-PH 0210251;%%
%
%\cite{Ibrahim:2007fb}
%\bibitem{Ibrahim:2007fb}
%  T.~Ibrahim and P.~Nath,
  %``CP violation from standard model to strings,''
  Rev.\ Mod.\ Phys.\  {\bf 80}, 577 (2008)
  [arXiv:0705.2008 [hep-ph]];\\
  %%CITATION = RMPHA,80,577;%%
%\cite{Ellis:2007kb}
%\bibitem{Ellis:2007kb}
  J.~R.~Ellis, J.~S.~Lee and A.~Pilaftsis,
  %``B-Meson Observables in the Maximally CP-Violating MSSM with Minimal Flavour
  %Violation,''
  Phys.\ Rev.\  D {\bf 76}, 115011 (2007)
  [arXiv:0708.2079 [hep-ph]].
  %%CITATION = PHRVA,D76,115011;%%



%%%%------------b physics
\bibitem{Amsler:2008zz}
  C.~Amsler {\it et al.}  [Particle Data Group],
  %``Review of particle physics,''
  Phys.\ Lett.\  B {\bf 667} (2008) 1.
  %%CITATION = PHLTA,B667,1;%%


%\cite{Buras:2005xt}
\bibitem{Buras:2005xt}
  For a review, see for example, A.~J.~Buras,
  %``Flavour physics and CP violation,''
  arXiv:hep-ph/0505175.
  %%CITATION = HEP-PH/0505175;%%

%%%%-------------- EW phase trans. in SM


\bibitem{Gavela:1993ts}
  M.~B.~Gavela, P.~Hernandez, J.~Orloff and O.~Pene,
  %``Standard Model CP-violation and Baryon asymmetry,''
  Mod.\ Phys.\ Lett.\  A {\bf 9}, 795 (1994)
  [arXiv:hep-ph/9312215];\\
  %%CITATION = MPLAE,A9,795;%%
%
%\cite{Gavela:1994dt}
%\bibitem{Gavela:1994dt}
  M.~B.~Gavela, P.~Hernandez, J.~Orloff, O.~Pene and C.~Quimbay,
  %``Standard model CP violation and baryon asymmetry. Part 2: Finite
  %temperature,''
  Nucl.\ Phys.\  B {\bf 430}, 382 (1994)
  [arXiv:hep-ph/9406289];\\
  %%CITATION = NUPHA,B430,382;%%
%
%\cite{Csikor:1998eu}
%\bibitem{Csikor:1998eu}
  F.~Csikor, Z.~Fodor and J.~Heitger,
  %``Endpoint of the hot electroweak phase transition,''
  Phys.\ Rev.\ Lett.\  {\bf 82}, 21 (1999)
  [arXiv:hep-ph/9809291].
  %%CITATION = HEP-PH 9809291;%%

%\cite{Riotto:1998bt}
\bibitem{Riotto:1998bt}
  For a review, see for example, \\
  A.~Riotto,
  %``Theories of baryogenesis,''
  arXiv:hep-ph/9807454;\\
  %%CITATION = HEP-PH 9807454;%%
%%% 
 %\cite{Bernreuther:2002uj}
% \bibitem{Bernreuther:2002uj}
  W.~Bernreuther,
  %``CP violation and baryogenesis,''
  Lect.\ Notes Phys.\  {\bf 591}, 237 (2002)
  [arXiv:hep-ph/0205279].
  %%CITATION = HEP-PH 0205279;%%


%---------------bounds EDM experimental--------------------

\bibitem{Regan:2002ta}
  B.~C.~Regan, E.~D.~Commins, C.~J.~Schmidt and D.~DeMille,
  %``New limit on the electron electric dipole moment,''
  Phys.\ Rev.\ Lett.\  {\bf 88}, 071805 (2002).
  %%CITATION = PRLTA,88,071805;%%

\bibitem{Baker:2006ts}
  C.~A.~Baker {\it et al.},
  %``An improved experimental limit on the electric dipole moment of the
  %neutron,''
  Phys.\ Rev.\ Lett.\  {\bf 97}, 131801 (2006)
  [arXiv:hep-ex/0602020].
  %%CITATION = PRLTA,97,131801;%%

\bibitem{Griffith:2009zz}
  W.~C.~Griffith, M.~D.~Swallows, T.~H.~Loftus, M.~V.~Romalis, B.~R.~Heckel and E.~N.~Fortson,
  %``Improved Limit on the Permanent Electric Dipole Moment of Hg-199,''
  Phys.\ Rev.\ Lett.\  {\bf 102}, 101601 (2009);\\
  %%CITATION = PRLTA,102,101601;%%
%
%\cite{Latha:2009nq}
%\bibitem{Latha:2009nq}
  K.~V.~P.~Latha, D.~Angom, B.~P.~Das and D.~Mukherjee,
  %``Probing CP violation with the electric dipole moment of atomic mercury,''
  arXiv:0902.4790 [physics.atom-ph].
  %%CITATION = ARXIV:0902.4790;%%


%\cite{Semertzidis:2003iq}
\bibitem{Semertzidis:2003iq}
  Y.~K.~Semertzidis {\it et al.}  [EDM Collaboration],
  %``A new method for a sensitive deuteron EDM experiment,''
  AIP Conf.\ Proc.\  {\bf 698}, 200 (2004)
  [arXiv:hep-ex/0308063];
  %%CITATION = APCPC,698,200;%%
%\cite{Semertzidis:2004uu}
%\bibitem{Semertzidis:2004uu}
%  Y.~K.~Semertzidis,
  %``Electric dipole moments of fundamental particles,''
  Nucl.\ Phys.\ Proc.\ Suppl.\  {\bf 131}, 244 (2004)
  [arXiv:hep-ex/0401016].
  %%CITATION = NUPHZ,131,244;%%



%--- SUSY-- CP Problem-----------
\bibitem{CP-Problem}
%\cite{Ellis:1982tk}
%\bibitem{Ellis:1982tk}
%  See, e.g.,
  J.~R.~Ellis, S.~Ferrara and D.~V.~Nanopoulos,
  %``CP Violation And Supersymmetry,''
  Phys.\ Lett.\ B {\bf 114}, 231 (1982);\\
  %%CITATION = PHLTA,B114,231;%%
%
%\cite{Abel:2005er}
%\bibitem{Abel:2005er}
  S.~Abel and O.~Lebedev,
  %``Neutron electron EDM correlations in supersymmetry 
  % and prospects for  EDM searches,''
  JHEP {\bf 0601}, 133 (2006)
  [arXiv:hep-ph/0508135].
  %%CITATION = HEP-PH 0508135;%
%\cite{Mercolli:2009ns}
%


%%%---------------EDM--------theoretical works/model dependence
\bibitem{EDM}
  T.~Ibrahim and P.~Nath,
  Phys.\ Rev.\ D {\bf 57} (1998) 478
  [Erratum-ibid.\ D {\bf 58} (1998) 019901, D {\bf 60} (1999) 079903,
  D {\bf 60} (1999) 119901] 
  [arXiv:hep-ph/9708456];\\
%
  M.~Brhlik, G.~J.~Good and G.~L.~Kane,
  %``Electric dipole moments do not require the CP-violating phases of
  %supersymmetry to be small,''
  Phys.\ Rev.\ D {\bf 59}, 115004 (1999)
  [arXiv:hep-ph/9810457];\\
%
  A.~Bartl, T.~Gajdosik, W.~Porod, P.~Stockinger and H.~Stremnitzer,
  %``Electron and neutron electric dipole moments in the constrained MSSM,''
  Phys.\ Rev.\  D {\bf 60} (1999) 073003
  [arXiv:hep-ph/9903402];\\
%
  S.~Yaser Ayazi and Y.~Farzan,
  % ``Reconciling large CP-violating phases with bounds on the electric dipole
  %moments in the MSSM,''
  Phys.\ Rev.\ D {\bf 74} (2006) 055008
  [arXiv:hep-ph/0605272];\\
%
%\cite{Olive:2005ru}
%\bibitem{Olive:2005ru}
  K.~A.~Olive, M.~Pospelov, A.~Ritz and Y.~Santoso,
  %``CP-odd phase correlations and electric dipole moments,''
  Phys.\ Rev.\  D {\bf 72}, 075001 (2005)
  [arXiv:hep-ph/0506106];\\
  %%CITATION = PHRVA,D72,075001;%
%
%--- MFV-----------
%\cite{Mercolli:2009ns}
%\bibitem{Mercolli:2009ns}
  L.~Mercolli and C.~Smith,
  %``EDM constraints on flavored CP-violating phases,''
  Nucl.\ Phys.\  B {\bf 817}, 1 (2009)
  [arXiv:0902.1949 [hep-ph]].
  %%CITATION = NUPHA,B817,1;%%



%%%---FURTHER -EDMs-----theory
 
\bibitem{Barger:2001nu}
  V.~D.~Barger, T.~Falk, T.~Han, J.~Jiang, T.~Li and T.~Plehn,
  %``CP-violating phases in SUSY, electric dipole moments, and linear
  %colliders,''
  Phys.\ Rev.\  D {\bf 64}, 056007 (2001)
  [arXiv:hep-ph/0101106].
  %%CITATION = PHRVA,D64,056007;%%


\bibitem{Bartl:2003ju}
  A.~Bartl, W.~Majerotto, W.~Porod and D.~Wyler,
  %``Effect of supersymmetric phases on lepton dipole moments and rare  lepton
  %decays,''
  Phys.\ Rev.\  D {\bf 68}, 053005 (2003)
  [arXiv:hep-ph/0306050].
  %%CITATION = PHRVA,D68,053005;%%

%\cite{Ellis:2008zy}
\bibitem{Ellis:2008zy}
% \bibitem{edm} 
 For a recent review see, for example,
 J.~R.~Ellis, J.~S.~Lee and A.~Pilaftsis,
  %``Electric Dipole Moments in the MSSM Reloaded,''
  JHEP {\bf 0810}, 049 (2008)
  [arXiv:0808.1819 [hep-ph]].
  %%CITATION = JHEPA,0810,049;%%

%\cite{Choi:2004rf}
\bibitem{Choi:2004rf}
For a connection of EDMs to ILC observables see, for example,
S.~Y.~Choi, M.~Drees and B.~Gaissmaier,
%``Systematic study of the impact of CP-violating phases of the MSSM on
%leptonic high-energy observables,''
Phys.\ Rev.\ D {\bf 70} (2004) 014010
[arXiv:hep-ph/0403054].



%%% -----------Higgs A-H transitions --early works
%cite{Pilaftsis:1997dr}
\bibitem{Pilaftsis}
  A.~Pilaftsis,
  %``Resonant CP violation induced by particle mixing in transition
  %amplitudes,''
  Nucl.\ Phys.\  B {\bf 504} (1997) 61
  [arXiv:hep-ph/9702393];
  %%CITATION = NUPHA,B504,61;%%
%
%\bibitem{PilaftsisAH}
%%\cite{Pilaftsis:1998pe}
%%\bibitem{Pilaftsis:1998pe}
%  A.~Pilaftsis,
  %``CP-odd tadpole renormalization of Higgs scalar-pseudoscalar mixing,''
  Phys.\ Rev.\  D {\bf 58}, 096010 (1998)
  [arXiv:hep-ph/9803297];
  %%CITATION = PHRVA,D58,096010;%%
%\cite{Pilaftsis:1998dd}
%\bibitem{Pilaftsis:1998dd}
%  A.~Pilaftsis,
  %``Higgs scalar-pseudoscalar mixing in the minimal supersymmetric standard
  %model,''
  Phys.\ Lett.\  B {\bf 435}, 88 (1998)
  [arXiv:hep-ph/9805373].
  %%CITATION = PHLTA,B435,88;%%


%%% ----------- the later results in  A-H mixings
\bibitem{HA}
%\cite{Gunion:1997aq}
%\bibitem{Gunion:1997aq}
  J.~F.~Gunion, B.~Grzadkowski, H.~E.~Haber and J.~Kalinowski,
  %``LEP limits on CP-violating non-minimal Higgs sectors,''
  Phys.\ Rev.\ Lett.\  {\bf 79} (1997) 982
  [arXiv:hep-ph/9704410];\\
  %%CITATION = PRLTA,79,982;%%
%\cite{Grzadkowski:1999ye}
%\bibitem{Grzadkowski:1999ye}
  B.~Grzadkowski, J.~F.~Gunion and J.~Kalinowski,
  %``Finding the CP-violating Higgs bosons at e+ e- colliders,''
  Phys.\ Rev.\  D {\bf 60} (1999) 075011
  [arXiv:hep-ph/9902308];\\
  %%CITATION = PHRVA,D60,075011;%%
%
%
%\cite{Demir:1999hj}
%\bibitem{Demir:1999hj}
  D.~A.~Demir,
  %``Effects of the supersymmetric phases on the neutral Higgs sector,''
  Phys.\ Rev.\  D {\bf 60} (1999) 055006
  [arXiv:hep-ph/9901389];\\
  %%CITATION = PHRVA,D60,055006;%%
%
%\cite{Choi:2000wz}
%\bibitem{Choi:2000wz}
  S.~Y.~Choi, M.~Drees and J.~S.~Lee,
  %``Loop corrections to the neutral Higgs boson sector of the MSSM with
  %explicit CP violation,''
  Phys.\ Lett.\  B {\bf 481} (2000) 57
  [arXiv:hep-ph/0002287];\\
  %%CITATION = PHLTA,B481,57;%%
%
%\cite{Ibrahim:2000qj}
%\bibitem{Ibrahim:2000qj}
  T.~Ibrahim and P.~Nath,
  %``Corrections to the Higgs boson masses and mixings from chargino, W and
  %charged Higgs exchange loops and large CP phases,''
  Phys.\ Rev.\  D {\bf 63}, 035009 (2001)
  [arXiv:hep-ph/0008237];\\
  %%CITATION = PHRVA,D63,035009;%%
%%%\cite{Ibrahim:2002zk}
%%%\bibitem{Ibrahim:2002zk}
%%%  T.~Ibrahim and P.~Nath,
  %``Neutralino exchange corrections to the Higgs boson mixings with  explicit
  %CP violation,''
  Phys.\ Rev.\  D {\bf 66}, 015005 (2002)
  [arXiv:hep-ph/0204092];\\
  %%CITATION = PHRVA,D66,015005;%%
%
%\bibitem{PW.explicitCP}
%\cite{Pilaftsis:1999qt}
%\bibitem{Pilaftsis:1999qt}
  A.~Pilaftsis and C.~E.~M.~Wagner,
  %``Higgs bosons in the minimal supersymmetric standard model with explicit CP
  %violation,''
  Nucl.\ Phys.\  B {\bf 553} (1999) 3
  [arXiv:hep-ph/9902371];\\
  %%CITATION = NUPHA,B553,3;%%
%
%\cite{Carena:2000yi}
%\bibitem{Carena:2000yi}
  M.~S.~Carena, J.~R.~Ellis, A.~Pilaftsis and C.~E.~M.~Wagner,
  %``Renormalization-group-improved effective potential for the MSSM Higgs
  %sector with explicit CP violation,''
  Nucl.\ Phys.\  B {\bf 586} (2000) 92
  [arXiv:hep-ph/0003180];\\
  %%CITATION = NUPHA,B586,92;%%
%
%\cite{Carena:2001fw}
%\bibitem{Carena:2001fw}
  M.~S.~Carena, J.~R.~Ellis, A.~Pilaftsis and C.~E.~M.~Wagner,
  %``Higgs-boson pole masses in the MSSM with explicit CP violation,''
  Nucl.\ Phys.\  B {\bf 625}, 345 (2002)
  [arXiv:hep-ph/0111245];\\
  %%CITATION = NUPHA,B625,345;%%
%\cite{Choi:2004kq}
%\bibitem{Choi:2004kq}
  S.~Y.~Choi, J.~Kalinowski, Y.~Liao and P.~M.~Zerwas,
  %``H / A Higgs mixing in CP-noninvariant supersymmetric theories,''
  Eur.\ Phys.\ J.\  C {\bf 40}, 555 (2005)
  [arXiv:hep-ph/0407347];\\
  %%CITATION = EPHJA,C40,555;%%
%
%\cite{Heinemeyer:2007aq}
%\bibitem{Heinemeyer:2007aq}
  S.~Heinemeyer, W.~Hollik, H.~Rzehak and G.~Weiglein,
  %``The Higgs sector of the complex MSSM at two-loop order: QCD
  %contributions,''
  Phys.\ Lett.\  B {\bf 652}, 300 (2007)
  [arXiv:0705.0746 [hep-ph]].
  %%CITATION = PHLTA,B652,300;%%



%%%-------------Benchmark scenarios CPX etc
\bibitem{Carena:2000ks}
%\cite{Carena:2000ks}
  M.~Carena, J.~R.~Ellis, A.~Pilaftsis and C.~E.~M.~Wagner,
  %``CP-violating MSSM Higgs bosons in the light of LEP 2,''
  Phys.\ Lett.\  B {\bf 495} (2000) 155
  [arXiv:hep-ph/0009212].
  %%CITATION = PHLTA,B495,155;%%

%%%-----------LEP bounds
%\cite{Barate:2003sz}
\bibitem{LEPbounds}
  R.~Barate {\it et al.}  [LEP Working Group for Higgs boson searches],
  %``Search for the standard model Higgs boson at LEP,''
  Phys.\ Lett.\  B {\bf 565}, 61 (2003)
  [arXiv:hep-ex/0306033];\\
  %%CITATION = PHLTA,B565,61;%%
%\cite{Schael:2006cr}
%\bibitem{Schael:2006cr}
  S.~Schael {\it et al.}  [ALEPH Collaboration],
  %``Search for neutral MSSM Higgs bosons at LEP,''
  Eur.\ Phys.\ J.\  C {\bf 47}, 547 (2006)
  [arXiv:hep-ex/0602042];\\
  %%CITATION = EPHJA,C47,547;%%
%
%\cite{Ghosh:2004wr}
%\bibitem{Ghosh:2004wr}
  D.~K.~Ghosh and S.~Moretti,
  %``Probing the light neutral Higgs boson scenario of the CP-violating MSSM
  %Higgs sector at the LHC,''
  Eur.\ Phys.\ J.\  C {\bf 42}, 341 (2005)
  [arXiv:hep-ph/0412365];\\
  %%CITATION = EPHJA,C42,341;%%
%
%\cite{Ghosh:2004cc}
%\bibitem{Ghosh:2004cc}
  D.~K.~Ghosh, R.~M.~Godbole and D.~P.~Roy,
  %``Probing the CP-violating light neutral Higgs in the charged Higgs decay  at
  %the LHC,''
  Phys.\ Lett.\  B {\bf 628}, 131 (2005)
  [arXiv:hep-ph/0412193];\\
  %%CITATION = PHLTA,B628,131;%%
%
%
%\cite{Bandyopadhyay:2007cp}
%\bibitem{Bandyopadhyay:2007cp}
  P.~Bandyopadhyay, A.~Datta, A.~Datta and B.~Mukhopadhyaya,
  %``Associated Higgs Production in CP-violating supersymmetry: probing the
  %`open hole' at the Large Hadron Collider,''
  Phys.\ Rev.\  D {\bf 78}, 015017 (2008)
  [arXiv:0710.3016 [hep-ph]];\\
  %%CITATION = PHRVA,D78,015017;%%
%
%\cite{Belyaev:2006rf}
%\bibitem{Belyaev:2006rf}
  A.~Belyaev, Q.~H.~Cao, D.~Nomura, K.~Tobe and C.~P.~Yuan,
  %``Light MSSM Higgs boson scenario and its test at hadron colliders,''
  Phys.\ Rev.\ Lett.\  {\bf 100}, 061801 (2008)
  [arXiv:hep-ph/0609079].
  %%CITATION = PRLTA,100,061801;%%

%%------------recent higgs review
\bibitem{Hreview}
For recent reviews and studies, see for example, \\
%\cite{Accomando:2006ga}
%\bibitem{Accomando:2006ga}
  E.~Accomando {\it et al.},
  %``Workshop on CP studies and non-standard Higgs physics,''
  arXiv:hep-ph/0608079;\\
%
%\cite{Godbole:2007cn}
%\bibitem{Godbole:2007cn}
  R.~M.~Godbole, D.~J.~Miller and M.~M.~Muhlleitner,
  %``Aspects of CP violation in the H ZZ coupling at the LHC,''
  JHEP {\bf 0712}, 031 (2007)
  [arXiv:0708.0458 [hep-ph]];\\
  %%CITATION = JHEPA,0712,031;%%
%
  %%CITATION = HEP-PH/0608079;%%
%\cite{Williams:2007dc}
%\bibitem{Williams:2007dc}
  K.~E.~Williams and G.~Weiglein,
  %``Precise predictions for h_a --> h_b h_c decays in the complex MSSM,''
  Phys.\ Lett.\  B {\bf 660} (2008) 217
  [arXiv:0710.5320 [hep-ph]];
  %%CITATION = PHLTA,B660,217;%%
%\cite{Williams:2007dg}
%\bibitem{Williams:2007dg}
%  K.~E.~Williams and G.~Weiglein,
  %``Higgs boson decays in the Complex MSSM,''
  arXiv:0710.5331 [hep-ph];\\
  %%CITATION = ARXIV:0710.5331;%%
%\bibitem{Dreiner:2007ay}
  H.~K.~Dreiner, O.~Kittel and F.~von der Pahlen,
  %``Disentangling CP phases in nearly degenerate resonances: neutralino
  %production via Higgs at a muon collider,''
  JHEP {\bf 0801}, 017 (2008)
  [arXiv:0711.2253 [hep-ph]];\\
  %%CITATION = JHEPA,0801,017;%%
%
%\cite{Kittel:2008be}
%\bibitem{Kittel:2008be}
  O.~Kittel and F.~von der Pahlen,
  %``CP-violating Higgs boson mixing in chargino production at the muon
  %collider,''
  JHEP {\bf 0808}, 030 (2008)
  [arXiv:0806.4534 [hep-ph]];\\
  %%CITATION = JHEPA,0808,030;%%
%
%\cite{Lee:2008eqa}
%\bibitem{Lee:2008eqa}
  J.~S.~Lee,
  %``Manifestations of CP Violation in the MSSM Higgs Sector,''
  AIP Conf.\ Proc.\  {\bf 1078}, 36 (2009)
  [arXiv:0808.2014 [hep-ph]];\\
  %%CITATION = APCPC,1078,36;%%
%
%\cite{Das:2008jp}
%\bibitem{Das:2008jp}
  S.~P.~Das, A.~Datta and M.~Drees,
  %``CP-violating Higgs at Tevatron,''
  AIP Conf.\ Proc.\  {\bf 1078}, 223 (2009)
  [arXiv:0809.2209 [hep-ph]];\\
  %%CITATION = APCPC,1078,223;%%
%
%\cite{Weiglein:2008zz}
%\bibitem{Weiglein:2008zz}
  G.~Weiglein,
  %``CP-violating loop effects in the MSSM Higgs sector,''
  Nucl.\ Phys.\ Proc.\ Suppl.\  {\bf 183}, 149 (2008);\\
  %%CITATION = NUPHZ,183,149;%%
%
%
%\bibitem{Bechtle:2008jh}
  P.~Bechtle, O.~Brein, S.~Heinemeyer, G.~Weiglein and K.~E.~Williams,
  %``HiggsBounds: Confronting Arbitrary Higgs Sectors with Exclusion Bounds from
  %LEP and the Tevatron,''
  Comput.\ Phys.\ Commun.\  {\bf 181}, 138 (2010)
  [arXiv:0811.4169 [hep-ph]];\\
  %%CITATION = CPHCB,181,138;%%
%
%
%\cite{Djouadi:2009nu}
%\bibitem{Djouadi:2009nu}
  A.~Djouadi and R.~M.~Godbole,
  %``Ewsb at LHC,''
  arXiv:0901.2030 [hep-ph];\\
  %%CITATION = ARXIV:0901.2030;%%
%
%\cite{Moretti:2007th}
%\bibitem{Moretti:2007th}
  S.~Moretti, S.~Munir and P.~Poulose,
  %``Explicit CP violation in the MSSM through Higgs --> gamma gamma,''
  Phys.\ Lett.\  B {\bf 649}, 206 (2007)
  [arXiv:hep-ph/0702242];\\
  %%CITATION = PHLTA,B649,206;%%
%\cite{Hesselbach:2009gw}
%\bibitem{Hesselbach:2009gw}
  S.~Hesselbach, S.~Moretti, S.~Munir and P.~Poulose,
  %``Explicit CP violation in the MSSM through gg-->H1-->gamma.gamma,''
  arXiv:0903.0747 [hep-ph].
  %%CITATION = ARXIV:0903.0747;%%


%
%%%%-------------- LHC-------------- 
\bibitem{LHC}
%\cite{Abdullin:1998pm}
%\bibitem{Abdullin:1998pm}
  S.~Abdullin {\it et al.}  [CMS Collaboration],
  %``Discovery potential for supersymmetry in CMS,''
  J.\ Phys.\ G {\bf 28}, 469 (2002)
  [arXiv:hep-ph/9806366];\\
  %%CITATION = JPHGB,G28,469;%%
%\cite{:1999fr}
%\bibitem{:1999fr}
 ATLAS collab.,  
  {\it ATLAS detector and physics performance. Technical design report.  Vol. 2},
  CERN–LHCC–99–15;\\
  %%CITATION = ATLAS-TDR-15;%%
%
%\cite{Weiglein:2004hn}
%\bibitem{Weiglein:2004hn}
  G.~Weiglein {\it et al.}  [LHC/LC Study Group],
  %``Physics interplay of the LHC and the ILC,''
  Phys.\ Rept.\  {\bf 426}, 47 (2006)
  [arXiv:hep-ph/0410364].
  %%CITATION = PRPLC,426,47;%%


%%%%-------------- ILC-------------- 
\bibitem{ILC}
%\cite{:2007sg}
%\bibitem{:2007sg}
  J.~Brau {\it et al.}  [ILC Collaboration],
  %``ILC Reference Design Report Volume 1 - Executive Summary,''
  arXiv:0712.1950 [physics.acc-ph];\\
  %%CITATION = ARXIV:0712.1950;%%
%
%\cite{Aguilar-Saavedra:2001rg}
%\bibitem{Aguilar-Saavedra:2001rg}
  J.~A.~Aguilar-Saavedra {\it et al.}  [ECFA/DESY LC Physics Working Group],
  %   ``TESLA Technical Design Report Part III: 
  %  Physics at an e+e- Linear Collider,''
  arXiv:hep-ph/0106315;\\
  %%CITATION = HEP-PH 0106315;%%
%
%\cite{Abe:2001nn}
%\bibitem{Abe:2001nn}
  T.~Abe {\it et al.}  [American Linear Collider Working Group],
  %   ``Linear collider physics resource book for Snowmass 2001. 1: 
  % Introduction,''
  %in {\it Proc. of the APS/DPF/DPB Summer Study on the Future of Particle 
  %Physics (Snowmass 2001) } ed. N.~Graf,
  arXiv:hep-ex/0106055;\\
  %%CITATION = HEP-EX 0106055;%%
%
%\cite{Abe:2001gc}
%\bibitem{Abe:2001gc}
  K.~Abe {\it et al.}  [ACFA Linear Collider Working Group],
  %``Particle physics experiments at JLC,''
  arXiv:hep-ph/0109166;\\
  %%CITATION = HEP-PH 0109166;%%
%\bibitem{Aguilar-Saavedra:2005pw}
  J.~A.~Aguilar-Saavedra {\it et al.},
  %``Supersymmetry parameter analysis: SPA convention and project,''
  Eur.\ Phys.\ J.\ C {\bf 46} (2006) 43
  [arXiv:hep-ph/0511344].
%

%%---------cp-even observables, top bottom masses
%%\cite{Ibrahim:2003ca}
%\bibitem{Ibrahim:2003ca}
%  T.~Ibrahim and P.~Nath,
%  %``SUSY QCD and SUSY electroweak loop corrections to b, t and tau masses
%  %including the effects of CP phases,''
%  Phys.\ Rev.\  D {\bf 67}, 095003 (2003)
%  [Erratum-ibid.\  D {\bf 68}, 019901 (2003)]
%  [arXiv:hep-ph/0301110].
%  %%CITATION = PHRVA,D67,095003;%% 

%---------cp-even observables, stop and sbottom,decays
%\cite{Bartl:2003he}
\bibitem{Bartl:2003he}
  A.~Bartl, S.~Hesselbach, K.~Hidaka, T.~Kernreiter and W.~Porod,
  %``Impact of CP phases on stop and sbottom searches,''
  Phys.\ Lett.\  B {\bf 573}, 153 (2003)
  [arXiv:hep-ph/0307317];
  %%CITATION = PHLTA,B573,153;%%
%\cite{Bartl:2003pd}
%\bibitem{Bartl:2003pd}
 % A.~Bartl, S.~Hesselbach, K.~Hidaka, T.~Kernreiter and W.~Porod,
  %``Top squarks and bottom squarks in the MSSM with complex parameters,''
  Phys.\ Rev.\  D {\bf 70}, 035003 (2004)
  [arXiv:hep-ph/0311338].
  %%CITATION = PHRVA,D70,035003;%%


%---------cp-even observables, stau 
%\cite{Bartl:2002uy}
\bibitem{Bartl:2002uy}
  A.~Bartl, K.~Hidaka, T.~Kernreiter and W.~Porod,
  %``Impact of CP phases on the search for sleptons stau and sneutrino/tau,''
  Phys.\ Lett.\  B {\bf 538} (2002) 137
  [arXiv:hep-ph/0204071];
  %%CITATION = PHLTA,B538,137;%%
%\cite{Bartl:2002bh}
%\bibitem{Bartl:2002bh}
%  A.~Bartl, K.~Hidaka, T.~Kernreiter and W.~Porod,
  %``Tau-Sleptons and Tau-Sneutrino in the MSSM with Complex Parameters,''
  Phys.\ Rev.\  D {\bf 66}, 115009 (2002)
  [arXiv:hep-ph/0207186].
  %%CITATION = PHRVA,D66,115009;%%

%---------cp-even observables, cross sections, 1. and 2. generation
%\cite{Alan:2007rp}
\bibitem{Alan:2007rp}
  For the impact of CP phases in the first and second generation
  squarks see, for example,
  A.~T.~Alan, K.~Cankocak and D.~A.~Demir,
  %``Squark pair production in the MSSM with explicit CP violation,''
  Phys.\ Rev.\  D {\bf 75}, 095002 (2007)
  [Erratum-ibid.\  D {\bf 76}, 119903 (2007)]
  [arXiv:hep-ph/0702289].
  %%CITATION = PHRVA,D75,095002;%%



%---------cp-even fermiion polarization with CP phases  
%\cite{Gajdosik:2004ed}
\bibitem{Gajdosik:2004ed}
  T.~Gajdosik, R.~M.~Godbole and S.~Kraml,
  %``Fermion polarization in sfermion decays as a probe of CP phases in the
  %MSSM,''
  JHEP {\bf 0409}, 051 (2004)
  [arXiv:hep-ph/0405167].
  %%CITATION = JHEPA,0409,051;%%
%---------CP conserving studies
%\cite{Boos:2003vf}
%\bibitem{Boos:2003vf}
%  E.~Boos, H.~U.~Martyn, G.~A.~Moortgat-Pick, M.~Sachwitz, A.~Sherstnev and P.~M.~Zerwas,
%  %``Polarisation in sfermion decays: Determining tan(beta) and trilinear
%  %couplings,''
%  Eur.\ Phys.\ J.\  C {\bf 30}, 395 (2003)
%  [arXiv:hep-ph/0303110].
  %%CITATION = EPHJA,C30,395;%%


%%%%------other CP-odd observables
\bibitem{otherCPodd}
%\cite{Choi:1998ub}
%\bibitem{Choi:1998ub}
  S.~Y.~Choi and M.~Drees,
  %``CP-violation through scalar tau oscillation,''
  Phys.\ Lett.\  B {\bf 435}, 356 (1998)
  [arXiv:hep-ph/9805474];\\
  %%CITATION = PHLTA,B435,356;%%
%
%\cite{Aoki:1998cq}
%\bibitem{Aoki:1998cq}
  M.~Aoki and N.~Oshimo,
  %``Decay rate asymmetry of top squark,''
  Mod.\ Phys.\ Lett.\  A {\bf 13}, 3225 (1998)
  [arXiv:hep-ph/9808217];\\
  %%CITATION = MPLAE,A13,3225;%%
%
%\cite{Yang:2002bj}
%\bibitem{Yang:2002bj}
  W.~M.~Yang and D.~S.~Du,
  %``CP asymmetry in tau slepton decay in the minimal supersymmetric  standard
  %model,''
  Phys.\ Rev.\  D {\bf 65}, 115005 (2002)
  [arXiv:hep-ph/0202049];\\
  %%CITATION = PHRVA,D65,115005;%%
%
%\cite{Ibrahim:2004gb}
%\bibitem{Ibrahim:2004gb}
  T.~Ibrahim and P.~Nath,
  %``Effective lagrangian for anti-q squark'(i) chi(j)+, anti-q squark'(i)
  %chi(j)0 interactions and fermionic decays of the squarks with CP  phases,''
  Phys.\ Rev.\  D {\bf 71}, 055007 (2005)
  [arXiv:hep-ph/0411272];\\
  %%CITATION = PHRVA,D71,055007;%%
%
%\cite{Rolbiecki:2007se}
%\bibitem{Rolbiecki:2007se}
  K.~Rolbiecki and J.~Kalinowski,
  %``CP violation at one loop in the polarization-independent chargino
  %production in e+e- collisions,''
  Phys.\ Rev.\  D {\bf 76}, 115006 (2007)
  [arXiv:0709.2994 [hep-ph]];
%\cite{Kalinowski:2008qm}
%\bibitem{Kalinowski:2008qm}
%  J.~Kalinowski and K.~Rolbiecki,
  %``CP violation in the chargino/neutralino sector of the MSSM,''
  Acta Phys.\ Polon.\  B {\bf 39}, 1585 (2008)
  [arXiv:0804.0549 [hep-ph]];
  %%CITATION = APPOA,B39,1585;%%
%\cite{Rolbiecki:2007qq}
%\bibitem{Rolbiecki:2007qq}
%  K.~Rolbiecki and J.~Kalinowski,
  %``CP violation in chargino production in e+e- collisions,''
  Acta Phys.\ Polon.\  B {\bf 38}, 3557 (2007)
  [arXiv:0710.3318 [hep-ph]];\\
  %%CITATION = APPOA,B38,3557;%%
%
  %%CITATION = PHRVA,D76,115006;%%
%\cite{Osland:2007xw}
%\bibitem{Osland:2007xw}
  P.~Osland and A.~Vereshagin,
  %``CP violation in unpolarized e+ e- --> charginos at one loop level,''
  Phys.\ Rev.\  D {\bf 76}, 036001 (2007)
  [arXiv:0704.2165 [hep-ph]];\\
  %%CITATION = PHRVA,D76,036001;%%
%\cite{Frank:2007zza}
%\bibitem{Frank:2007zza}
  M.~Frank and I.~Turan,
  %``CP Asymmetry in Charged Higgs Decays to Chargino-Neutralino,''
  Phys.\ Rev.\  D {\bf 76}, 076008 (2007)
  [arXiv:0708.0026 [hep-ph]];\\
  %%CITATION = PHRVA,D76,076008;%%
%\cite{Kittel:2008be}
%\bibitem{Kittel:2008be}
  O.~Kittel and F.~von der Pahlen,
  %``CP-violating Higgs boson mixing in chargino production at the muon
  %collider,''
  JHEP {\bf 0808}, 030 (2008)
  [arXiv:0806.4534 [hep-ph]];\\
  %%CITATION = JHEPA,0808,030;%%
%\cite{Dreiner:2007ay}
%\bibitem{Dreiner:2007ay}
  H.~K.~Dreiner, O.~Kittel and F.~von der Pahlen,
  %``Disentangling CP phases in nearly degenerate resonances: neutralino
  %production via Higgs at a muon collider,''
  JHEP {\bf 0801}, 017 (2008)
  [arXiv:0711.2253 [hep-ph]];\\
  %%CITATION = JHEPA,0801,017;%%
%
%\bibitem{Christova:2006fb}
  E.~Christova, H.~Eberl, E.~Ginina and W.~Majerotto,
  %``CP violation in charged Higgs decays in the MSSM,''
  JHEP {\bf 0702}, 075 (2007)
  [arXiv:hep-ph/0612088];
  %%CITATION = JHEPA,0702,075;%%
%
%\cite{Christova:2008jv}
%\bibitem{Christova:2008jv}
  E.~Christova, H.~Eberl, E.~Ginina and W.~Majerotto,
  %``CP violation in $H^\pm t$ production at the LHC,''
  Phys.\ Rev.\  D {\bf 79}, 096005 (2009)
  [arXiv:0812.4392 [hep-ph]];\\
  %%CITATION = PHRVA,D79,096005;%%
%\cite{Christova:2008fw}
%\bibitem{Christova:2008fw}
  E.~Christova, H.~Eberl and E.~Ginina,
  %``CP violation in associated production of charged Higgs boson and top quark
  %at the LHC,''
  arXiv:0812.1129 [hep-ph];\\
  %%CITATION = ARXIV:0812.1129;%%
%\cite{Eberl:2005ay}
%\bibitem{Eberl:2005ay}
  H.~Eberl, T.~Gajdosik, W.~Majerotto and B.~Schrausser,
  %``CP-violating asymmetry in chargino decay into neutralino and W boson,''
  Phys.\ Lett.\  B {\bf 618}, 171 (2005)
  [arXiv:hep-ph/0502112];\\
  %%CITATION = PHLTA,B618,171;%%
%\cite{Yang:2002am}
%\bibitem{Yang:2002am}
  W.~M.~Yang and D.~S.~Du,
  %``CP violation in chargino decays in the MSSM,''
  Phys.\ Rev.\  D {\bf 67}, 055004 (2003)
  [arXiv:hep-ph/0211453].
  %%CITATION = PHRVA,D67,055004;%%

%%%-------------Triple products-basics---------------------
\bibitem{tripleprods}
%\bibitem{valencia} 
G.~Valencia, 
%``Constructing CP odd observables,''
%arXiv:hep-ph/9411441.
arXiv:hep-ph/9411441;\\
%%CITATION = HEP-PH 9411441;%%
G.C.~Branco, L.~Lavoura, and J.P.~Silva, {\em CP violation}, 
Oxford University Press, New York, 1999.



%%%-----Singeled out references for Triple products in Neutralino 2 body decays--------

%\cite{Bartl:2003tr}
\bibitem{Bartl:2003tr}
  A.~Bartl, H.~Fraas, O.~Kittel and W.~Majerotto,
  %``CP asymmetries in neutralino production in e+ e- collisions,''
  Phys.\ Rev.\  D {\bf 69}, 035007 (2004)
  [arXiv:hep-ph/0308141].
  %%CITATION = PHRVA,D69,035007;%%

%%%-----Triple products in Neutralino 2 body decays (originating from sfermions)----
%\cite{Bartl:2003ck}
\bibitem{Bartl:2003ck}
  A.~Bartl, H.~Fraas, T.~Kernreiter and O.~Kittel,
  %``T-odd correlations in the decay of scalar fermions,''
  Eur.\ Phys.\ J.\  C {\bf 33}, 433 (2004)
  [arXiv:hep-ph/0306304].
  %%CITATION = EPHJA,C33,433;%%


%%%-----Triple products in Neutralino 2 body decays-----------
\bibitem{NEUT2}
For further studies with neutralino two-body decays at the ILC, see\\
%\cite{Bartl:2003gr}
%\bibitem{Bartl:2003gr}
  A.~Bartl, T.~Kernreiter and O.~Kittel,
  %``A CP asymmetry in e+ e- --> neutralino(i) neutralino(j) -->  neutralino(j)
  %tau stau(k) with tau polarization,''
  Phys.\ Lett.\  B {\bf 578}, 341 (2004)
  [arXiv:hep-ph/0309340];\\
  %%CITATION = PHLTA,B578,341;%%
%
%\cite{Choi:2003pq}
%\bibitem{Choi:2003pq}
  S.~Y.~Choi, M.~Drees, B.~Gaissmaier and J.~Song,
  %``Analysis of CP violation in neutralino decays to tau sleptons,''
  Phys.\ Rev.\  D {\bf 69}, 035008 (2004)
  [arXiv:hep-ph/0310284];\\
  %%CITATION = PHRVA,D69,035008;%%
%
%\cite{Bartl:2004ut}
%\bibitem{Bartl:2004ut}
  A.~Bartl, H.~Fraas, O.~Kittel and W.~Majerotto,
  %``CP sensitive observables in e+ e- --> neutralino(i) neutralino(j) and
  %neutralino decay into Z boson,''
  Eur.\ Phys.\ J.\ C {\bf 36}, 233 (2004)
  [arXiv:hep-ph/0402016];
%\cite{Bartl:2003tt}
%\bibitem{Bartl:2003tt}
%  A.~Bartl, H.~Fraas, O.~Kittel and W.~Majerotto,
  %``CP asymmetries in e+ e- --> neutralino/i neutralino/j,''
  [arXiv:hep-ph/0308143];\\
  %%CITATION = HEP-PH/0308143;%%
%
%\cite{Bartl:2003kn}
%\bibitem{Bartl:2003kn}
  A.~Bartl, H.~Fraas, T.~Kernreiter, O.~Kittel and W.~Majerotto,
  %``Impact of beam polarization on CP asymmetries in neutralino pair
  %production,''
  arXiv:hep-ph/0310011;\\
  %%CITATION = HEP-PH/0310011;%%
%
%
%\cite{Aguilar-Saavedra:2004dz}
%\bibitem{Aguilar-Saavedra:2004dz}
  J.~A.~Aguilar-Saavedra,
  % ``CP violation in neutralino(1) neutralino(2) production at a linear
  %collider,''
  Nucl.\ Phys.\ B {\bf 697} (2004) 207
  [arXiv:hep-ph/0404104];\\
  %%CITATION = HEP-PH 0404104;%%
%
  %%CITATION = HEP-PH 0402016;%%
%\cite{Choi:2003fs}
%\bibitem{Choi:2003fs}
  S.~Y.~Choi and Y.~G.~Kim,
  %``Analysis of the neutralino system in two-body decays of neutralinos,''
  Phys.\ Rev.\  D {\bf 69}, 015011 (2004)
  [arXiv:hep-ph/0311037];\\
  %%CITATION = PHRVA,D69,015011;%%
%
%\cite{Bartl:2009pg}
%\bibitem{Bartl:2009pg}
  A.~Bartl, K.~Hohenwarter-Sodek, T.~Kernreiter, O.~Kittel and M.~Terwort,
  %``CP-sensitive spin-spin correlations in neutralino production at the ILC,''
  JHEP {\bf 0907}, 054 (2009)
  [arXiv:0905.1782 [hep-ph]].
  %%CITATION = JHEPA,0907,054;%%


%%%-----Triple products in Neutralino 3 body decays----
\bibitem{NEUT3}
For studies with neutralino three-body decays at the ILC, see\\
%\cite{Kizukuri:1990iy}
%\bibitem{Kizukuri:1990iy}
Y.~Kizukuri and N.~Oshimo,
%``T Odd Asymmetry Mediated By Neutralino In E+ E- Annihilation,''
Phys.\ Lett.\ B {\bf 249} (1990) 449;\\
%%CITATION = PHLTA,B249,449;%%
%
%\cite{Choi:1999cc}
%\bibitem{Choi:1999cc}
  S.~Y.~Choi, H.~S.~Song and W.~Y.~Song,
  %``CP phases in correlated production and decay of neutralinos in the  minimal
  %supersymmetric standard model,''
  Phys.\ Rev.\  D {\bf 61}, 075004 (2000)
  [arXiv:hep-ph/9907474];\\
  %%CITATION = PHRVA,D61,075004;%%
%
%\cite{Bartl:2004jj}
%\bibitem{Bartl:2004jj}
  A.~Bartl, H.~Fraas, S.~Hesselbach, 
  K.~Hohenwarter-Sodek and G.~A.~Moortgat-Pick,
  %``A T-odd asymmetry in neutralino production and decay,''
  JHEP {\bf 0408}, 038 (2004)
  [arXiv:hep-ph/0406190];\\
  %%CITATION = HEP-PH 0406190;%%
%\cite{Choi:2005gt}
%\bibitem{Choi:2005gt}
  S.~Y.~Choi, B.~C.~Chung, J.~Kalinowski, Y.~G.~Kim and K.~Rolbiecki,
  %``Analysis of the neutralino system in three-body leptonic decays of
  %neutralinos,''
  Eur.\ Phys.\ J.\  C {\bf 46}, 511 (2006)
  [arXiv:hep-ph/0504122].
  %%CITATION = EPHJA,C46,511;%%

%\cite{Kittel:2004rp}
\bibitem{Kittel:2004rp}
  O.~Kittel,
  %``CP violation in production and decay of supersymmetric particles,''
  arXiv:hep-ph/0504183.
  %%CITATION = HEP-PH/0504183;%%


%%%-----Triple products in Chargino 2 body decays----
\bibitem{CHAR2}
For studies with chargino two-body decays at the ILC, see\\
%\cite{Choi:2000ta}
%\bibitem{Choi:2000ta}
  S.~Y.~Choi, A.~Djouadi, M.~Guchait, J.~Kalinowski, H.~S.~Song and P.~M.~Zerwas,
  %``Reconstructing the chargino system at e+ e- linear colliders,''
  Eur.\ Phys.\ J.\ C {\bf 14}, 535 (2000)
  [arXiv:hep-ph/0002033];\\
  %%CITATION = HEP-PH 0002033;%%
%
%\cite{Bartl:2004vi}
%\bibitem{Bartl:2004vi}
  A.~Bartl, H.~Fraas, O.~Kittel and W.~Majerotto,
  %``CP violation in chargino production and decay into sneutrino,''
  Phys.\ Lett.\  B {\bf 598}, 76 (2004)
  [arXiv:hep-ph/0406309];\\
  %%CITATION = PHLTA,B598,76;%%
%
%\cite{Kittel:2004kd}
%\bibitem{Kittel:2004kd}
  O.~Kittel, A.~Bartl, H.~Fraas and W.~Majerotto,
  %``CP sensitive observables in chargino production and decay into a W
  %boson,''
  Phys.\ Rev.\  D {\bf 70}, 115005 (2004)
  [arXiv:hep-ph/0410054];\\
  %%CITATION = PHRVA,D70,115005;%%
%
%\cite{AguilarSaavedra:2004ru}
%\bibitem{AguilarSaavedra:2004ru}
  J.~A.~Aguilar-Saavedra,
  %``Sneutrino cascade decays sneutrino/e --> e- chargino(1)+ --> e- f  anti-f'
  %neutralino(1) as a probe of chargino spin properties and CP  violation,''
  Nucl.\ Phys.\  B {\bf 717}, 119 (2005)
  [arXiv:hep-ph/0410068];\\
  %%CITATION = NUPHA,B717,119;%%
%
%\cite{Bartl:2008fu}
%\bibitem{Bartl:2008fu}
  A.~Bartl, K.~Hohenwarter-Sodek, T.~Kernreiter, O.~Kittel and M.~Terwort,
  %``CP observables with spin-spin correlations in chargino production,''
  Nucl.\ Phys.\  B {\bf 802}, 77 (2008)
  [arXiv:0802.3592 [hep-ph]].
  %%CITATION = NUPHA,B802,77;%%

%%%-----Triple products in Chargino 3 body decays----
\bibitem{CHAR3}
For studies with chargino three-body decays at the ILC, see\\
%\cite{Kizukuri:1993vh}
%\bibitem{Kizukuri:1993vh}
  Y.~Kizukuri and N.~Oshimo,
  %``T-odd asymmetry in chargino pair production processes,''
  arXiv:hep-ph/9310224;\\
  %%CITATION = HEP-PH/9310224;%%
%
%\cite{Bartl:2006yv}
%\bibitem{Bartl:2006yv}
  A.~Bartl, H.~Fraas, S.~Hesselbach, K.~Hohenwarter-Sodek, T.~Kernreiter and G.~Moortgat-Pick,
  %``CP asymmetries in chargino production and decay: The three-body decay
  %case,''
  Eur.\ Phys.\ J.\  C {\bf 51}, 149 (2007)
  [arXiv:hep-ph/0608065].
  %%CITATION = EPHJA,C51,149;%%

%%%---------Transverse beam pol in....
\bibitem{Trans}
%\cite{MoortgatPick:2005cw}
%\bibitem{MoortgatPick:2005cw}
  G.~A.~Moortgat-Pick {\it et al.},
  %``The role of polarized positrons and electrons in revealing fundamental
  %interactions at the linear collider,''
  Phys.\ Rept.\  {\bf 460}, 131 (2008)
  [arXiv:hep-ph/0507011];\\
  %%CITATION = PRPLC,460,131;%%
%
%\cite{Bartl:2004xy}
%\bibitem{Bartl:2004xy}
  A.~Bartl, K.~Hohenwarter-Sodek, T.~Kernreiter and H.~Rud,
  % ``CP sensitive observables in chargino production with transverse e+-  beam
  %polarization,''
  Eur.\ Phys.\ J.\ C {\bf 36} (2004) 515
  [arXiv:hep-ph/0403265];\\
  %%CITATION = HEP-PH 0403265;%%
%
%\cite{Bartl:2005uh}
%\bibitem{Bartl:2005uh}
  A.~Bartl, H.~Fraas, S.~Hesselbach, K.~Hohenwarter-Sodek, 
  T.~Kernreiter and G.~A.~Moortgat-Pick,
  %``CP-odd observables in neutralino production with transverse 
  % e+ and e- beam polarization,''
  JHEP {\bf 0601}, 170 (2006)
  [arXiv:hep-ph/0510029];\\
  %%CITATION = HEP-PH 0510029;%%
%
%\cite{Choi:2006vh}
%\bibitem{Choi:2006vh}
  S.~Y.~Choi, M.~Drees and J.~Song,
  %``Neutralino production and decay at an e+ e- linear collider with
  %transversely polarized beams,''
  JHEP {\bf 0609}, 064 (2006)
  [arXiv:hep-ph/0602131];\\
  %%CITATION = JHEPA,0609,064;%%
%
%\cite{Bartl:2006bn}
%\bibitem{Bartl:2006bn}
  A.~Bartl, H.~Fraas, K.~Hohenwarter-Sodek, T.~Kernreiter, G.~Moortgat-Pick and A.~Wagner,
  %``Selectron production at an e- e- linear collider with transversely
  %polarized beams,''
  Phys.\ Lett.\  B {\bf 644}, 165 (2007)
  [arXiv:hep-ph/0610431];\\
  %%CITATION = PHLTA,B644,165;%%
%
%\cite{Bartl:2007qy}
%\bibitem{Bartl:2007qy}
  A.~Bartl, K.~Hohenwarter-Sodek, T.~Kernreiter and O.~Kittel,
  %``CP asymmetries with Longitudinal and Transverse Beam Polarizations in
  %Neutralino Production and Decay into the Z^0 Boson at the ILC,''
  JHEP {\bf 0709}, 079 (2007)
  [arXiv:0706.3822 [hep-ph]].
  %%CITATION = JHEPA,0709,079;%%

%%%%%------triple products proceedings
\bibitem{CPreview}
%\cite{Kittel:2009fg}
%\bibitem{Kittel:2009fg}
  O.~Kittel,
  %``SUSY CP phases and asymmetries at colliders,''
  J.\ Phys.\ Conf.\ Ser.\  {\bf 171}, 012094 (2009)
  [arXiv:0904.3241 [hep-ph]];\\
  %%CITATION = 00462,171,012094;%%
%
%\cite{Kraml:2007pr}
%\bibitem{Kraml:2007pr}
  S.~Kraml,
  %``CP violation in SUSY,''
  arXiv:0710.5117 [hep-ph];\\
  %%CITATION = ARXIV:0710.5117;%%
%
%\cite{Hesselbach:2007dq}
%\bibitem{Hesselbach:2007dq}
  S.~Hesselbach,
  %``CP Violation in SUSY Particle Production and Decay,''
%{\it In the Proceedings of 2007 International Linear Collider Workshop (LCWS07 and ILC07), Hamburg, Germany, 30 May - 3 Jun 2007, pp SUS11}
  arXiv:0709.2679 [hep-ph];
  %%CITATION = ECONF,C0705302,SUS11;%%
%\cite{Hesselbach:2004sp}
%\bibitem{Hesselbach:2004sp}
%  S.~Hesselbach,
  %``Effects of CP-violating phases in supersymmetry,''
  Acta Phys.\ Polon.\  B {\bf 35}, 2739 (2004)
  [arXiv:hep-ph/0410174];\\
  %%CITATION = APPOA,B35,2739;%%
%
%\cite{Bartl:2004iy}
%\bibitem{Bartl:2004iy}
  A.~Bartl and S.~Hesselbach,
  %``Effects of CP phases on the phenomenology of SUSY particles,''
  arXiv:hep-ph/0410237.
  %%CITATION = HEP-PH/0410237;%%


%%%%%--------triple product in stop decay

%~\cite{Bartl:2004jr}
\bibitem{Bartl:2004jr}
  A.~Bartl, E.~Christova, K.~Hohenwarter-Sodek and T.~Kernreiter,
  %``Triple product correlations in top squark decays,''
  Phys.\ Rev.\  D {\bf 70}, 095007 (2004)
  [arXiv:hep-ph/0409060].
  %%CITATION = PHRVA,D70,095007;%%

%%%%%--------triple product in sbottom decay
%\cite{Bartl:2006hh}
\bibitem{Bartl:2006hh}
  A.~Bartl, E.~Christova, K.~Hohenwarter-Sodek and T.~Kernreiter,
  %``CP asymmetries in scalar bottom quark decays,''
  JHEP {\bf 0611}, 076 (2006)
  [arXiv:hep-ph/0610234].
  %%CITATION = JHEPA,0611,076;%%


%\cite{Ellis:2008hq}
\bibitem{Ellis:2008hq}
  J.~Ellis, F.~Moortgat, G.~Moortgat-Pick, J.~M.~Smillie and J.~Tattersall,
  %``Measurement of CP Violation in Stop Cascade Decays at the LHC,''
  Eur.\ Phys.\ J.\  C {\bf 60}, 633 (2009)
  [arXiv:0809.1607 [hep-ph]].
  %%CITATION = EPHJA,C60,633;%%

 
%\cite{Langacker:2007ur}
\bibitem{Langacker:2007ur}
  P.~Langacker, G.~Paz, L.~T.~Wang and I.~Yavin,
  %``A T-odd observable sensitive to CP violating phases in squark decay,''
  JHEP {\bf 0707}, 055 (2007)
  [arXiv:hep-ph/0702068].
  %%CITATION = JHEPA,0707,055;%%

%%%%%--------spin formalism Haber
\bibitem{Haber:1994pe}
  H.~E.~Haber,
  %``Spin formalism and applications to new physics searches,''
  arXiv:hep-ph/9405376;\\
  %%CITATION = HEP-PH/9405376;%%
%
%\cite{Dreiner:2008tw}
%\bibitem{Dreiner:2008tw}
  H.~K.~Dreiner, H.~E.~Haber and S.~P.~Martin,
  %``Two-component spinor techniques and Feynman rules for quantum field theory
  %and supersymmetry,''
  arXiv:0812.1594 [hep-ph].
  %%CITATION = ARXIV:0812.1594;%%

%%%---------triple products in three-body decay stop
%\cite{Bartl:2002hi}
\bibitem{Bartl:2002hi}
  A.~Bartl, T.~Kernreiter and W.~Porod,
  %``A CP sensitive asymmetry in the three--body decay stop_1 -> b tau+
  %tau-sneutrino,''
  Phys.\ Lett.\  B {\bf 538} (2002) 59
  [arXiv:hep-ph/0202198];\\
  %%CITATION = PHLTA,B538,59;%%
%\cite{Kiers:2006aq}
%\bibitem{Kiers:2006aq}
  K.~Kiers, A.~Szynkman and D.~London,
  %``CP violation in supersymmetric theories: stop(2) --> stop(1) tau- tau+,''
  Phys.\ Rev.\  D {\bf 74}, 035004 (2006)
  [arXiv:hep-ph/0605123];
  %%CITATION = PHRVA,D74,035004;%%
%\cite{Szynkman:2007uc}
%\bibitem{Szynkman:2007uc}
%  A.~Szynkman, K.~Kiers and D.~London,
  %``CP Violation in Supersymmetric Theories: stop2->stop1 H H, stop2->stop1 Z
  %Z, stop2->stop1 W+ W-, stop2->stop1 Z H,''
  Phys.\ Rev.\  D {\bf 75}, 075009 (2007)
  [arXiv:hep-ph/0701165].
  %%CITATION = PHRVA,D75,075009;%%


%%%%%-------triple products in neutralino 3-body decays
%%%-----Triple products in Neutralino 3 body decays (originating from sfermions)--
%\cite{AguilarSaavedra:2004hu}
\bibitem{AguilarSaavedra:2004hu}
  J.~A.~Aguilar-Saavedra,
  %``CP violation in selectron cascade decays selectron(L) --> e  neutralino(2)
  %--> e neutralino(1) mu+ mu-,''
  Phys.\ Lett.\  B {\bf 596}, 247 (2004)
  [arXiv:hep-ph/0403243].
  %%CITATION = PHLTA,B596,247;%%


%%%%%----optimal observables------
\bibitem{optimal} 
%\bibitem{Atwood:1991ka}
  D.~Atwood and A.~Soni,
  %``Analysis For Magnetic Moment And Electric Dipole Moment Form-Factors Of The
  %Top Quark Via E+ E- $\to$ T Anti-T,''
  Phys.\ Rev.\  D {\bf 45}, 2405 (1992);\\
  %%CITATION = PHRVA,D45,2405;%%
%\cite{Diehl:1993br}
%\bibitem{Diehl:1993br}
  M.~Diehl and O.~Nachtmann,
  %``Optimal Observables For The Measurement Of Three Gauge Boson Couplings In
  %E+ E- $\to$ W+ W-,''
  Z.\ Phys.\  C {\bf 62}, 397 (1994);\\
  %%CITATION = ZEPYA,C62,397;%%
%\cite{Grzadkowski:1995rx}
%\bibitem{Grzadkowski:1995rx}
  B.~Grzadkowski and J.~F.~Gunion,
  %``Using decay angle correlations to detect CP violation in the neutral Higgs
  %sector,''
  Phys.\ Lett.\  B {\bf 350}, 218 (1995)
  [arXiv:hep-ph/9501339].
  %%CITATION = PHLTA,B350,218;%%


%\cite{AguilarSaavedra:2006fy}
\bibitem{AguilarSaavedra:2006fy}
  J.~A.~Aguilar-Saavedra, J.~Carvalho, N.~F.~Castro, F.~Veloso and A.~Onofre,
  %``Probing anomalous W t b couplings in top pair decays,''
  Eur.\ Phys.\ J.\  C {\bf 50}, 519 (2007)
  [arXiv:hep-ph/0605190];
  %%CITATION = EPHJA,C50,519;%%
%\cite{AguilarSaavedra:2007rs}
%\bibitem{AguilarSaavedra:2007rs}
%  J.~A.~Aguilar-Saavedra, J.~Carvalho, N.~F.~Castro, A.~Onofre and F.~Veloso,
  %``ATLAS sensitivity to Wtb anomalous couplings in top quark decays,''
  Eur.\ Phys.\ J.\  C {\bf 53}, 689 (2008)
  [arXiv:0705.3041 [hep-ph]].
  %%CITATION = EPHJA,C53,689;%%


\bibitem{Lee:2007gn}
  J.~S.~Lee, M.~Carena, J.~Ellis, A.~Pilaftsis and C.~E.~M.~Wagner,
  %``CPsuperH2.0: an Improved Computational Tool for Higgs Phenomenology in the
  %MSSM with Explicit CP Violation,''
  Comput.\ Phys.\ Commun.\  {\bf 180}, 312 (2009)
  [arXiv:0712.2360 [hep-ph]].
  %%CITATION = CPHCB,180,312;%%

\bibitem{Leptogen}
%\bibitem{Deppisch:2002vz}
  F.~Deppisch, H.~Pas, A.~Redelbach, R.~Ruckl and Y.~Shimizu,
  %``Probing the Majorana mass scale of right-handed neutrinos in mSUGRA,''
  Eur.\ Phys.\ J.\  C {\bf 28}, 365 (2003)
  [arXiv:hep-ph/0206122], and references therein;\\
  %%CITATION = EPHJA,C28,365;%%
%
%
%\bibitem{Deppisch:2005rv}
  F.~Deppisch, H.~Pas, A.~Redelbach and R.~Ruckl,
  %``Constraints on SUSY seesaw parameters from leptogenesis and lepton  flavor
  %violation,''
  Phys.\ Rev.\  D {\bf 73}, 033004 (2006)
  [arXiv:hep-ph/0511062], and references therein;\\
  %%CITATION = PHRVA,D73,033004;%%
%
%\cite{Chung:2009eb}
%\bibitem{Chung:2009eb}
  D.~J.~H.~Chung, B.~Garbrecht, M.~J.~Ramsey-Musolf,
  %``Low Scale Leptogenesis from Non-Leptonic CP-Phases,''
  arXiv:0904.1591 [hep-ph].
  %%CITATION = ARXIV:0904.1591;%%

\bibitem{reviewLeptogen}
For recent reviews on Leptogenesis, see for example,\\
%
%\cite{Pilaftsis:2009pk}
%\bibitem{Pilaftsis:2009pk}
  A.~Pilaftsis,
  %``The Little Review on Leptogenesis,''
  J.\ Phys.\ Conf.\ Ser.\  {\bf 171}, 012017 (2009)
  [arXiv:0904.1182 [hep-ph]];\\
  %%CITATION = 00462,171,012017;%%
%
%\cite{Davidson:2008bu}
%\bibitem{Davidson:2008bu}
  S.~Davidson, E.~Nardi and Y.~Nir,
  %``Leptogenesis,''
  Phys.\ Rept.\  {\bf 466}, 105 (2008)
  [arXiv:0802.2962 [hep-ph]];\\
  %%CITATION = PRPLC,466,105;%%
%
%\cite{Pascoli:2006ci}
%\bibitem{Pascoli:2006ci}
  S.~Pascoli, S.~T.~Petcov and A.~Riotto,
  %``Leptogenesis and low energy CP violation in neutrino physics,''
  Nucl.\ Phys.\  B {\bf 774}, 1 (2007)
  [arXiv:hep-ph/0611338];\\
  %%CITATION = NUPHA,B774,1;%%
%
%\cite{Buchmuller:2004tu}
%\bibitem{Buchmuller:2004tu}
  W.~Buchmuller, P.~Di Bari and M.~Plumacher,
  %``Some Aspects of Thermal Leptogenesis,''
  New J.\ Phys.\  {\bf 6}, 105 (2004)
  [arXiv:hep-ph/0406014].
  %%CITATION = NJOPF,6,105;%%


%--------Stop production at NLO
%
\bibitem{NLOstop}
%\cite{Beenakker:1997ut}
%\bibitem{Beenakker:1997ut}
  W.~Beenakker, M.~Kramer, T.~Plehn, M.~Spira and P.~M.~Zerwas,
%  %``Stop production at hadron colliders,''
  Nucl.\ Phys.\  B {\bf 515}, 3 (1998)
  [arXiv:hep-ph/9710451];\\
%  %%CITATION = NUPHA,B515,3;%%
%
%%\cite{Hollik:2007wf}
%\bibitem{Hollik:2007wf}
  W.~Hollik, M.~Kollar and M.~K.~Trenkel,
%  %``Hadronic production of top-squark pairs with electroweak NLO
%  %contributions,''
  JHEP {\bf 0802}, 018 (2008)
  [arXiv:0712.0287 [hep-ph]];\\
%  %%CITATION = JHEPA,0802,018;%%
%
%
%%\cite{Beccaria:2008mi}
%\bibitem{Beccaria:2008mi}
  M.~Beccaria, G.~Macorini, L.~Panizzi, F.~M.~Renard and C.~Verzegnassi,
%  %``Stop-antistop and sbottom-antisbottom production at LHC: a one-loop 
%%   search for model parameters dependence,''
  Int.\ J.\ Mod.\ Phys.\  A {\bf 23}, 4779 (2008)
  [arXiv:0804.1252 [hep-ph]].
%
%%% this one withdrawn by authors
%%; arXiv:0710.5357 [hep-ph].
%  %%CITATION = IMPAE,A23,4779;%%


\bibitem{Stelzer:1994ta}
  T.~Stelzer and W.~F.~Long,
  %``Automatic generation of tree level helicity amplitudes,''
  Comput.\ Phys.\ Commun.\  {\bf 81}, 357 (1994)
  [arXiv:hep-ph/9401258].
  %%CITATION = CPHCB,81,357;%%


%--------Stop production at NNLO

%\cite{Langenfeld:2009eg}
\bibitem{Langenfeld:2009eg}
  U.~Langenfeld and S.~O.~Moch,
  %``Higher-order soft corrections to squark hadro-production,''
  Phys.\ Lett.\  B {\bf 675}, 210 (2009)
  [arXiv:0901.0802 [hep-ph]];\\
  %%CITATION = PHLTA,B675,210;%%
%
%%\cite{Kulesza:2008jb}
%\bibitem{Kulesza:2008jb}
  A.~Kulesza and L.~Motyka,
  %``Threshold resummation for squark-antisquark and gluino-pair 
  % production at the LHC,''
  Phys.\ Rev.\ Lett.\  {\bf 102}, 111802 (2009)
  [arXiv:0807.2405 [hep-ph]].
%  %%CITATION = PRLTA,102,111802;%%

%--------PYTHIA 6.4
%\cite{Sjostrand:2006za}
\bibitem{Sjostrand:2006za}
  T.~Sjostrand, S.~Mrenna and P.~Skands,
  %``PYTHIA 6.4 Physics and Manual,''
  JHEP {\bf 0605}, 026 (2006)
  [arXiv:hep-ph/0603175].
  %%CITATION = JHEPA,0605,026;%%



%%%---------Higher order electroweak neutralino mass and BR

%\cite{Oller:2005xg}
\bibitem{Oller:2005xg}
  W.~Oller, H.~Eberl and W.~Majerotto,
  %``Precise predictions for chargino and neutralino pair production in e+  e-
  %annihilation,''
  Phys.\ Rev.\  D {\bf 71} (2005) 115002
  [arXiv:hep-ph/0504109];
  %%CITATION = PHRVA,D71,115002;%%
%W.~Oller, H.~Eberl and W.~Majerotto,
  %``Full one-loop corrections to neutralino pair production in e+ e-
  %annihilation,''
  Phys.\ Lett.\  B {\bf 590} (2004) 273
  [arXiv:hep-ph/0402134];\\
  %%CITATION = PHLTA,B590,273;%%
%
%\cite{Fritzsche:2004nf}
%\bibitem{Fritzsche:2004nf}
  T.~Fritzsche and W.~Hollik,
  %``One-loop calculations for SUSY processes,''
  Nucl.\ Phys.\ Proc.\ Suppl.\  {\bf 135}, 102 (2004)
  [arXiv:hep-ph/0407095].
  %%CITATION = NUPHZ,135,102;%%


%\cite{Drees:2006um}
\bibitem{Drees:2006um}
  M.~Drees, W.~Hollik and Q.~Xu,
  %``One-loop calculations of the decay of the next-to-lightest neutralino in
  %the MSSM,''
  JHEP {\bf 0702} (2007) 032
  [arXiv:hep-ph/0610267].
  %%CITATION = JHEPA,0702,032;%%


%%%---------Divers



%\cite{Bartl:1986hp}
\bibitem{Bartl:1986hp}
  A.~Bartl, H.~Fraas and W.~Majerotto,
  %``PRODUCTION AND DECAY OF NEUTRALINOS IN e+ e- ANNIHILATION,''
  Nucl.\ Phys.\ B {\bf 278} (1986) 1.
  %%CITATION = NUPHA,B278,1;%%



\bibitem{Byckling}  E. Byckling, K. Kajantie, {\it Particle Kinematics},
                   John Wiley\& Sons, 1973.

%-------------narrow width
\bibitem{narrowwidth}
%\cite{Hagiwara:2005wg}
%\bibitem{Hagiwara:2005wg}
  K.~Hagiwara {\it et al.},
  %``Supersymmetry simulations with off-shell effects for LHC and ILC,''
  Phys.\ Rev.\  D {\bf 73} (2006) 055005
  [arXiv:hep-ph/0512260];\\
  %%CITATION = PHRVA,D73,055005;%%
%
%\cite{Berdine:2007uv}
%\bibitem{Berdine:2007uv}
  D.~Berdine, N.~Kauer and D.~Rainwater,
  %``Breakdown of the Narrow Width Approximation for New Physics,''
  Phys.\ Rev.\ Lett.\  {\bf 99}, 111601 (2007)
  [arXiv:hep-ph/0703058]; \\
  %%CITATION = PRLTA,99,111601;%%
%%\cite{Kauer:2007zc}
%%\bibitem{Kauer:2007zc}
  N.~Kauer,
  %``Narrow-width approximation limitations,''
  Phys.\ Lett.\  B {\bf 649}, 413 (2007)
  [arXiv:hep-ph/0703077];\\
  %%CITATION = PHLTA,B649,413;%%
%\cite{Kauer:2007nt}
%\bibitem{Kauer:2007nt}
%  N.~Kauer,
  %``A threshold-improved narrow-width approximation for BSM physics,''
  JHEP {\bf 0804}, 055 (2008)
  [arXiv:0708.1161 [hep-ph]];\\
  %%CITATION = JHEPA,0804,055;%%
%\cite{Uhlemann:2008pm}
%\bibitem{Uhlemann:2008pm}
  C.~F.~Uhlemann and N.~Kauer,
  %``Narrow-width approximation accuracy,''
  Nucl.\ Phys.\  B {\bf 814}, 195 (2009)
  [arXiv:0807.4112 [hep-ph]];\\
  %%CITATION = NUPHA,B814,195;%%
%\cite{Gigg:2008yc}
%\bibitem{Gigg:2008yc}
  M.~A.~Gigg and P.~Richardson,
  %``Simulation of Finite Width Effects in Physics Beyond the Standard Model,''
  arXiv:0805.3037 [hep-ph].
  %%CITATION = ARXIV:0805.3037;%%


%%%%%--------spin formalism Kawasaki

%\cite{Kawasaki:1973hf}
\bibitem{Kawasaki:1973hf}
  S.~Kawasaki, T.~Shirafuji and S.~Y.~Tsai,
  %``Productions and decays of short-lived particles in e+ e- colliding beam
  %experiments,''
  Prog.\ Theor.\ Phys.\  {\bf 49} (1973) 1656.
  %%CITATION = PTPKA,49,1656;%%


%%%%%--------spin formalism gudi
\bibitem{gudi}
For chargino/neutralino production in the spin-density matrix formalism, see\\
%\cite{MoortgatPick:1998sk}
%\bibitem{MoortgatPick:1998sk}
  G.~A.~Moortgat-Pick, H.~Fraas, A.~Bartl and W.~Majerotto,
  %``Spin Correlations in Production and Decay of Charginos,''
  Eur.\ Phys.\ J.\  C {\bf 7}, 113 (1999)
  [arXiv:hep-ph/9804306];
  %%CITATION = EPHJA,C7,113;%%
%
%\bibitem{Moortgat-Pick:1999di}
%  G.~A.~Moortgat-Pick, H.~Fraas, A.~Bartl and W.~Majerotto,
  %``Polarization and spin effects in neutralino production and decay,''
  Eur.\ Phys.\ J.\ C {\bf 9}, 521 (1999)
  [Erratum-ibid.\ C {\bf 9}, 549 (1999)]
  [arXiv:hep-ph/9903220].
  %%CITATION = HEP-PH 9903220;%%


\bibitem{private}
J.~A.~Aguilar-Saavedra, private discussions.

%-----------------------------------------------------------------------------
\end{thebibliography}
\end{document}